\documentclass[twocolumn,numberedappendix,appendixfloats,twocolappendix]{aastex6}

\usepackage{graphicx,amsmath,amsfonts,amssymb,subfigure,hyperref}

\newcommand{\fsps}{\texttt{FSPS}}
\newcommand{\emcee}{\texttt{emcee}}
\newcommand{\pythonfsps}{\texttt{python-fsps}}

\newcommand{\mname}{\textsc{Prospector-$\alpha$}}
\newcommand{\scipy}{\texttt{scipy}}
\newcommand{\prospector}{\texttt{Prospector}}
\newcommand{\cloudy}{\texttt{CLOUDY}}
\newcommand{\magphys}{\texttt{MAGPHYS}}
\newcommand{\dustone}{\hat{\tau}_{\lambda,1}}
\newcommand{\dusttwo}{\hat{\tau}_{\lambda,2}}
\newcommand{\didx}{n}
\newcommand{\halpha}{H$\alpha$}
\newcommand{\hbeta}{H$\beta$}
\newcommand{\hdelta}{H$\delta$}
\newcommand{\dn}{D$_{\mathrm{n}}$4000}
\newcommand{\oiii}{{\sc [O~iii]}}
\newcommand{\oii}{{\sc [O~ii]}}
\newcommand{\nii}{{\sc [N~ii]}}
\newcommand{\umin}{U$_\mathrm{min}$}
\newcommand{\gammae}{$\gamma_{\mathrm{e}}$}
\newcommand{\qpah}{Q$_{\mathrm{PAH}}$}
\newcommand{\herschel}{\textit{Herschel}}
\newcommand{\spitzer}{\textit{Spitzer}}
\newcommand{\angstrom}{\mbox{\normalfont\AA}}

\newcommand{\lir}{{L$_{\mathrm{IR}}$}}
\newcommand{\luv}{{L$_{\mathrm{UV}}$}}
\newcommand{\lpah}{{L$_{\mathrm{PAH}}$}}

\begin{document}

\title{Deriving Physical Properties from Broadband Photometry with Prospector: Description of the Model and a Demonstration of its Accuracy using 129 Galaxies in the Local Universe}

\author{Joel Leja\altaffilmark{1}, Benjamin D. Johnson\altaffilmark{1}, Charlie Conroy\altaffilmark{1}, Pieter G. van Dokkum\altaffilmark{2}, Nell Byler\altaffilmark{3}}

\altaffiltext{1}{Harvard-Smithsonian Center for Astrophysics, 60 Garden St., Cambridge, MA 02138, USA}
\altaffiltext{2}{Department of Astronomy, Yale University, New Haven, CT 06511, USA}
\altaffiltext{3}{Department of Astronomy, University of Washington, Seattle, WA 98185, USA}

\begin{abstract}
Broadband photometry of galaxies measures an unresolved mix of complex stellar populations, gas, and dust. Interpreting these data is a challenge for models: many studies have shown that properties derived from modeling galaxy photometry are uncertain by a factor of two or more, and yet answering key questions in the field now requires higher accuracy than this. Here, we present a new model framework specifically designed for these complexities. Our model, \mname{}, includes dust attenuation and re-radiation, a flexible attenuation curve, nebular emission, stellar metallicity, and a 6-component nonparametric star formation history. The flexibility and range of the parameter space, coupled with MCMC sampling within the \prospector{} inference framework, is designed to provide unbiased parameters and realistic error bars. We assess the accuracy of the model with aperture-matched optical spectroscopy, which was excluded from the fits. We compare spectral features predicted solely from fits to the broadband photometry to the observed spectral features. Our model predicts \halpha{} luminosities with a scatter of $\sim$0.18 dex and an offset of $\sim$0.1 dex across a wide range of morphological types and stellar masses. This agreement is remarkable, as the \halpha{} luminosity is dependent on accurate star formation rates, dust attenuation, and stellar metallicities. The model also accurately predicts dust-sensitive Balmer decrements, spectroscopic stellar metallicities, PAH mass fractions, and the age- and metallicity-sensitive features \dn{} and \hdelta{}. Although the model passes all these tests, we caution that we have not yet assessed its performance at higher redshift or the accuracy of recovered stellar masses. 
\end{abstract}

\section{Introduction}
One of the primary goals of galaxy evolution studies is to understand how galaxies assembled their stars over time. The most straightforward way to accomplish this is to measure star formation rates (SFRs) and stellar masses as a function of redshift for representative samples of the galaxy population. This has now been done by many studies out to high redshift, both for the stellar mass function (e.g., {Marchesini} {et~al.} 2009; {Muzzin} {et~al.} 2013; {Ilbert} {et~al.} 2013; {Moustakas} {et~al.} 2013; {Davidzon} {et~al.} 2013; {Tomczak} {et~al.} 2014) and for the SFR-mass relationship (e.g., {Brinchmann} {et~al.} 2004; {Noeske} {et~al.} 2007; {Daddi} {et~al.} 2007; {Peng} {et~al.} 2010; {Karim} {et~al.} 2011; {Whitaker} {et~al.} 2012; {Speagle} {et~al.} 2014; {Whitaker} {et~al.} 2014; {Ilbert} {et~al.} 2015).

The star-forming sequence and stellar mass function are derivative-integral pairs, and as a consistency check, the stellar mass growth rates as a function of mass and redshift implied by each can be compared. This approach minimizes systematics by using star formation rates and masses derived from the same data sets, and in contrast to techniques which integrate over mass, is largely insensitive to the mass completeness limits ({Leja} {et~al.} 2015; {Tomczak} {et~al.} 2016). This technique finds broad consistency at $z < 1$ ({Bell} {et~al.} 2007; {Weinmann} {et~al.} 2012): however, at $z = 1 - 3$, the observed star formation rates and stellar masses are systematically incompatible with one another by a factor of $\sim$2 ({Leja} {et~al.} 2015; {Tomczak} {et~al.} 2016; {Contini} {et~al.} 2016). The most straightforward explanation is that these systematics are arise from the models used to convert observed fluxes into stellar masses and/or SFRs.

Compilations of galaxy star formation rates and masses in the literature tell a more complex story.  Comparisons of observed star formation rate densities to the evolution of the stellar mass density show systematic differences of a factor of $1.5-2$ ({Madau} \& {Dickinson} 2014; {Yu} \& {Wang} 2016). Empirical abundance matching models can appear able to reconcile these two measurements (e.g., {Behroozi}, {Wechsler}, \&  {Conroy} 2013; {Moster}, {Naab}, \& {White} 2013).  These models contain a greater degree of complexity, which may be important in reconciling SFRs and stellar mass densities, but some of these additional model ingredients are yet to be thoroughly tested.  Part of the challenge associated with interpreting literature compilations is that systematic uncertainties vary across techniques, and so combining literature results likely washes away substantial systematic offsets.

The apparent inconsistencies in observational quantities make it difficult for simulations of galaxy evolution to simultaneously satisfy observed SFR and mass constraints. Hydrodynamical simulations (e.g., {Genel} {et~al.} 2014; {Torrey} {et~al.} 2014; {Furlong} {et~al.} 2015), semi-analytical models (e.g., {Mitchell} {et~al.} 2014; {Henriques} {et~al.} 2015), and pure analytic models (e.g., {Lilly} {et~al.} 2013; {Dekel} \& {Burkert} 2014) are all systematically offset from observations of masses, SFRs, or both at the level of $0.2-0.5$ dex. This is a particularly interesting comparison because simulated star formation rates and stellar masses are self-consistent by definition. 

This systematic factor of two uncertainty in basic galaxy properties has direct implications for our understanding of how galaxies assemble their stars. Massive galaxies are thought to double in stellar mass from $z = 2$ to the present epoch (e.g., {van Dokkum} {et~al.} 2010; {Patel} {et~al.} 2013a). It is not possible to assess the uncertainty in this result without accurate models, nor in the evolution (or lack thereof) of the massive end of the mass function (e.g. {Moustakas} {et~al.} 2013). Milky Way-mass galaxies are thought to grow by a factor of $\sim$5 since $z = 2$, and M31-mass galaxies by $\sim$3, with most of the growth occurring between $z = 1$ and $z = 2$ ({van Dokkum} {et~al.} 2013; {Patel} {et~al.} 2013b; {Papovich} {et~al.} 2015). Systematic adjustment of stellar masses and star formation rates at $z = 1 - 3$ could change these results significantly.

Furthermore, in order to convert dynamical measurements into the dark matter fractions within galaxies, it is critical to know the stellar masses to within a factor of two (e.g., {Cappellari} {et~al.} 2012). A systematic factor of two uncertainty in stellar mass results in no constraint for the dark matter fraction within the average quiescent ({van de Sande} {et~al.} 2015) or star-forming ({Wuyts} {et~al.} 2016) galaxy at $z = 1 - 3$.

In order to advance the field, basic stellar mass and star formation rate estimates must be improved. The dominant source of uncertainty in stellar populations fitting is no longer instrumental noise, sample size, or sample selection, but instead are the degeneracies and limitations in synthesizing and fitting complex stellar populations to spectral energy distributions (SEDs) ({Conroy}, {Gunn}, \& {White} 2009; {Wuyts} {et~al.} 2009; {Behroozi}, {Conroy}, \&  {Wechsler} 2010; {Walcher} {et~al.} 2011; {Mobasher} {et~al.} 2015; {Santini} {et~al.} 2015).

These parameter degeneracies arise because galaxy SEDs are the result of many complex physical variables, and it is challenging to uniquely determine these variables in individual galaxies with only broadband photometry. The age-dust-metallicity degeneracy is a well-known example of this ({Bell} \& {de Jong} 2001), and uncertainty in these parameters can result in systematic changes to both stellar mass and SFR estimates. Another example is the shape of the dust attenuation curve, which is thought to change from galaxy to galaxy based on both dust geometry and composition ({Witt} \& {Gordon} 2000). Since the effect of extinction is strongest at blue wavelengths, there is a strong degeneracy between the shape of the attenuation curve and the total column density of dust, with most studies requiring spectra or infrared photometry to separate the two ({Johnson} {et~al.} 2007; {Wild} {et~al.} 2011; {Reddy} {et~al.} 2015). These parameter degeneracies complicate comparisons between different codes, because the outputs are strongly dependent on the priors in each SED fitting routine. The prior of a model parameter is the probability distribution which expresses the belief about the distribution of a parameter before the observations are fit: priors can either be explicitly applied in the likelihood calculation, or implicitly applied in the chosen parameterization of the variable, e.g. by using a decaying $\tau$ model to fit star formation histories. Complicating the situation further, the priors for each model are often not clearly defined or explored.

There are also substantial uncertainties in the underlying stellar populations synthesis models. One challenge is the difficulty in sourcing accurate spectra for short-lived phases of stellar evolution and nonsolar stellar metallicities. For example, the contribution of thermally-pulsing asymptotic giant branch stars (TP-AGB) may or may not dominate the near-IR emission of galaxies of intermediate age galaxies ({Maraston} 2005; {Marigo} {et~al.} 2008; {Conroy} \& {Gunn} 2010). This uncertainty alone can result in systematic differences of a factor of two in both stellar mass and age estimates ({Bruzual} 2007), though recent analysis imply that the data favor a relatively low TP-AGB contribution (e.g., {Kriek} {et~al.} 2010).

Galaxy SED-fitting routines thus far have largely used simple models to derive stellar masses, with fixed or discrete dust attenuation curves, limited or fixed stellar metallicities, simple parameterized star formation histories (SFHs), and simple chi-squared minimization routines ({Bolzonella}, {Miralles}, \&  {Pell{\'o}} 2000; {Brammer}, {van Dokkum}, \&  {Coppi} 2008; {Kriek} {et~al.} 2009). Yet, observed galaxies show significant variation in stellar metallicity ({Tremonti} {et~al.} 2004) and attenuation curves ({Reddy} {et~al.} 2015; {Salmon} {et~al.} 2016), and neglecting this variation can bias the parameters determined from SED fitting ({Mitchell} {et~al.} 2013). SFH recovery is critical to accurate recovery of stellar masses ({Lee} {et~al.} 2009; {Maraston} {et~al.} 2010; {Pforr}, {Maraston}, \& {Tonini} 2012), and simple exponentially declining SFHs fail to reproduce the mass function of galaxies at earlier epochs by several orders of magnitude ({Wuyts} {et~al.} 2011b). They also fit simulated SFHs poorly, leaving significant biases in stellar mass estimates ({Simha} {et~al.} 2014). Simple chi-squared fitting routines cannot properly estimate the uncertainties when there are substantial degeneracies, as is the case when fitting broadband photometry of galaxies. Progress has been made in fitting SEDs with nonparametric SFHs (e.g., STECMAP ({Ocvirk} {et~al.} 2006), VESPA ({Tojeiro} {et~al.} 2007), LITTLE THINGS ({Zhang} {et~al.} 2012), CSI ({Kelson} 2014; {Dressler} {et~al.} 2016)); however, this practice is not yet widespread, as nonparametric SFHs require either very high signal-to-noise data or a fitting algorithm which can handle significant degeneracies.

It has also been shown that the SFRs derived from fitting the UV-NIR SED are often biased low for highly star-forming galaxies ({Wuyts} {et~al.} 2011a), largely because they cannot account for star formation obscured by dust. SFRs are instead often derived separately, with simple ``recipes'' to convert observed UV, MIR, or emission line fluxes into SFRs (e.g., {Kennicutt} 1998). This approach also suffers from known biases: the conversion from emission line fluxes to SFRs is affected by the ionization state of the gas, extra dust attenuation towards HII regions, and stellar metallicity. Inferring total IR luminosities with only MIR broadband photometry can be strongly affected by galaxy-to-galaxy variability in polycyclic aromatic hydrocarbon (PAH) emission in the MIR ({Draine} {et~al.} 2007) and by AGN contribution ({Kirkpatrick} {et~al.} 2015). The interpretation of \lir{} itself is complicated by the contribution of evolved stars to the IR luminosity ({Cortese} {et~al.} 2008; {Hayward} {et~al.} 2014; {Utomo} {et~al.} 2014). UV luminosities are sensitive to stellar metallicity and recent SFH and very sensitive to the amount of dust attenuation, which can be estimated from the UV slope $\beta$ but with serious limitations ({Viaene} {et~al.} 2016).

Many of the issues in estimating SFRs with ``recipes'' can be avoided by instead performing a full-SED fit, which consolidates all available information regarding the physical condition of the stars, dust, and gas. Parameter degeneracies can be marginalized over with Bayesian inference techniques and physically motivated priors. Some codes have begun to take advantage of Bayesian approaches in fitting SEDs: \texttt{CIGALE} ({Burgarella}, {Buat}, \&  {Iglesias-P{\'a}ramo} 2005; {Noll} {et~al.} 2009), \magphys{} ({da Cunha}, {Charlot}, \&  {Elbaz} 2008), \texttt{GALMC} ({Acquaviva}, {Gawiser}, \&  {Guaita} 2011), \texttt{BAYESED} ({Han} \& {Han} 2014), and \texttt{BEAGLE} ({Chevallard} \& {Charlot} 2016). Yet due to the complexity in choosing fit parameters, priors, and stellar population synthesis techniques, SED fitting codes with multiple, degenerate parameters often produce different solutions for key physical parameters like stellar mass and star formation rates ({Santini} {et~al.} 2015). This motivates a thorough investigation into the accuracy of the posteriors in SFRs, stellar masses, and other key physical parameters determined from broadband photometry.

In this paper, we create and test the \mname{} model within the \prospector{} inference framework (Johnson et al. in prep) using the Flexible Stellar Populations Synthesis (\fsps{}) stellar populations code ({Conroy} {et~al.} 2009). The \mname{} model includes many of the advances mentioned above in a single, self-consistent framework. The model is fit to the broadband photometry from the {Brown} {et~al.} (2014) spectral atlas, which is unique in that it has optical spectra which are aperture-matched to the photometry. These data allow us to probe the accuracy of our model fits by comparing posterior probability functions (i.e., probability distributions for parameters of interest after the data are taken into account) for diagnostic spectral features to the observed features from the aperture-matched spectroscopy. This posterior check is a strong test of the output star formation rates, star formation history, dust attenuation model, and stellar metallicities derived from fitting the \mname{} model to broadband photometry.

In Section \ref{section:data}, the galaxy sample is described, along with the available broadband photometric and spectroscopic observations. In Section \ref{section:features}, the \mname{} model is introduced, and the \prospector{} sampling procedure is described. In Section \ref{section:results}, the comparison between the \mname{} model posteriors and the observed spectroscopic quantities is shown, including the \halpha{} and \hbeta{} luminosities, Balmer decrements, \dn{}, \hdelta{} absorption, and stellar metallicities. In Section \ref{section:discussion}, the implications of the posterior checks are discussed, and past and future changes to the \mname{} model are discussion in Section \ref{section:future}. The conclusion is presented in Section \ref{section:conclusion}. A WMAP9 cosmology ({Hinshaw} {et~al.} 2013) and a {Chabrier} (2003) initial mass function (IMF) are adopted for all relevant calculations.

\section{Data}
\label{section:data}
\begin{figure*}[th!]
\begin{center}
\includegraphics[width=0.95\linewidth]{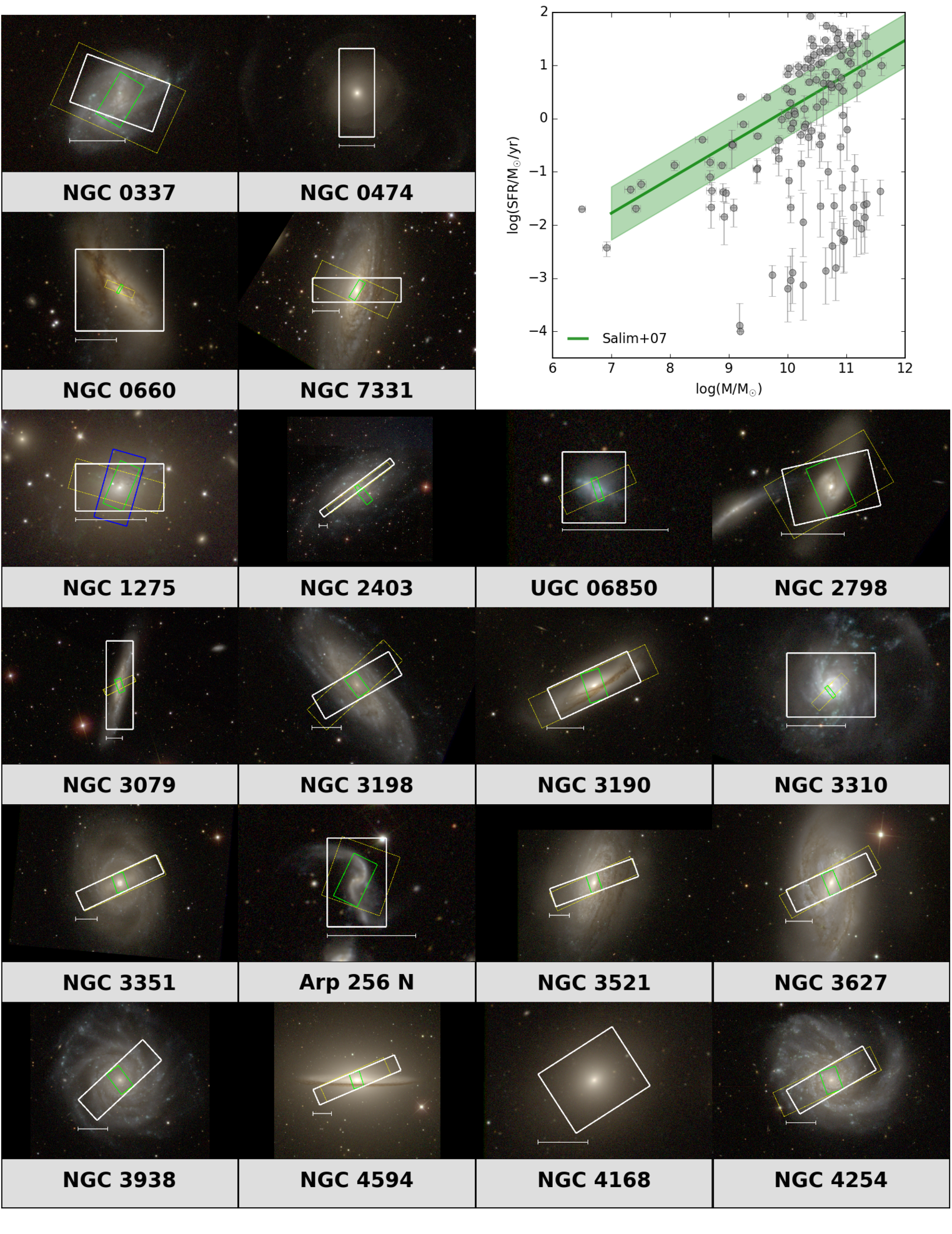}
\caption{Colorized images of a subset of galaxies in the sample, chosen to illustrate the diversity of morphologies and optical colors. The cutouts are taken from the online catalog provided by {Brown} {et~al.} (2014). The apertures used for both the photometry and optical spectra are shown in white, the Akari (when available) apertures are shown in blue, and the \spitzer{} SL/LL apertures (when available) are shown in green/yellow. The location of each galaxy on the star-forming sequence (from the \mname{} fits) is shown in the upper-right. The white bar denotes an angular size of 1'. The star-forming sequence from {Salim} {et~al.} (2007) is shown for reference. We note that the {Brown} {et~al.} (2014) catalog is not a volume-limited sample.}
\label{fig:intro}
\end{center}
\end{figure*}

\subsection{Broadband Photometry}
We use broadband photometry and spectra from the {Brown} {et~al.} (2014) spectral atlas of 129 local galaxies. There are 26 photometric bands available from a range of surveys and missions, including {\it Swift}/UVOT, {\it Galaxy Evolution Explorer} (GALEX), the Sloan Digital Sky Survey (SDSS), the Two Micron All Sky Survey (2MASS), \spitzer{}, and the {\it Wide-field Infrared Space Explorer} (WISE), giving wavelength coverage from the far-UV to the mid infrared (MIR).

Figure \ref{fig:intro} shows color cutouts of the sample from the online data catalog provided by {Brown} {et~al.} (2014), along with the star formation rates and stellar masses relative to the star-forming sequence. The sample covers a wide range of morphological types and stellar masses. The sample selection of the {Brown} {et~al.} (2014) catalog is based on the availability of the aperture-matched spectra, and is not volume-complete.

Considerable care was taken with the absolute calibration of the photometry: the atlas includes corrections for foreground dust extinction, SDSS photometric calibration, scattered light in IRAC imaging, and errors in the pre-launch WISE W4 filter curve ({Brown} {et~al.} 2014). The photometry is aperture-matched across all filters.

{\it Swift}/UVOT U and V bands and \spitzer{}/IRS Blue and Red Peak Up imaging channels are not included in the fits. These spectral regions are well-covered by SDSS optical imaging and \spitzer{}/MIPS and WISE imaging, respectively.

For 26 of these galaxies, we add \herschel{} PACS and SPIRE imaging from the KINGFISH survey ({Kennicutt} {et~al.} 2011), spanning 70-500 $\mu$m. Galaxy emission at these wavelengths is dominated by thermal dust emission, which in the \mname{} model, is self-consistently modeled with the stellar emission. The effect of the \herschel{} photometry on the model posteriors is explored in Appendix \ref{section:herschel}.
\subsection{Spectra}
We utilize optical spectra from multiple ground-based telescopes, mid-infrared \spitzer{} spectra ({Houck} {et~al.} 2004), and {\it Akari} spectra ({Onaka} {et~al.} 2007), all provided in the {Brown} {et~al.} (2014) spectral atlas. The optical spectra largely come from {Moustakas} \& {Kennicutt} (2006) and {Moustakas} {et~al.} (2010), with some additional spectra from {Kennicutt} (1992) and {Gavazzi} {et~al.} (2004). The resolution of the optical spectra is
is $R \sim 650$ and the wavelength coverage spans 3650 to 6900 \angstrom{}.

The assembly of these spectra are described in detail in {Brown} {et~al.} (2014) and in references therein. In brief, the optical spectra use drift-scan techniques, allowing them to be aperture-matched to the photometry. The spatial matching between the optical spectroscopy and the photometry is a key feature of the atlas: these two data products describe the same stellar populations without the uncertainty typically introduced by aperture corrections. The spectra have also been flux-normalized to be consistent with the photometry. These features allow direct comparison of the posteriors from fitting the photometry to the observed spectral features.

We measure the strength of key diagnostic emission and absorption features from the spectra, and compare these to expectations from fits to the photometry. For each galaxy, we measure the following features: \hdelta{}, \oiii{} 4959,\oiii{} 5007, \hbeta{},\nii{} 6549, \nii{} 6583, \halpha{}, and \dn{}. To ensure uniform quality, these spectral quantities are only measured when the spectra in question have a minimum resolution of $R = 400$ at \halpha{} wavelengths. This removes IRAS 08572+3915, NGC 4860, and IRAS 17208-0014 from the spectral analysis.

\subsubsection{Measuring Optical Emission Line Fluxes}
\label{section:optical_fluxes}
In order to extract emission line fluxes, we perform simultaneous fits to the \oiii{} 4959, \oiii{} 5007, \hbeta{}, \nii{} 6549, \nii{} 6583, and \halpha{} features. The simultaneous fits allow overlapping lines to be properly separated. Each emission and absorption feature is modeled with a Gaussian, with a variable amplitude and width. The center of each line is fixed relative to one another, but the overall redshift of the model is allowed to vary around the fiducial redshift provided by the {Brown} {et~al.} (2014) catalog. For doublet emission lines (\oiii{} and \nii{}), the relative amplitudes of the emission lines are fixed to their fundamental ratios, and the line widths are also fixed to be equal.

The best-fit \mname{} spectrum (see Section \ref{section:spec_residuals}) is adopted as the stellar continuum. A simple approach to measuring emission line strength from spectra in the case of mixed absorption and emission is to separately measure the underlying absorption in the model stellar continuum, and add that to the observed emission line strength. However, the observed spectra often show gas emission resolved separately from stellar absorption, especially when the width of the stellar absorption line profile is dominated by age-dependent stellar broadening. In these cases, the line emission fills the center of the line, but the wings of the absorption are not filled with emission. To account for this complex line profile, the underlying stellar continuum is subtracted from the observed spectra, and the emission line flux is measured from the residuals.

Before subtracting the stellar continuum model, it must first be smoothed to the appropriate resolution and normalized to the observed spectrum. The observed spectral resolution is controlled by instrumental broadening, atmospheric effects, and the intrinsic line-of-sight velocity dispersion. The appropriate smoothing resolution is determined by performing a first-pass fit to the observed spectrum with the Gaussian model for absorption and emission lines. The two emission lines with the highest equivalent width are selected, and the average of their line-widths is used to smooth the model spectrum. This implicitly assumes the gas and stars have similar velocity dispersions. The normalization is done by fitting two separate linear components for the continuum: one between 4700 \angstrom{} $ < \lambda < 5100 $ \angstrom{}, and one between 6350 \angstrom{} $< \lambda < 6680$ \angstrom{}. This is done for both the model spectrum from the best-fit to the photometry and the observed spectrum, and the ratio between the linear components is used to renormalize the model spectrum separately in each wavelength range. Since the model is fit to the photometry, and the observed photometry and observed spectra have consistent normalizations by design ({Brown} {et~al.} 2014), this normalization factor is typically $<$5\%. 

The {Brown} {et~al.} (2014) spectral atlas does not include error estimates for the spectral fluxes. To estimate errors, the smoothed, normalized stellar continuum model is subtracted from the observed spectra, with known emission lines masked. A Gaussian is fit to the histogram of the residuals. The error on the spectra fluxes is taken to be the width of the Gaussian. This error is typically between 3-15\% of the flux. This provides a conservative estimate of the true instrumental noise, as it also includes inevitable mismatches between the model spectrum and the observed spectrum.

To derive errors on the line fluxes, a bootstrap analysis is performed: for each pixel, the flux is perturbed with a random value drawn from a Gaussian centered at zero, with a width equal to the errors derived above, then the line fluxes are re-fit. This is performed 100 times for each galaxy, and the final line fluxes are taken to be the 50th percentile of this distribution, with the +/- 1$\sigma$ errors taken to be the 84th and 16th percentiles, respectively. For \halpha{} and \hbeta{}, an additional error due to uncertainty in the absorption correction is added, equal to the 1$\sigma$ variance in the underlying model stellar absorption as sampled from the model posteriors.

\subsubsection{\dn{}}
\dn{} is measured using the {Balogh} {et~al.} (1999) definition: the ratio of the average flux density F$_{\nu}$ in the narrow wavelength bands 3850-3950 $\angstrom{}$ and 4000-4100 $\angstrom{}$. This is narrower and thus less sensitive to reddening effects than the original {Bruzual} 1983; {Hamilton} 1985 definition. This definition is used to measure \dn{} in both the observed and the model spectra, and both are convolved to a common resolution of 200 km/s before measurement. This is done to remove any potential systematics from different spectral resolutions between the model and the observed spectrum.

\subsubsection{\hdelta{}}
\hdelta{} is measured separately from the other line features. This is done for two reasons: (1) the \hdelta{} emission and absorption are often of comparable size, resulting in a complex line morphology with a weak net signal (often poorly fit by a Gaussian), and (2) because nearby CN features affect simple linear continuum fits, which introduces a dependence on the detailed elemental abundance ratios ({Prochaska} {et~al.} 2007). The relatively low S/N of the optical spectra also make the measurement of the intrinsically weak \hdelta{} feature challenging.

To measure \hdelta{}, fixed wavelength indices are defined over which to measure the continuum and the strength of the \hdelta{} feature (e.g., {Worthey} {et~al.} 1994; {Worthey} \& {Ottaviani} 1997; {Nelan} {et~al.} 2005). Defining these indices is challenging, since the intrinsic \hdelta{} absorption line profile is sensitive to the age of the composite stellar populations ({Prochaska} {et~al.} 2007). For example, K-dwarf stars have \hdelta{} FWHMs of $\sim$1-2 \AA, while A stars have FWHMs of $\sim$20\AA, with the wings of the absorption profile extending $\sim$50 \AA{} from the center ({Prochaska} {et~al.} 2007).  

In order to accurately measure the strength of the \hdelta{} feature for a variety of line profile shapes, we thus adopt a wide and narrow set of indices. The narrow set of indices measures \hdelta{} as the average flux between 4095 \angstrom{} $< \lambda <$ 4114 \angstrom{}, the blue continuum as the average flux between 4072 \angstrom{} $< \lambda <4095$ \angstrom{}, and the red continuum as the average flux between 4114 \angstrom{} $< \lambda <4130$ \angstrom{}. The broad set of indices measures \hdelta{} as the average flux between 4082 \angstrom{} $< \lambda <4124$ \angstrom{}, the blue continuum as the average flux between 4042 \angstrom{} $< \lambda <4082$ \angstrom{}, and the red continuum as the average flux between 4124 \angstrom{} $< \lambda <4151$ \angstrom{}.

In accordance with the age dependence of the line profile, the wide index is used to measure the strength of the \hdelta{} feature for galaxies with a half-mass assembly time of $<8$ Gyr, while the narrow index is adopted for galaxies with a half-mass assembly time of $>8$ Gyr. The half-mass assembly time is measured directly from the model posteriors for each galaxy. This classification scheme qualitatively correlates the width of the \hdelta{} feature in the observed spectra. The same index to measure the \hdelta{} strength is used in both the model and the observations, and the model measurement is performed at the same 200 km/s resolution. In the observed spectra, this method of measuring line fluxes inherently measures the sum of \hdelta{} absorption and emission. Errors are derived with the same bootstrapping method described in Section \ref{section:optical_fluxes}.

\begin{figure*}[ht!]
\begin{center}
\includegraphics[width=0.95\linewidth]{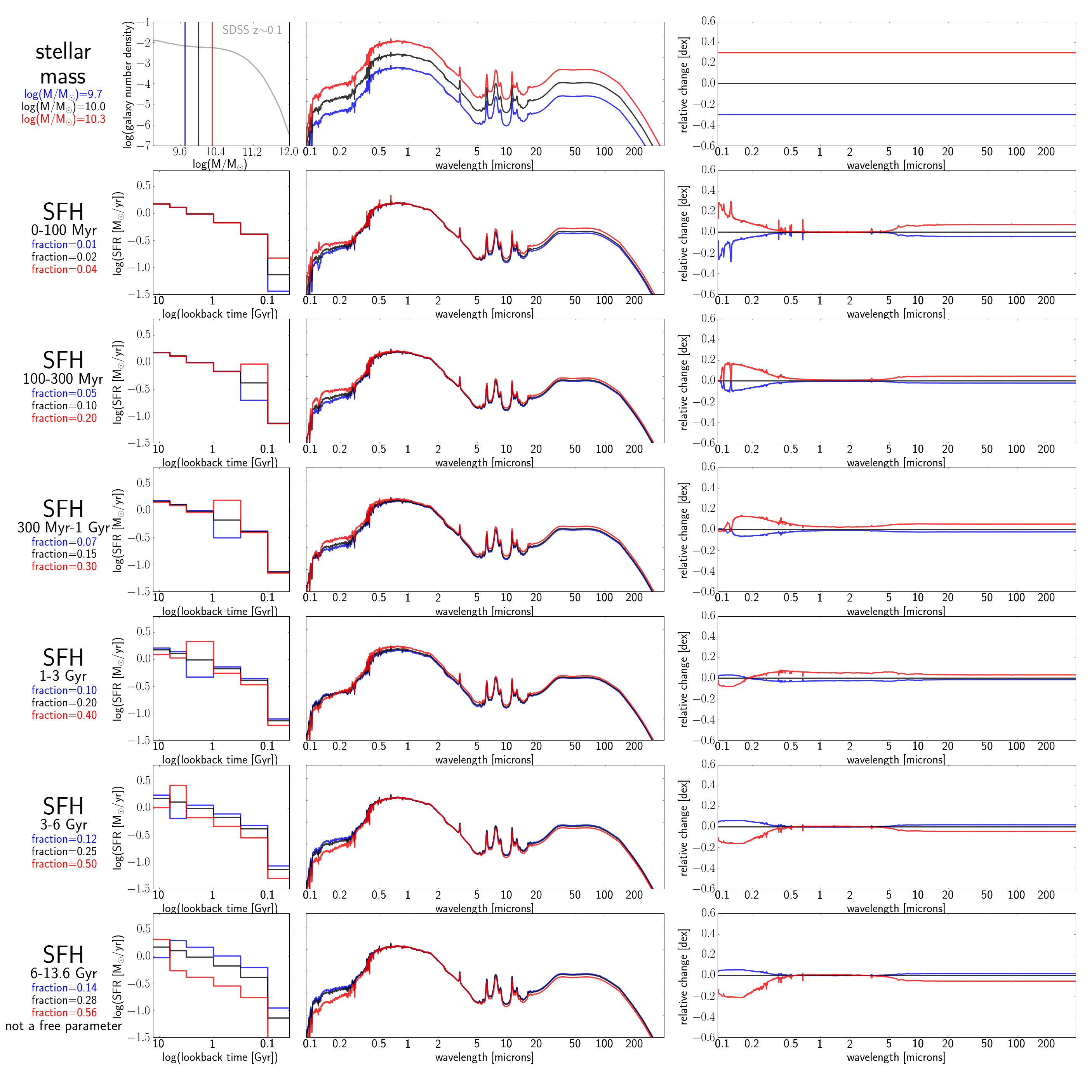}
\caption{Six of the thirteen free parameters in the \mname{} model. For each parameter, we show from left to right: a panel showing the physical significance of the given parameter, a model SED with three different values of the given parameter, and the relative change in the SED for these three different values. The first panel shows the stellar mass function for stellar mass, and the full history of star formation for the five SFH bins. We note that the sixth SFH bin is not a free parameter, as it is inferred from the values of the other five bins (see Section \ref{section:sfh} for further details).}
\label{fig:model_diagram1}
\end{center}
\end{figure*}

\begin{figure*}[h!]
\begin{center}
\includegraphics[width=0.95\linewidth]{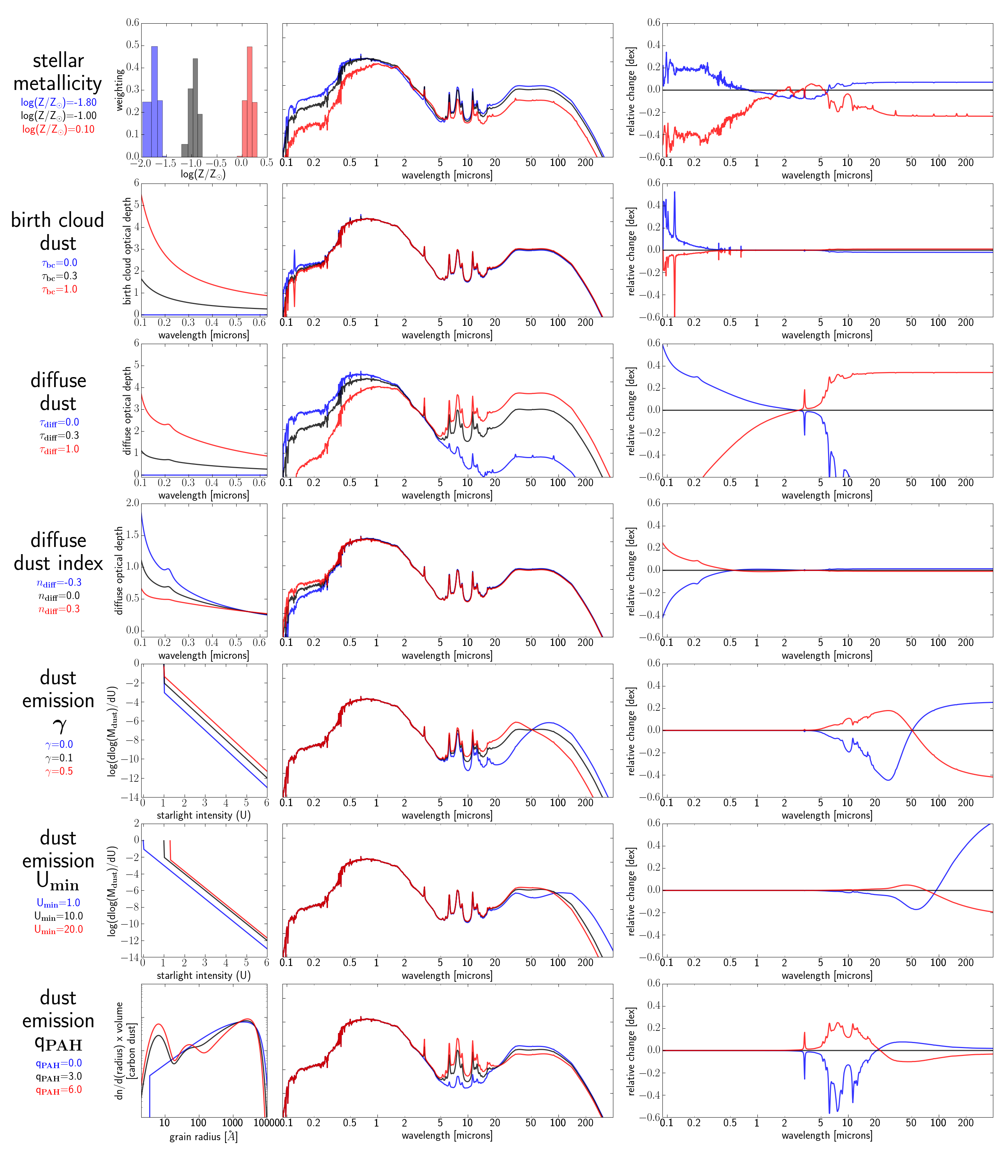}
\caption{Seven of the thirteen free parameters in the \mname{} model. For each parameter, we show from left to right: a panel showing the physical significance of the given parameter, a model SED with three different values of the given parameter, and the relative change in the SED for three different values of the given parameter. For the stellar metallicity, the first panel shows the triangular weighting scheme used in the metallicity distribution function. For the dust absorption parameters, the first panel shows the optical depth for a single component of the dust model. For the dust emission parameters, the first panel shows either the distribution of starlight intensity incident on the dust or the grain size distribution for carbon dust.}
\label{fig:model_diagram2}
\end{center}
\end{figure*}

\section{Definition of the Physical Model and Inference of Model Posteriors from SEDs}
The process of generating and fitting a galaxy model to an observed SED is described here. Section \ref{section:features} describes the relevant fixed and free parameters in our physical model for galaxies, \mname{}, and the associated priors for each parameter. Section \ref{section:methodology} describes the methodology used to fit this model to the observed galaxy photometry.

\subsection{Main Features of the \mname{} Model}
\label{section:features}
Here the salient features of our physical model for galaxies are described, including all relevant free and fixed parameters, and the priors for each free parameter. A simple illustration of the effect of each model parameter on the SED is shown in Figs. \ref{fig:model_diagram1} and \ref{fig:model_diagram2}. The physical origin of each parameter is described, and the change in the SED from varying each free parameter while holding the others fixed is shown. The example SED in these figures is a relatively quiescent galaxy, similar to the Milky Way (sSFR $\sim 1 \times 10^{-11}$ yr$^{-1}$): we note that the effect of some parameters (e.g. birth-cloud dust, stellar metallicity) is highly dependent on the sSFR.

For stellar population synthesis, the Flexible Stellar Population Synthesis (\fsps{}) package\footnote{\url{https://github.com/cconroy20/fsps}} is used ({Conroy} {et~al.} 2009). The \pythonfsps{} \footnote{\url{https://github.com/dfm/python-fsps}} package is used to communicate with \fsps{} in a python interface. All of the parameters and choices described below are implemented in the \fsps{} source code. We use the default SPS parameters in \fsps{}\footnote{Github commit hash \texttt{77fb45841476d44b9106aed021aa1e2c76773c30}}, except the MILES stellar library is adopted. 

\subsubsection{Stellar Mass and Star Formation History}
\label{section:sfh}
\begin{figure}[ht!]
\begin{center}
\includegraphics[width=0.95\linewidth]{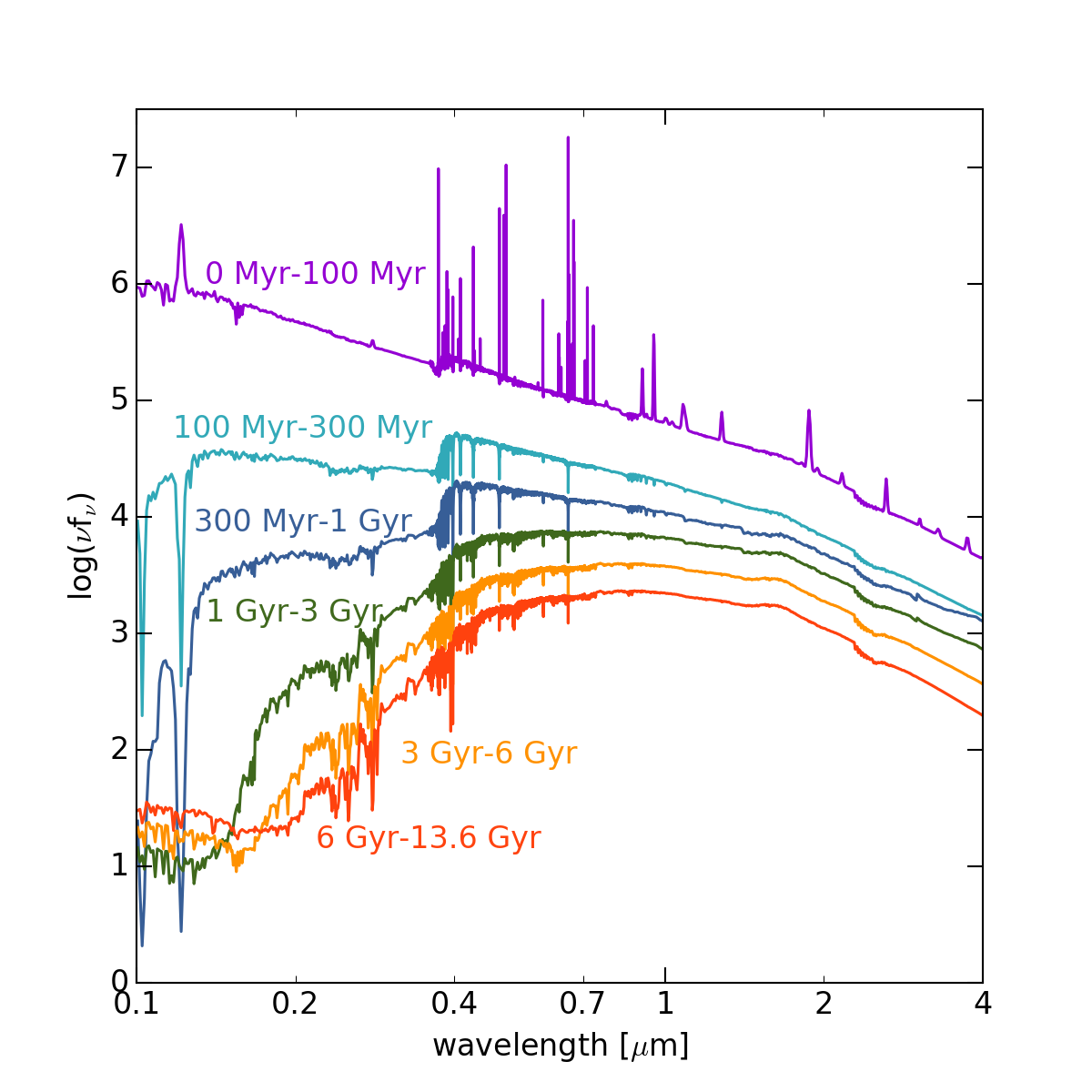}
\caption{The SED for each of the six bins of the non-parametric SFH are shown here. Each spectrum has 1 solar mass of stars formed at t=0. The youngest bin is the only bin which produces nebular emission lines.}
\label{fig:templates}
\end{center}
\end{figure}
Stellar mass, defined as the mass of existing stars and stellar remnants, is a free parameter in the \mname{} model. Stellar emission powers all sources of radiation in our model (gas, dust, and stars); thus, the stellar mass sets the overall normalization of the SED. Since the normalization constant is defined in mass instead of luminosity, it often has covariances with the star formation history parameters. A flat prior is adopted on the logarithm of the stellar mass, $5 <$log(M/M${_\odot})< 14$, and a {Chabrier} (2003) initial mass function is used.

A nonparametric star formation history prescription is adopted in the \mname{} model. Using a nonparametric SFH has the great advantage of avoiding the poorly-quantified systematics introduced by using a parameterized SFH ({Conroy} 2013): for example, exponentially declining $\tau$ models inherently couple the early-time to late-time star formation history ({Simha} {et~al.} 2014). These systematics are particularly challenging when determining proper errors on the derived SFH.

In the literature, nonparametric SFHs have primarily been used when performing full-spectrum fitting to high-resolution, high S/N data, e.g. STECMAP ({Ocvirk} {et~al.} 2006) or VESPA ({Tojeiro} {et~al.} 2007), or both low-resolution spectra and photometry ({Kelson} {et~al.} 2014; {Dressler} {et~al.} 2016). This is because, when performing classical minimization routines, non-parametric SFHs are very flexible and have too many degenerate parameters, leading to poorly-behaved and noisy best-fit solutions. Bayesian statistics are more suitable for this problem, particularly when coupled with an MCMC sampler which can fully explore the posterior probability for poorly-constrained parameters. Using a non-parametric SFH with Bayesian statistics will result in clean, accurate errors on SFH parameters compared to previous approaches, at the expense of longer computational time.

Fundamentally, the selection of the number, size, and location of SFH bins is a tradeoff between the required computational time and maximizing the amount of SFH information extracted from the data. This is challenging for a model which is fit to a large sample of galaxies with diverse SFHs, as the recoverability of the SFH given a set of data is highly dependent on the SFH of the galaxy itself (e.g. {Tojeiro} {et~al.} 2007). To inform our selection of the SFH bins, we use a procedure which is described in detail in Johnson et al. 2016 in prep. Here, it is briefly summarized. 

The goal is to estimate the fractional uncertainty on the SFH bins given a set of data (i.e., broadband filters) and the associated noise parameters. The fractional uncertainty on each bin is used to select the bin spacing and size. This is done with the Cramer-Rao bound, which is the matrix inverse of the Fisher information matrix. The Fisher information matrix measures the fractional curvature of the likelihood function near the location of maximum likelihood: small curvature implies a broad posterior, and thus more uncertainty in the parameters. In practice, the SFH recoverability is estimated assuming only a single metallicity and no dust. These simplifying assumptions mean that this procedure will specify a lower limit on the uncertainties in the bin sizes. 

As a simple estimate, two exponentially declining star formation histories are used: a star-forming galaxy with $\tau$ = 1 Gyr observed after t = 1 Gyr, and a quiescent galaxy with $\tau$ = 1 Gyr observed after 10 Gyr. This procedure suggests that the six time bins shown in the bottom six panels of Figure \ref{fig:model_diagram1} recover all SFH information for the chosen model galaxies with the {Brown} {et~al.} (2014) photometry. The spectrum of each SFH bin is shown separately in Figure \ref{fig:templates}. The model star formation rate is constant within each bin.

The most straightforward way to fit these bins to the data would be to fit directly for the mass in each bin. In practice, the fractional mass in each bin (i.e., the ``shape" of the SFH) is used instead, and the mass normalization, M, is fit as a separate parameter. This separation helps MCMC convergence by restricting the SFH parameter space to a smaller, well-defined volume, and also provides cleaner posteriors. We fit for the weight in each bin, $f_n$, such that
\begin{equation}
m_n = \frac{t_n f_n}{\sum\limits_N t_n f_n},
\end{equation}
where $m_n$ is the fractional mass in each bin, $N$ is the number of bins, and $t_n$ is the amount of time in each bin in years. The meaning of $f_n$ is more clear in the following equation:
\begin{equation}
\mathrm{SFR}_n = \mathrm{M} \frac{m_n}{t_n} = \mathrm{M} \frac{f_n}{\sum\limits_N t_n f_n},
\end{equation}
where M is the total mass and SFR$_n$ is the star formation rate in a bin. The weights $f_n$ are thus proportional to the specific star formation rate (and the star formation rate) within each time bin. This maps directly into the prediction of physical quantities such as the age of the stellar populations and the strength of spectral absorption and emission features. We require that $f_n \geq 0$, i.e. no bin can have a negative amount of stars formed.

In order to ensure a unique mapping between the weights $f_n$ and the shape of the resulting SFH, the weights are constrained during the fitting process such that
\begin{equation}
\sum_n f_n = 1.
\end{equation}
In practice, given the sampling algorithm, applying this constraint directly would be highly inefficient. This is because in each sampling step, the sampler chooses a new set of parameters without taking constraints into account, then tests whether the parameters are within the constraints: for this constraint, the probability of randomly chosen parameters which fulfill this constraint is vanishingly small.

To alleviate this inefficiency, we instead allow $N-1$ bin fractions to be free model parameters with the constraint that
\begin{equation}
\sum_{n-1} f_n \leq 1.
\end{equation}
The $N$th bin fraction is then determined by
\begin{equation}
f_N = 1 - \sum_1^{N-1} f_n.
\end{equation}
These constraints means that, when the data have little to no constraining power over the SFH, the fractional variables $f_n$ will follow a flat Dirichlet distribution. We have verified numerically that this puts a Dirichlet prior on both the $N$-1 explicit fractional variables and the $N$th implicit fractional variable. The marginal probability distribution for a single $f_n$, which corresponds to the effective Dirichlet prior on each $f_n$, is:
\begin{equation}
\mathrm{PDF} = (1-x)^{N-2} \frac{\Gamma(N)}{\Gamma(N-1)}.
\end{equation}
which is a Beta distribution. So for the star formation rate in a single bin with $N$=6 bins, the prior probability distribution is p$(f_n) \propto (1-f_n)^4$ with $0 \leq f_n \leq 1$.

The upshot of the above parameterization for the star formation history is that there is a roughly Gaussian-shaped prior on the logarithm of the specific star formation rate in each time bin, i.e. on log(SFR$_{\mathrm{bin}}$ / M$_{\mathrm{tot}}$). The center of this prior is at log(1/t$_{\mathrm{univ}}$) $\sim$ -10.1 yr$^{-1}$ , which is a constant star formation rate. The FWHM of the Gaussian prior is roughly 1.5 dex. Thus, in the absence of strong evidence from the data (where ``strong" is determined by the FWHM of the prior), \mname{} will assign a constant SFR(t) to a galaxy.

A clear advantage of the nonparametric SFH formulation over libraries of star formation histories from simulations is that the prior on sSFR(t) can be explicitly written down as has been done here. The star formation history posterior for an individual galaxy is a combination of the prior and the constraining power of the data. By knowing the prior, it is straightforward to discern how much of the posterior comes directly from the data, and how much of it comes from assumptions that are built-in to every SFH formulation.

\subsubsection{Stellar Metallicity}
Stellar metallicity affects the optical to NIR flux ratio (see Figure \ref{fig:model_diagram2}), which is important in determining ages, dust attenuation, and masses ({Bell} \& {de Jong} 2001; {Mitchell} {et~al.} 2013). The stellar metallicity also helps to set the normalization and shape of the SED blueward of the Lyman limit, which is used to generate emission line fluxes.

When fitting galaxy SEDs, it is important to interpolate between metallicities rather than using a discrete set of metallicities, else the resulting stellar mass estimates will be biased ({Mitchell} {et~al.} 2013). We refine this approach further and use a metallicity distribution function rather than interpolated metallicities. Simple interpolation between stellar spectra of different metallicities will introduce non-physical features in the interpolated spectra, because stellar SEDs have a nonlinear response to changes in metallicity. The metallicity distribution function avoids interpolation, producing more physically motivated SEDs.

Stellar metallicity does not vary with age in \mname{}, and there is no age-metallicity relationship implemented. This is in contrast to expectations in galaxies, where an age-metallicity relationship is expected due to the build-up of heavy elements with the passing of cosmic time. In testing, we have found that implementing an age-metallicity relationship can impart a unique shape to the UV-NIR SED which cannot be replicated by a fixed stellar metallicity, though this depends on both the assumed star formation history and metal enrichment history. For example, quiescent galaxies are less sensitive to an age-metallicity relationship, as their SFR timescales are shorter. The complexity of implementation combined with the poor constraining power of broadband photometry prevent inclusion of an age-metallicity relationship in \mname{}. Ideally, implementation would be guided by performing a detailed comparison of stellar metallicities recovered by \mname{} to robust spectroscopic metallicities from a sample of galaxies with a wide range in sSFR. This is a topic for future work.

The range in metallicity in our model is determined by the coverage of the Padova isochrones ({Marigo} \& {Girardi} 2007; {Marigo} {et~al.} 2008) in \fsps{}, which extend from $-2.00 <$ log(Z/Z$_{\odot}) < 0.20$. The metallicity distribution function is implemented by summing simple stellar populations at discrete metallicity values, using a triangular weighting scheme. This is illustrated in Figure \ref{fig:model_diagram2}. The metallicity is specified in units of log(Z/Z$_{\odot}$), where log(Z/Z$_{\odot}$) defines the center of the triangular weighting scheme. A flat prior over $-2.0 <$ log(Z/Z$_{\odot}) < 0.2$ is used in the \mname{} model.

\subsubsection{Dust Attenuation}
\label{section:dustattenuation}
We use the two-component {Charlot} \& {Fall} (2000) dust attenuation model, which postulates separate birth-cloud and diffuse dust screens. The birth-cloud dust simulates the embedding of young stars in molecular clouds and HII regions, the net effect of which is extra attenuation towards young stars. We allow the normalization of the birth-cloud and diffuse dust components to vary separately in the fits, along with a power-law index describing the shape of the attenuation curve for the diffuse dust component. The variance of the SED with these parameters is shown in Fig. \ref{fig:model_diagram2}.

It is difficult to achieve tight posteriors on dust attenuation parameters from UV-NIR broadband photometry alone, due to the well-known age-dust degeneracy ({Papovich}, {Dickinson}, \&  {Ferguson} 2001) (see Section \ref{section:posteriors} for further discussion). This degeneracy can be broken either by directly measuring the dust emission with MIR or FIR broadband filters ({Burgarella} {et~al.} 2005), or with excellent photometric coverage of the age-sensitive \dn{} break ({MacArthur} {et~al.} 2004). In the {Brown} {et~al.} (2014) sample, the MIR is sampled via WISE and \spitzer{} photometry, and for a limited subsample, the FIR is measured with \herschel{} photometry.

The birth-cloud component ($\dustone{}$) in our model has an attenuation curve that scales with wavelength as $\lambda^{-1}$, and attenuates stellar emission only from stars formed in the last $t = 10$ Myr, the typical timescale for the disruption of a molecular cloud ({Blitz} \& {Shu} 1980). It also attenuates nebular emission. It has the following functional form:
\begin{equation}
\dustone = {\hat{\tau}_{1}} (\lambda/5500\: \angstrom{})^{-1.0}.
\end{equation}
The diffuse component ($\dusttwo{}$) has a variable attenuation curve, and attenuates all stellar and nebular emission from the galaxy. For the wavelength scaling, we use the following prescription from {Noll} {et~al.} (2009):
\begin{equation}
\dusttwo = \frac{{\hat{\tau}_{2}}}{4.05} [k^{'}(\lambda) + D(\lambda)] \left(\frac{\lambda}{\lambda_V}\right)^n.
\end{equation}
The free parameters in this equation are $\dusttwo{}$, which controls the normalization of the diffuse dust, and \didx{}, the diffuse dust attenuation index. $k'(\lambda$) is the (fixed) {Calzetti} {et~al.} (2000) attenuation curve. D($\lambda$) is a Lorentzian-like Drude profile describing the UV dust bump. We tie the strength of the UV dust absorption bump to the best-fit diffuse dust attenuation index, following the results of {Kriek} \& {Conroy} (2013).
We adopt a flat prior over $0< \dusttwo{} <4.0$, a flat prior over $0<\dustone{} <4.0$, and a flat prior over $-2.2 < \didx{} < 0.4$. The upper limit on $\didx{}$ is chosen to disallow a flat attenuation curve, which would cause $\dusttwo{}$ to be nearly fully degenerate with the normalization of the SED.

The similar effects of $\dustone{}$ and $\dusttwo{}$ on the stellar SED means that they are often degenerate (though this is SFH-dependent). It is important to distinguish between these parameters to properly predict emission line equivalent widths. Previous work suggests that the total optical depth towards nebular emission lines is $\sim$twice that of the stellar component ({Calzetti}, {Kinney}, \&  {Storchi-Bergmann} 1994; {Price} {et~al.} 2014), though the exact value varies with star formation rate and galaxy inclination ({Wild} {et~al.} 2011). Translating this result into the {Charlot} \& {Fall} (2000) model implies that $\dustone{} \sim \dusttwo{}$, as $\dustone{}$ applies to the entire galaxy, whereas star-forming regions only experience attenuation from $\dusttwo{}$. We adopt a joint prior on the ratio of the two: $0 < (\dustone{}/\dusttwo{}) < 2.0$, which allows some reasonable variation around the fiducial results in the literature. 

\subsubsection{Dust Emission}
\label{section:dustemission}
To model the dust emission from galaxies, we assume energy balance, where all starlight attenuated by the dust is re-emitted in the infrared ({da Cunha} {et~al.} 2008). This assumption means that the MIR and FIR photometry are additional constraints both on the total amount of dust attenuation, and on the dust-free stellar SED. Simultaneous modeling of the dust and stellar emission is key to accurate SFRs, particularly for galaxies with lower sSFRs where dust-heating via old stellar populations becomes more important (e.g., {Utomo} {et~al.} 2014). While a useful tool, energy balance cannot be used to determine \lir{} from the UV-MIR SED alone without making some assumptions about the shape of the infrared SED (see Appendix \ref{section:herschel} for further discussion).

We use the {Draine} \& {Li} (2007) dust emission templates to describe the shape of the infrared SED. This model is based on the silicate-graphic-PAH model of interstellar dust ({Mathis}, {Rumpl}, \& {Nordsieck} 1977; {Draine} \& {Lee} 1984). The size distribution of carbonaceous and silicate grains is calibrated to be consistent with the wavelength-dependent extinction in the Milky Way. We note that in our model, the shape of the dust emission is modeled independently from the shape of the dust attenuation curve. This is a reasonable implementation because the shape of the {\it attenuation} (not extinction) curve is largely set by the geometry and column depth of the dust, rather than the composition of the dust ({Chevallard} {et~al.} 2013). In the future, it may be useful to tie the strength of the 2175 \AA{} extinction bump to the strength of the PAH emission, as it has long been suggested that these features both arise from the same dust grains ({Stecher} \& {Donn} 1965; {Draine} 1989).

The {Draine} \& {Li} (2007) model has three free parameters controlling the shape of the IR SED: \umin{}, \gammae{}, and \qpah{}. The effect of these three parameters on the resulting SED is demonstrated in Figure \ref{fig:model_diagram2}.

\umin{} and \gammae{} together control the shape and location of the thermal dust emission bump in the infrared SED. Physically, they describe the fraction of dust mass exposed to the distribution of starlight intensities U, which is parameterized in the {Draine} \& {Li} (2007) model by a combination of a delta function and a power-law distribution over \umin{} $< U < U_{\mathrm{max}}$:
\begin{align}
\frac{dM_{\mathrm{dust}}}{dU} =& (1-\gamma_{\mathrm{e}})M_{\mathrm{dust}} \delta(U-U_{min})+\\
& \gamma_{\mathrm{e}} M_{\mathrm{dust}} \frac{(\alpha-1)}{(U_{min}^{1-\alpha}-U_{\mathrm{max}}^{1-\alpha})}U^{-\alpha} \notag,
\end{align}
with $dM_{\mathrm{dust}}$ as the mass of dust heated by starlight intensities between $U$ and $U + dU$, $M_{\mathrm{dust}}$ is the total dust mass, and $\alpha$ is a power-law index. {Draine} \& {Li} (2007) finds that MIR photometry of galaxies in the SINGS survey ({Kennicutt} {et~al.} 2003) is well-reproduced with $\alpha = 2$ and $U_{\mathrm{max}} = 10^6$, which we adopt here. $M_{\mathrm{dust}}$ is determined by the normalization of the SED.

Specifically, \umin{} represents the minimum starlight intensity to which the dust mass is exposed, and \gammae{} represents the fraction of dust mass which is exposed to this minimum starlight intensity. The minimum starlight intensity roughly represents the intensity experienced by dust in the general diffuse ISM. The inclusion of this delta function at the minimum starlight intensity is the major difference between the {Draine} \& {Li} (2007) and {Dale} \& {Helou} (2002) models for $dM_{\mathrm{dust}}/dU$.

The final free parameter, \qpah{}, describes the fraction of total dust mass that is in polycyclic aromatic hydrocarbons (PAHs). This parameter is particularly important because a substantial fraction or a majority of the MIR emission comes from strong PAH emission features. 

The PAH ionized fractions and absorption cross-sections in {Draine} \& {Li} (2007) are updated from the {Li} \& {Draine} (2001) model by using observations of SINGS galaxies with 5-38 $\mu$m \spitzer{} IRS spectra ({Smith} {et~al.} 2007). {Draine} \& {Li} (2007) provides infrared spectral templates for $0.1 < $\qpah{}$ < 4.58$. We note that with the excellent MIR coverage of the {Brown} {et~al.} (2014) spectral atlas, \qpah{} can be very well-determined photometrically (see Appendix \ref{appendix:mock_tests} for further details).

We adopt a flat prior of $0.1<$ \umin{} $<15$ and $0.0< $\gammae{}$ < 0.15$. For \qpah{}, we extrapolate the \spitzer{} IRS spectra out past \qpah{}=4.58, and adopt a flat prior over $0.1 < $\qpah{}$ < 7.0$. The extrapolation is linear in both flux and \qpah{}. In brief, the adopted priors are consistent with both the SINGS sample ({Draine} {et~al.} 2007) and the {Brown} {et~al.} (2014) galaxies with \herschel{} photometry, and adopting more permissive priors would bias the \lir{}, SFR, and dust attenuation in galaxies without \herschel{} photometry. We discuss the origin and effects of these priors further in Appendix \ref{section:herschel}.

\subsubsection{Nebular Emission}
\label{section:nebemission}
We generate nebular continuum and line emission using the \cloudy{} ({Ferland} {et~al.} 1998, 2013) implementation within \fsps{}. This is described in detail by {Byler} {et~al.} (2016), and summarized briefly here.

The \cloudy{} implementation in \fsps{} is a grid in the following parameters: (1) simple stellar population (SSP) age, (2) SSP and gas-phase metallicity, and (3) the ionization parameter, $U$, which is the ratio of ionizing photons to the total hydrogen density. Before the SSPs within FSPS are combined into composite stellar populations, the number of ionizing photons in each SSP is calculated and removed from the SSP SEDs. The energy removed from the SSPs is then rerouted into line luminosities and nebular continua. 
This is only done for SSPs with ages $<10$ Myr, as the ionizing contribution from older SSPs is very small. This also implicitly assumes an ionizing photon escape fraction of zero.

Within the \mname{} model, the gas-phase metallicity is set equal to the model stellar metallicity, and the ionization parameter $U$ is fixed to 0.01 in all fits. We note that the ionization parameter has very little effect on the Balmer emission line luminosities (e.g., \halpha{} or \hbeta{}), though it has a strong effect on the luminosity of forbidden lines (\oii{}, \oiii{}, etc). In principle, it could be fit as well, though broadband photometry alone provides very little constraining power. For the {Brown} {et~al.} (2014) sample, the contribution to the broadband photometry from emission lines is relatively low ($\lesssim$5\%) due to the relatively low sSFRs, so $U$ is fixed for simplicity. In the future, it may be useful to allow $U$ to vary, perhaps with a joint prior between sSFR and $U$, as suggested by recent work which finds a strong relationship between the two (e.g., {Dickey} {et~al.} 2016). This would be particularly relevant for galaxies with high sSFR, where the nebular contribution to the broadband flux can become significant (for typical star-forming galaxies at $z\sim1-2$, it is 5-10\% or greater in certain rest-frame optical bands; {Pacifici} {et~al.} 2015).

In the \mname{} model, after production, emission lines fluxes experience attenuation from both the birth-cloud and diffuse dust component (see Section \ref{section:dustattenuation}). The emission line equivalent widths are unaffected by attenuation from the diffuse dust component, as it attenuates the stellar emission equally. However, the emission line equivalent widths are sensitive to the birth-cloud attenuation, since it only attenuates emission from young stars and nebular emission. Both nebular continuum and line emission are added to the full SED model, and their effects are included in the broadband photometry.

\subsubsection{Example Model Posteriors}
\label{section:posteriors}
\begin{figure*}[th!]
\begin{center}
\includegraphics[width=0.95\linewidth]{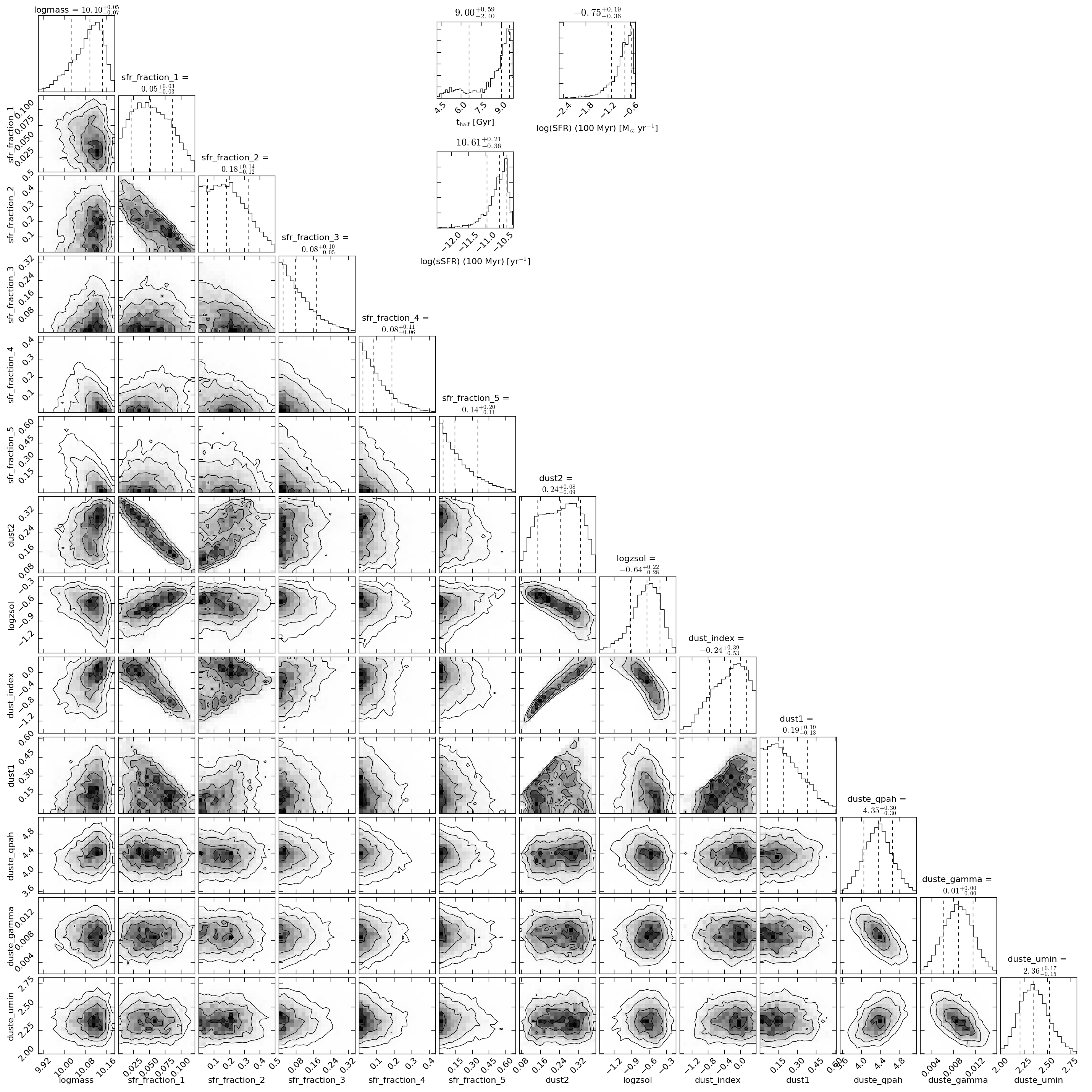}
\caption{A ``corner" plot, where the posterior probability distribution functions are shown for each model parameter along the diagonal panels, and the correlation between model parameter posteriors are shown along the columns. Above each probability distribution, the median of the parameter posterior is printed, along with the $\pm1 \sigma$ error bars. In the upper right panels, the posterior probability functions for the half-mass time, the SFR, and the sSFR are also shown. The object shown is NGC 0628, a mildly star-forming galaxy. This object was chosen to best illustrate important parameter degeneracies, including the dust-metallicity-SFR degeneracy (dust2-logzsol-sfr\_fraction1, respectively), and the amount of dust and the shape of the attenuation curve (dust2 and dust\_index).}
\label{fig:corner_plot}
\end{center}
\end{figure*}

Here we describe the typical posteriors for an application of the \mname{} model to the {Brown} {et~al.} (2014) photometry. Figure \ref{fig:corner_plot} shows a ``corner"\footnote{Made with code adopted from https://github.com/dfm/corner.py} plot for NGC 0628, a mildly star-forming galaxy chosen to illustrate several important parameter degeneracies. 

The panels along the diagonal show the posterior probability functions for each model parameter. The median of each posterior is printed above the diagonals, with the $\pm$1$\sigma$ credible intervals also indicated. Most model parameters have Gaussian-like posteriors, and the exceptions in the SFH and dust parameters are a result of explicit or implicit priors in the \mname{} model. The fractional star formation history parameters largely have p$(f_n) \propto (1-f_n)^4$, as a result of the implicit prior set by the SFH implementation described in Section \ref{section:sfh}. This is not true for the youngest SFH time bin, where the evidence from the photometry is strong enough to reshape the posterior into a Gaussian. The birth-cloud dust parameter posterior, dust1, is truncated where dust1 = dust2, as a result of the dust priors described in Section \ref{section:dustattenuation}. Finally, the dust\_index parameter, which controls the shape of the dust attenuation curve, is against the prior limit described in Section \ref{section:dustattenuation}.

The panels in each column show the correlation between the posteriors for each model parameter. These illustrate several degeneracies which are important in understanding the limited precision for certain parameters, including the dust-metallicity-SFR degeneracy (dust2-logzsol-sfr\_fraction1, respectively), and the amount of dust and the shape of the attenuation curve (dust2 and dust\_index). The parameters controlling the shape of the dust emission (duste\_gamma, duste\_qpah, and duste\_umin) are also mildly degenerate with one another, due to the limited sampling of the IR SED. These degeneracies are accurately explored by \prospector{} due to the on-the-fly model generation and the MCMC sampling.

\begin{figure*}[th!]
\begin{center}
\includegraphics[width=0.96\linewidth]{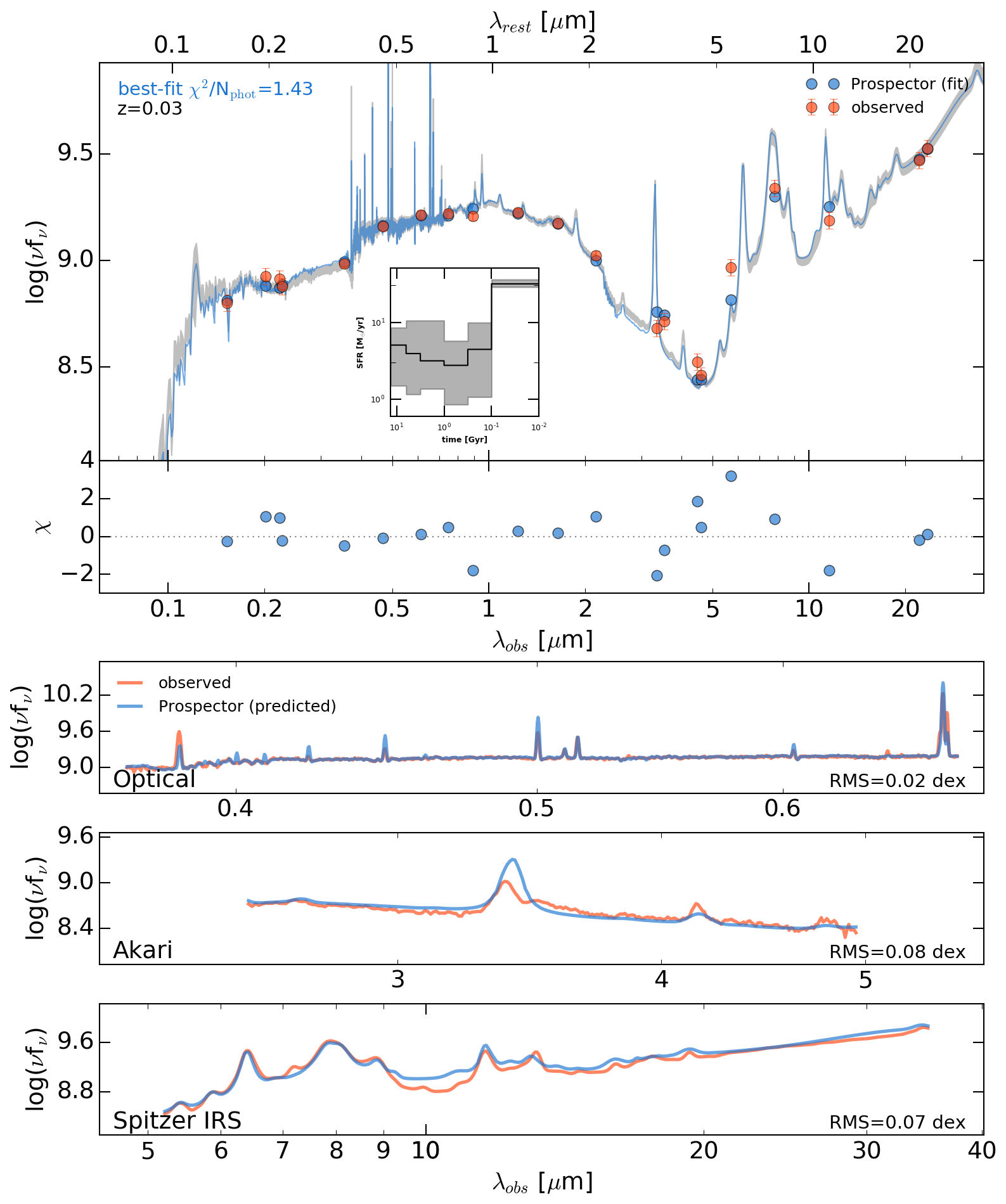}
\caption{Broadband photometry of a star-forming galaxy (top panel), NGC 6090, along with the best-fit photometry and spectrum of the \mname{} model. The 16th and 84th percentiles of the model spectrum are shaded in grey. The \mname{} model is fit to the broadband photometry only. The 16th, 50th, and 84th percentiles of the star formation history are in the inset panel. The photometric residuals are in the lower attached panel. Comparisons between the best-fit model spectrum and the optical, Akari, and \spitzer{} IRS spectra are shown in the lower three panels. For the optical and Akari spectra, the models are smoothed to the resolution of the data; for the \spitzer{} IRS spectra, the data are smoothed to the resolution of the models. The disagreement in the emission line fluxes at {\sc [O~ii 3727]}, {\sc [O~iii 4959]} and {\sc [O~iii 5007]} is expected, as the ionization parameter is fixed in the \mname{} model (see Section \ref{section:nebemission}).}
\label{fig:starforming_sed}
\end{center}
\end{figure*}

\begin{figure*}[th!]
\begin{center}
\includegraphics[width=0.95\linewidth,trim={0 5.6cm 0 0}]{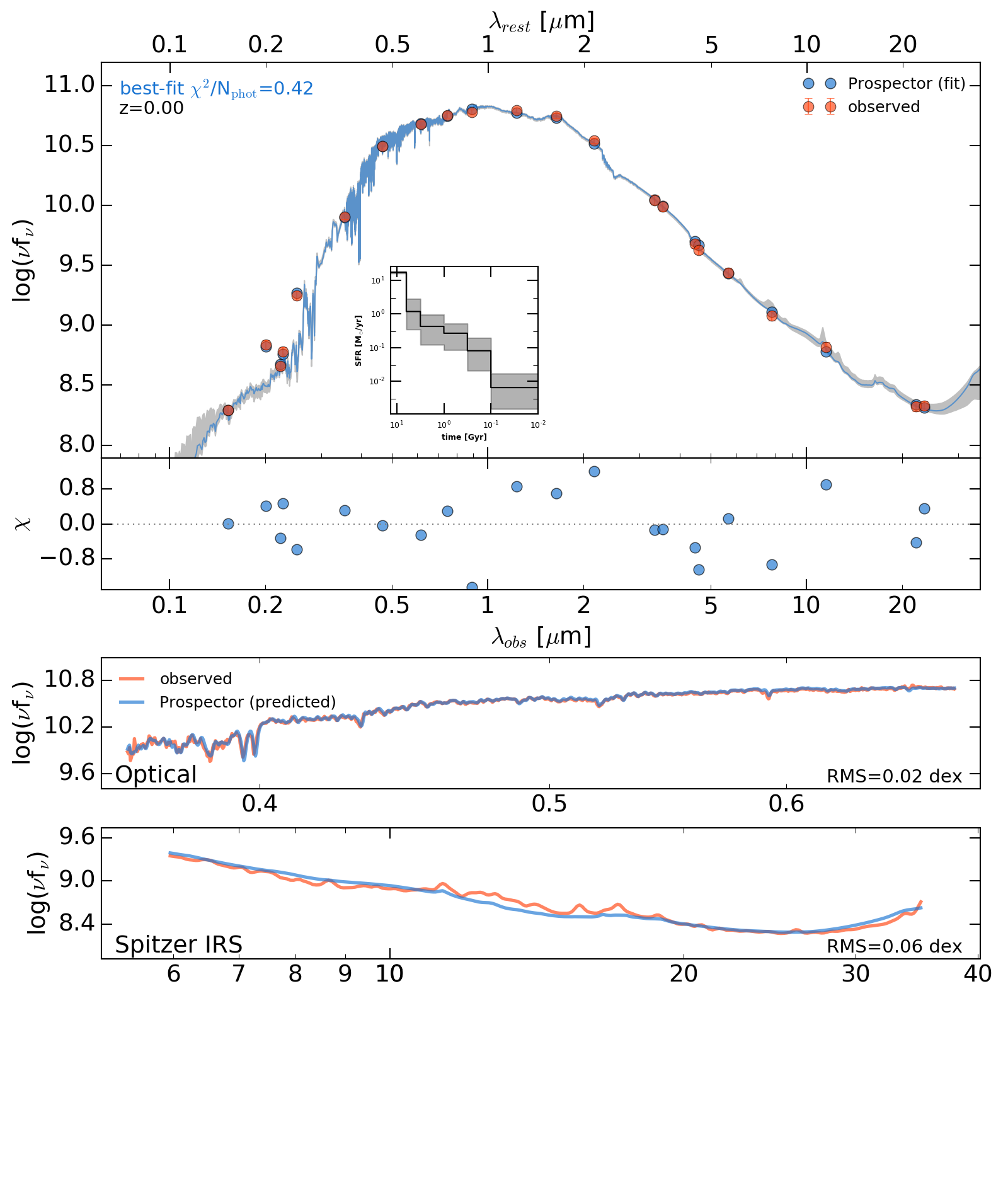}
\caption{Broadband photometry of a quiescent galaxy (top panel), NGC 4125, along with the best-fit photometry and spectrum of the \mname{} model. The 16th and 84th percentiles of the model spectrum are shaded in grey. The \mname{} model is fit to the broadband photometry only. The 16th, 50th, and 84th percentiles of the star formation history are in the inset panel. The photometric residuals are in the lower attached panel. Comparisons between the best-fit model spectrum and the optical and \spitzer{} IRS spectra are shown in the lower two panels. For the optical spectra, the models are smoothed to the resolution of the data; for the \spitzer{} IRS spectra, the data are smoothed to the resolution of the models.}
\label{fig:quiescent_sed}
\end{center}
\end{figure*}

\subsection{Fitting Procedure}
\label{section:methodology}
In this section we describe the algorithms used to generate and fit model spectral energy distributions (SEDs) to the observed photometry.
\subsubsection{Inference Framework}
We construct the \mname{} galaxy model within \prospector{}\footnote{\url{https://github.com/bd-j/prospector}. The build used in this work can be found under Github commit hash \texttt{0c98d45af32456c12d28306c4d80214179ad47c4}}. \prospector{} is a Python package which serves as a framework to bring together a galaxy model with a noise model in a posterior probability function. That probability function is then passed to algorithms which either maximize or sample the posterior probability function. \prospector{} has a highly flexible, modular design: input observations can be photometric, spectroscopic, or both; the sampling / minimizing procedure can be changed to suit the problem at hand; and the likelihood function, physical model, and priors are cleanly separated from one another.

We adopt a simple chi-squared noise model, i.e. we model the photometric noise with Gaussian independent errors of known dispersion. We note that \prospector{} has a number of more sophisticated noise models available, including estimation of covariant noise between sets of photometric filters, and estimating an overall jitter term from the available photometric data.

\subsubsection{Minimization and Sampling}
We use a two-step process to find the posterior probability distribution function for our model. The first step is a simple minimization scheme intended to find the best-fit model parameters. The second step is an MCMC sampler, which is initialized in the region of the best-fit parameters from the previous step. 

The minimization step uses the Powell minimization scheme from the \scipy{} python package. We take advantage of multiprocessing to perform this process. For N available processors, N sets of initial parameter values are generated randomly within the constraints of the priors, and then each processor performs an individual Powell minimization scheme for a specified number of iterations, or to a specified likelihood tolerance, whichever is reached first. The best-fitting model parameters from the Powell routine are used as the initial conditions for the next step.

We use the \emcee{} package ({Foreman-Mackey} {et~al.} 2013) for the sampling step. \emcee{} is a Python-based
MCMC sampler based on an affine-invariant algorithm. In brief, \emcee{} uses an ensemble of ÕwalkersÕ to explore probability space. Each walker will attempt a specified number of steps in parameter space. A step is attempted by first proposing a new position, and then calculating the likelihood of the new position based on the data. The walker has a chance to move to the new position based on the ratio of the likelihood of its current position to that of the new position. After a specified number of steps, the algorithm is completed and the position and likelihood of each walker as a function of step number is written out. These positions and likelihoods are used to generate the posterior probability distribution function for each model parameter.

To initialize \emcee{}, we generate walker positions in a Gaussian ball around the best-fitting model parameters from the Powell minimization phase. The dispersion of the Gaussian ball is customized for each parameter. The dispersions are chosen to represent the typical uncertainty in each parameter. If the first minimization process returned a set of best-fit parameters close to the global maximum, then convergence is swift and the probability space around the global maximum will be well-sampled. If the initial conditions were far from the global maximum, the walkers will attempt to reach and explore the global maximum before the process is completed. In some cases the process completes before the walkers have reached an equilibrium around the global maximum (before they have Óburned inÓ), and the resulting parameter probability distributions will not be useful. We visually inspect plots of the walker location in each parameter as a function of iteration step for each galaxy to ensure they are burned in. If the MCMC process is not burned in, the fit is repeated with more iterations until equilibrium is achieved. Future versions of \prospector{} will have quantitative criteria to assess convergence.

It is critical that we use an MCMC algorithm to explore our model posterior, as a standard grid approach with 13 parameters and sufficient grid spacing would take up a prohibitive amount of memory. One way to limit the amount of grid space to explore is to use generating functions to sample reasonable combinations of parameters, and create static template libraries from this sampling of parameter space. However, this method necessarily results in opaque priors. On-the-fly stellar populations generation with \fsps{} combined with MCMC sampling allows \prospector{} to explore model parameter space free from the complex priors from pre-generated template libraries, at the expense of increased computational time.

\section{Evaluating the Quality of the Model Fits}
\label{section:results}
In this section we discuss the results of fitting the photometry from the {Brown} {et~al.} (2014) sample with the \mname{} model. Section \ref{section:phot_residuals} describes the residuals from the fits to the photometry. Section \ref{section:spec_residuals} describes overall trends in the residuals between the best-fit model spectra and the observed spectra; the spectra are {\it not} included in the fits, so this is an excellent test of the \mname{} model. The remaining sections quantify the predictive value of specific features or combinations of features in the observed spectra from fits to the photometry: \halpha{} and \hbeta{} emission (Section \ref{section:halpha}), Balmer decrements (Section \ref{section:bdec}), \dn{} (Section \ref{section:dn}), \hdelta{} absorption (Section \ref{section:hdelta}), and stellar metallicity (Section \ref{section:metallicity}). The spectral quantities presented from the \mname{} model are calculated from the median of the model posterior, and the error bars are the 16th and 84th percentiles of the model posteriors.
\subsection{Photometric Residuals}
\label{section:phot_residuals}

\begin{figure*}[ht!]
\begin{center}
\includegraphics[width=0.95\linewidth]{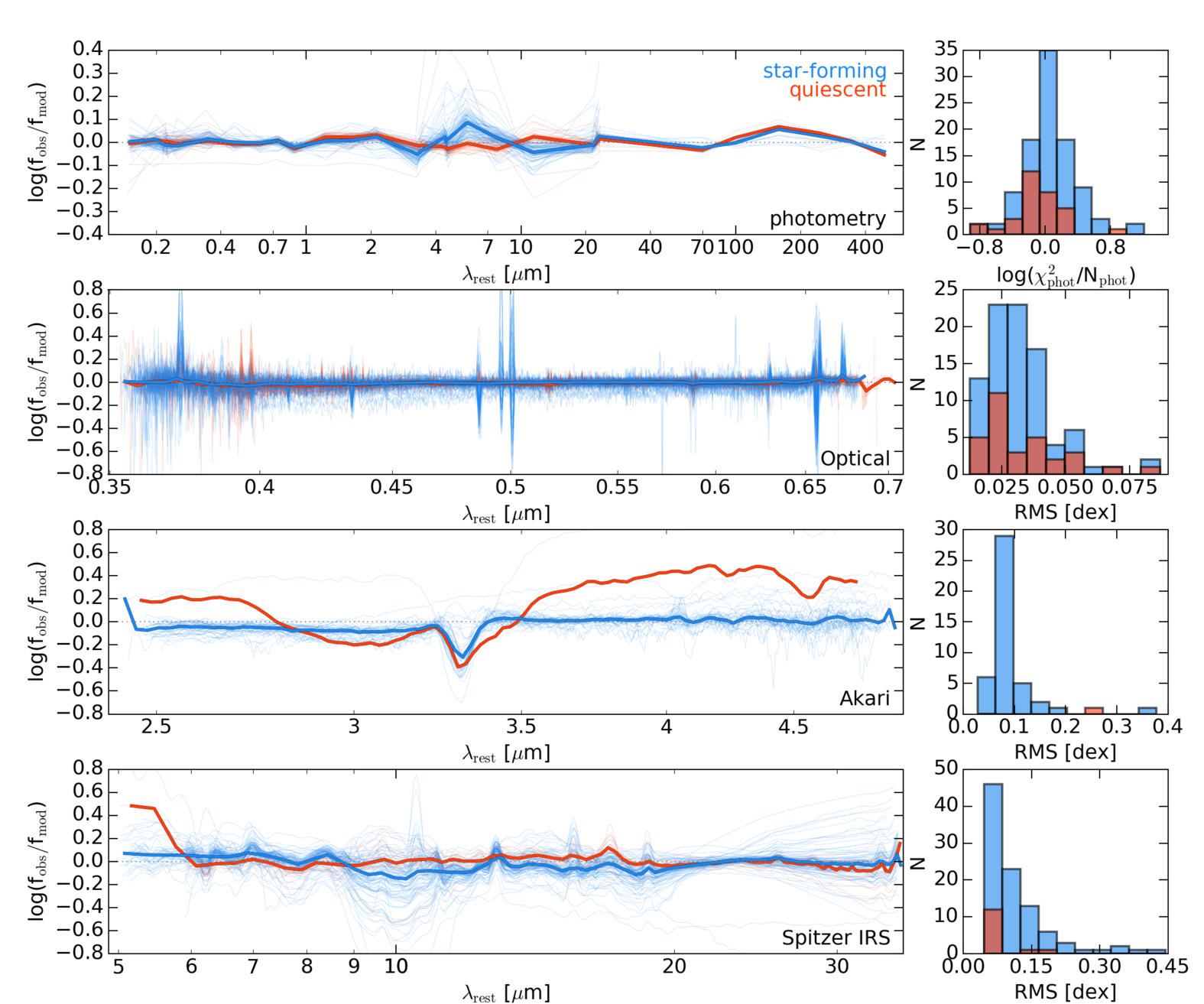}
\caption{Residuals between the observations and model predictions as a function of rest-frame wavelength. Star-forming galaxies (defined as having sSFR(\mname{}) $> 10^{-11}$ yr$^{-1}$) are shown in blue, while quiescent galaxies are shown in red. The residuals for individual galaxies are shown as light lines, while the medians over star-forming and quiescent galaxies are shown separately as heavy lines. The panels show, from top to bottom, the residuals for broadband photometry, the optical spectra, the Akari spectra, and the \spitzer{} IRS spectra. The panels on the right-hand side show histograms of the $\chi^2$ and RMS values. Strong emission lines are masked when calculating the optical RMS. Systematic features are clearly visible: for example, large residuals at 3.3$\mu$m in the Akari spectra suggests that the {Draine} \& {Li} (2007) model overpredicts the 3.3$\mu$m PAH emission line strength.}
\label{fig:median_residuals}
\end{center}
\end{figure*}

In Figures \ref{fig:starforming_sed} and \ref{fig:quiescent_sed} we show example model fits to the photometry of a star-forming and a quiescent galaxy respectively, along with the photometric residuals, the best-fit and marginalized SFHs, and a comparison between the best-fit model spectrum and the observed spectra.

We show the photometric and spectral residuals for the entire catalog in Figure \ref{fig:median_residuals}. The residuals from star-forming and quiescent galaxies are separated, where star-forming is defined by an sSFR cut of 10$^{-11}$ yr$^{-1}$.

The photometric residuals are lowest at well-studied optical and NIR wavelengths, and increase in the UV and MIR bands. A significant minority of the galaxies show IR colors and emission line ratios indicative of AGN activity ({Brown} {et~al.} 2014). This may contribute to the residuals at MIR wavelengths by adding significant amounts of non-stellar emission. The larger MIR residuals also reflect the well-known difficulty of modeling diffuse dust, PAHs, and circumstellar disks heated by AGB stars, star formation and AGN ({Mentuch} {et~al.} 2009; {Lange} {et~al.} 2016).

\subsection{Spectral Residuals}
\label{section:spec_residuals}
Here we describe the residuals between the observed spectra and the model spectra. The model spectra here are generated from the posteriors of the fit to the broadband photometry; the observed spectra are not fit.

The spectral residuals in the optical are flat and have RMS deviations at the $<$10\% level for the majority of both quiescent and star-forming galaxies. The strongest residuals are in emission lines, particularly the \oii{} and \oiii{} lines. This is due to their sensitivity to the ionization parameter ({Kewley} {et~al.} 2013). The ionization parameter is fixed in the \mname{} model. This has a negligible effect on the fit for most galaxies, as forbidden lines generally contribute a small fraction of the flux in the broadband photometry, with the exception of highly star-forming galaxies ({Pacifici} {et~al.} 2015).

The Akari and \spitzer{} IRS spectral residuals are also relatively flat with wavelength. There is substantial variation in the absorption feature at 10$\mu$m, which is likely related to AGN activity (e.g., {Nenkova} {et~al.} 2008a, 2008b. The dominant residuals come from PAH emission features. This is not unexpected, as there is significant galaxy-to-galaxy variation in the strength of PAH features ({Smith} {et~al.} 2007; {O'Dowd} {et~al.} 2009). Specifically, the strength of many of the PAH emission features at $6 < \lambda < 20$ $\mu$m show substantial variation, while the 3.3 $\mu$m feature is strongly overpredicted. As discussed in Section \ref{section:dustemission}, the {Draine} \& {Li} (2007) dust emission model is calibrated to \spitzer{} IRS 5-38$\mu$m spectra of SINGS galaxies, and so the 3.3 $\mu$m feature was not in the calibration range of the {Draine} \& {Li} (2007) model. The systematic differences in the remaining PAH features may be related to sample selection differences between the {Brown} {et~al.} (2014) spectral catalog and the SINGS survey ({Kennicutt} {et~al.} 2003). The {Brown} {et~al.} (2014) selection is based on the availability of the drift-scan spectroscopy and is thus highly heterogeneous, with significant diversity in stellar masses and sSFRs (see Figure \ref{fig:intro}), whereas the SINGS survey targets ``normal" ($\sim$L$^*$) star-forming galaxies ({Kennicutt} {et~al.} 2003). The correlation between the strength of the PAH features and the variation in \qpah{} in the \mname{} fits is discussed further in Section \ref{section:pah}.

There are also substantial residuals in the absorption feature at 10$\mu$m for star-forming galaxies. This is likely related to AGN activity (e.g., {Nenkova} {et~al.} 2008a); see the discussion in Section \ref{section:future} for more details.

\subsection{Balmer Emission Lines}
\label{section:halpha}

\begin{figure*}[ht!]
\begin{center}
\includegraphics[width=0.9\linewidth]{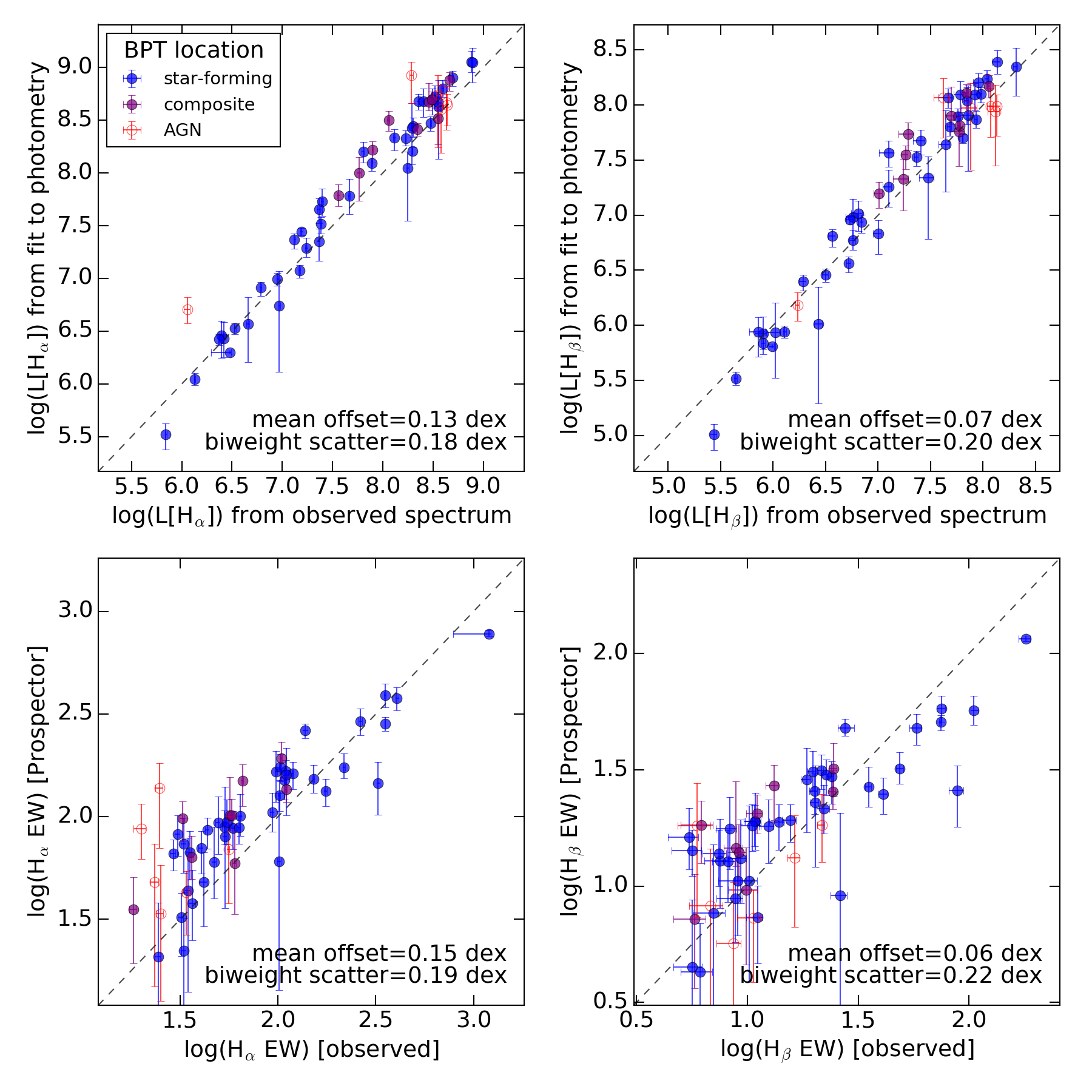}
\caption{Predicted \halpha{} and \hbeta{} from fitting the \mname{} model to the broadband photometry, compared to observed values from the spectrum. The increase in scatter from \halpha{} to \hbeta{} may be related to the intrinsically weaker \hbeta{} emission being more affected by uncertainty in the underlying stellar absorption. Only galaxies with S/N (\halpha{}, \hbeta{}) $> 5$ are shown. Galaxies are color-coded by their BPT classification. The equivalent widths are in \angstrom{}, while the luminosities are in L$_{\odot}$.}
\label{fig:balmer_lineflux}
\end{center}
\end{figure*}

The upper panels of Figure \ref{fig:balmer_lineflux} shows the observed \halpha{} and \hbeta{} luminosities versus the model \halpha{} and \hbeta{} luminosities, as predicted by the \mname{} model. The points are color-coded by their position on the BPT diagram, which is a diagnostic for their primary excitation mechanism ({Baldwin}, {Phillips}, \&  {Terlevich} 1981). Only galaxies with observed \halpha{} and \hbeta{} S/N $>$ 5 are shown. We note that these cuts preferentially select galaxies with higher sSFRs: the average \mname{} sSFR for the entire sample is 3 x 10$^{-10}$ Gyr$^{-1}$, with a range of 3 x 10$^{-14}$ yr$^{-1}$ to 6 x 10$^{-9}$ yr$^{-1}$, while after the S/N cuts, the average sSFR is 6 x 10$^{-10}$ yr$^{-1}$, with a range of 6 x 10$^{-11}$ yr$^{-1}$ to 6 x 10$^{-9}$ yr$^{-1}$. 

There is excellent agreement between the models and observations, with a small offset of $\sim$ 0.13 (0.07) dex and $\sim$ 0.18 (0.2) dex scatter in \halpha{} (\hbeta{}) across a wide range of galaxy types and stellar masses (6.3 $<$ log M/M$_{\odot} < 11.4$). The good agreement between predicted and observed \halpha{} suggests that galaxy star formation rates do not strongly vary over 100 Myr timescales.

Figure \ref{fig:balmer_lineflux} also shows the same comparison in equivalent width (EW). The EW and luminosity comparisons roughly measure the ability of the fitter to recover sSFR and SFR, respectively. The EWs show a similar amount of scatter and offset to the luminosity comparison, suggesting that sSFR and SFR are equally well-recovered.

\hbeta{} is recovered with a similar accuracy to \halpha{}, with slightly more scatter. One possible source of this extra scatter is errors in the model dust attenuation curve, which would have a larger effect on \hbeta{} than \halpha{} since it is at bluer wavelengths. Errors in the absorption corrections are another possibility: since \hbeta{} line emission is intrinsically weaker than \halpha{}, the absorption correction makes up a larger percentage of the obsered flux.

The excellent agreement between observed \halpha{} and \halpha{} from \mname{} fits to the broadband photometry is a key result of the paper, and we spend much of the remainder of the manuscript discussing it. In Section \ref{section:balmer_line_discussion}, we show that the primary determinants of the \halpha{} luminosity from a galaxy are the star formation rate over the last 10 Myr, the dust attenuation, and the stellar metallicity. We discuss the correlation between the \halpha{} and \hbeta{} residuals in Section \ref{section:halpha_resid}, and show that this means the dominant source of error in predicting the Balmer lines is not dust reddening, but rather in the overall normalization of the Balmer lines. We also use the spread between the predictions and the observations to show that the width of the model \halpha{} posteriors is accurate to within $\sim$20\%. In Appendix \ref{appendix:magphys}, we demonstrate that an SED-fitting code of similar complexity, \magphys{}, shows considerably more scatter in the model-observational \halpha{} predictions.

\subsection{\dn{}}
\label{section:dn}

\begin{figure}[t!]
\begin{center}
\includegraphics[width=\linewidth]{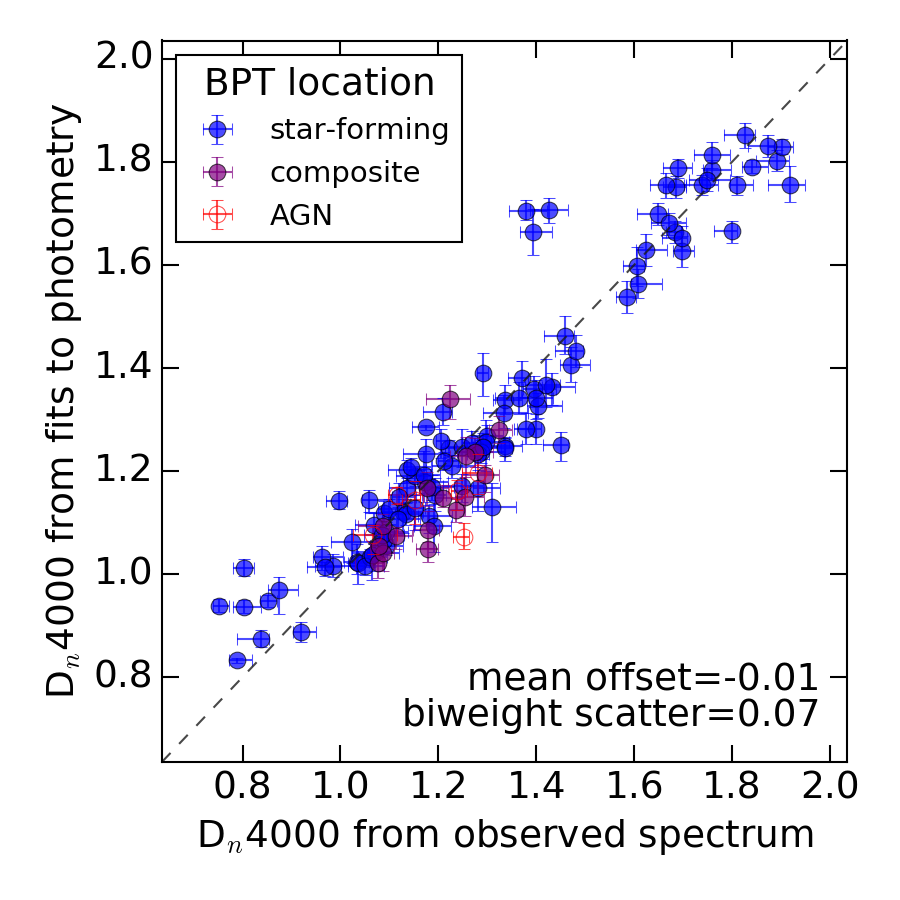}
\caption{Observed \dn{} compared to predictions from the \mname{} fits to the broadband photometry. The \dn{} feature is sensitive to a combination of stellar age and metallicity. The accuracy and precision of this comparison validates the accuracy of the underlying model SFH and metallicity parameters. Galaxies are color-coded by their BPT classification.}
\label{fig:dn}
\end{center}
\end{figure}

Figure \ref{fig:dn} shows the \dn{} measured from optical spectroscopy versus the \dn{} predicted from fits to the photometry. \dn{} is a classic spectral indicator of both the stellar metallicity and stellar age of a galaxy ({Bruzual} 1983; {Hamilton} 1985; {Balogh} {et~al.} 1999). The \mname{} model recovers \dn{} from the broadband photometry alone across a wide range of stellar ages and metallicities.

The \mname{} model fits slightly underpredict the age and/or metallicities of galaxies with larger \dn{}. This is likely related to the maximum galaxy age in the \mname{} model, which, due to the spacing of the SFH bins, is $\sim$10 Gyr (representing the average stellar age when all SFR occurs in the oldest SFH bin).

Appendix \ref{appendix:mock_tests} explores the ability of the \mname{} model to recover \dn{} in mock tests. \dn{} is recovered with no bias and very little scatter in the mock tests. This is not unexpected, as the SFH priors in the mock tests are the same as the SFH priors in the \mname{} model; however, this is likely not true when fitting real galaxies. Since the SFH at $\gtrsim$ 1 Gyr recovered from broadband photometry alone can be very prior-dependent, it may be useful to adjust the priors on the SFH parameters in future analyses to minimize the residuals in plots such as Figure \ref{fig:dn}.
\subsection{Balmer Decrements}
\label{section:bdec}

\begin{figure}[t!]
\begin{center}
\includegraphics[width=\linewidth]{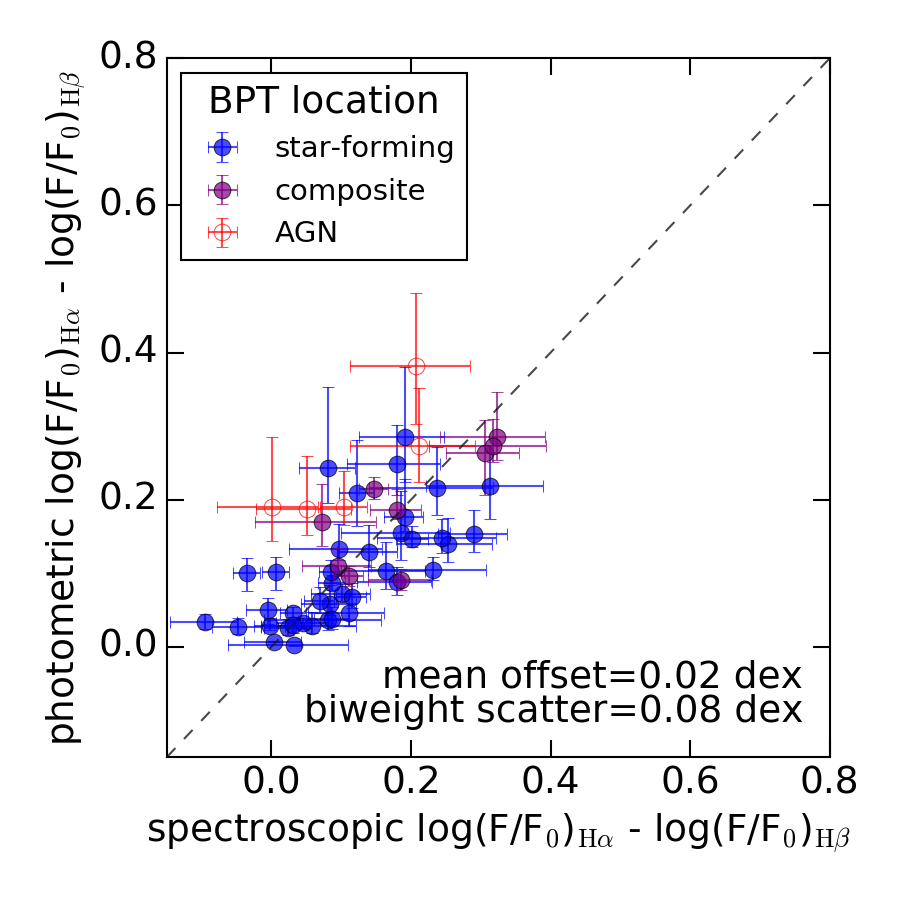}
\caption{Predicted reddening towards HII regions from fitting the \mname{} model to the broadband photometry compared to the observed reddening. The reddening is calculated from the observed Balmer decrement for each galaxy, and converted into the difference in dust attenuation towards the birth-cloud regions between \halpha{} wavelengths and \hbeta{} wavelengths. Classically, accurate nebular dust attenuation measurements have required expensive spectroscopic observations; this comparison suggests they may be recovered accurately from the photometry instead. Galaxies are color-coded by their BPT classification. Only galaxies with S/N (\halpha{}, \hbeta{}) $> 5$ are shown.}
\label{fig:reddening}
\end{center}
\end{figure}

The Balmer decrement is the observed ratio of \halpha{} / \hbeta{} emission line fluxes. For case B recombination, the intrinsic ratio of these fluxes is set to 2.86 by atomic physics. Any deviation from this ratio can be converted directly into the reddening along lines of sight to HII regions (e.g., {Osterbrock} 1989; {Calzetti} {et~al.} 2000; {Brinchmann} {et~al.} 2004).

In Figure \ref{fig:reddening}, we show the observed Balmer decrements compared to the Balmer decrements predicted from fitting the \mname{} physical model to the photometry. The units of the plot are the logarithm of the amount of reddening between \halpha{} and \hbeta{}. The S/N and EW cuts are identical to the \halpha{} and \hbeta{} comparisons. The model Balmer decrements are calculated directly from the posteriors for the model dust parameters; all three dust parameters ($\dustone{},$ $\dusttwo{}$, and \didx{}) are used to calculate the reddening towards HII regions. The dust reddening towards AGN is systematically overpredicted, which is expected; \mname{} fits the excess MIR emission caused by the AGN-heated dust by adding an excess of dust to the stellar model. 

Overall, we find excellent agreement between the spectroscopic reddening measurement and the reddening inferred by the photometric model. This is a particularly encouraging result, because the effect of the dust parameters on the observed photometry is often highly degenerate (see Figure \ref{fig:model_diagram2}). The sample size of studies that rely on expensive spectroscopic Balmer decrements for reddening measurements can be greatly expanded by using full SED fits to estimate the reddening from simple broadband photometry (e.g., {Gallazzi} {et~al.} 2005; {Shivaei} {et~al.} 2016b).

\subsection{\hdelta{} Absorption}
\label{section:hdelta}

\begin{figure*}[t!]
\begin{center}
\includegraphics[width=0.4\linewidth]{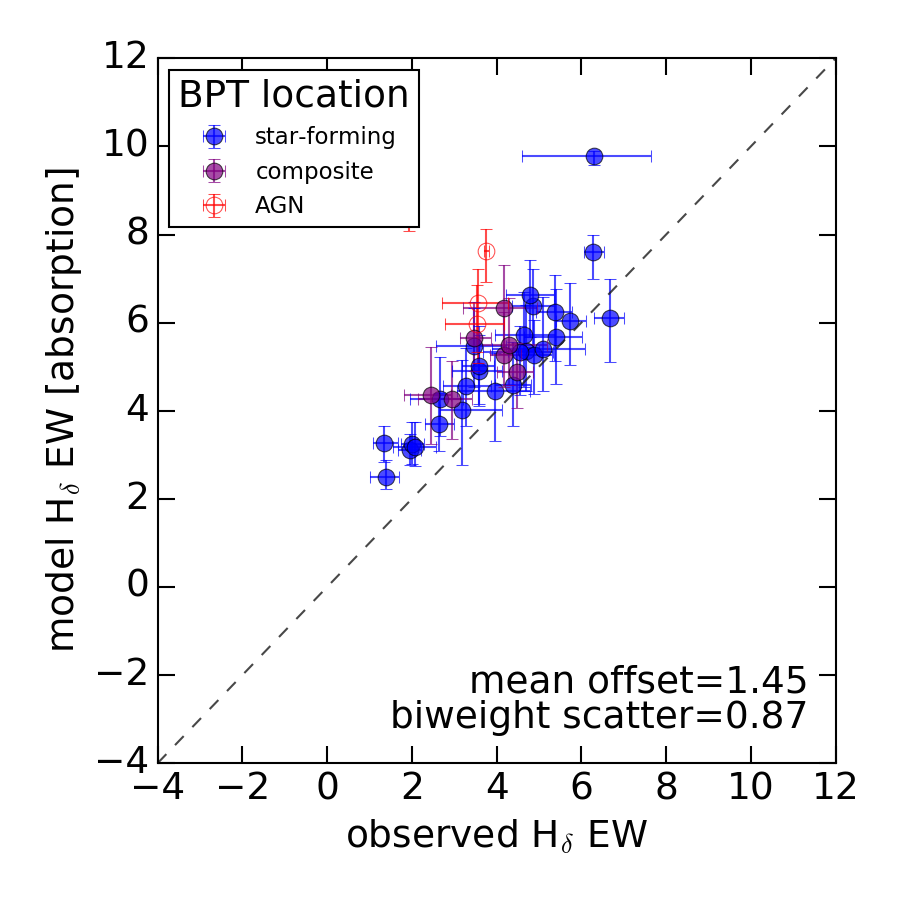}
\includegraphics[width=0.4\linewidth]{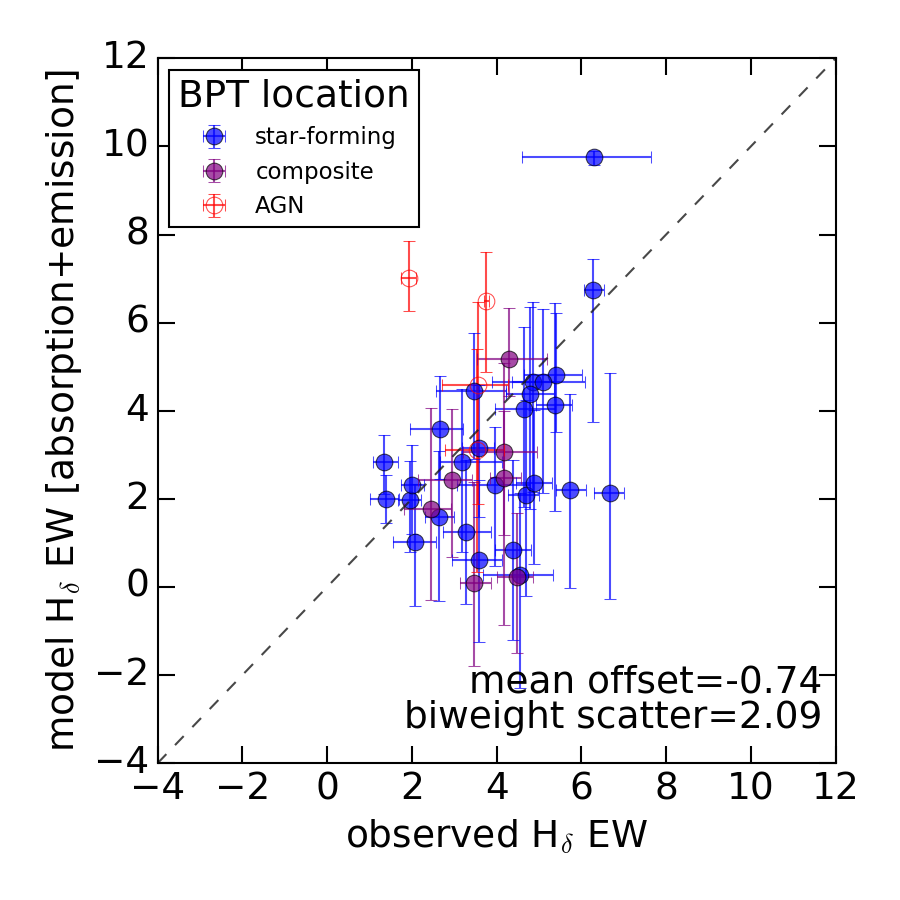}
\caption{\hdelta{} EWs measured from the observed spectra compared to predictions from \mname{} fits to the broadband photometry. The \hdelta{} feature is, in most galaxies, a mixture of absorption and emission, both of which are modeled by \prospector{}. The left panel compares to only the model absorption, while the right panel compares to the modeled emission+absorption. Here, higher values of EW indicate more absorption. The absorption-only models show good correlation, though they are systematically stronger than the observed values. Adding the expected nebular emission removes this offset at the expense of added scatter. Galaxies are color-coded by their BPT classification. Only galaxies with S/N \hdelta{} absorption $> 5$ are shown. The equivalent widths are in \angstrom{}.}
\label{fig:hdelta}
\end{center}
\end{figure*}

\hdelta{} is a hydrogen Balmer line that will appear in either emission or absorption, depending on the mix of stellar populations in the observed galaxy. \hdelta{} absorption is a well-known proxy for stellar age, as it measures the temperature of the main-sequence turnoff ({Worthey} \& {Ottaviani} 1997; {Kauffmann} {et~al.} 2003). To a lesser extent, it is also affected by stellar metallicity ({Worthey} \& {Ottaviani} 1997).

In Figure \ref{fig:hdelta}, we show the observed \hdelta{} EW versus the \mname{} \hdelta{} EW marginalized over the posterior of the fit parameters. Both an absorption-only model and an absorption+emission model are compared. We select galaxies where the \hdelta{} feature has an S/N $>$ 5. The \hdelta{} absorption is stronger than the emission for all points shown.

Figure \ref{fig:hdelta} shows reasonable agreement between the absorption-only model and observed \hdelta{} strength (left panel), with a small scatter of 0.87 in EW. There is a systematic offset such that the absorption-only model produces stronger absorption than is observed. When adding in the expected nebular emission (right panel), this bias is reduced and instead a small offset is produced in the other direction, such that the feature is systematically weaker than observed. This overestimation of the strength of \hdelta{} nebular emission is consistent with the mild overestimation of \halpha{} and \hbeta{} nebular emission in Figure \ref{fig:balmer_lineflux}. The enhanced scatter when using the absorption+emission model is not unexpected, given the size of the model error bars. The large model error bars come from the model is marginalizing over both emission and absorption in a regime where both are nearly of equal strength, and the fact that uncertainties in the dust attenuation model affect \hdelta{} more strongly than \halpha{} or \hbeta{}, since it is at a bluer wavelength.
\subsection{Stellar Metallicity}
\label{section:metallicity}

\begin{figure}[t!]
\begin{center}
\includegraphics[width=0.9\linewidth]{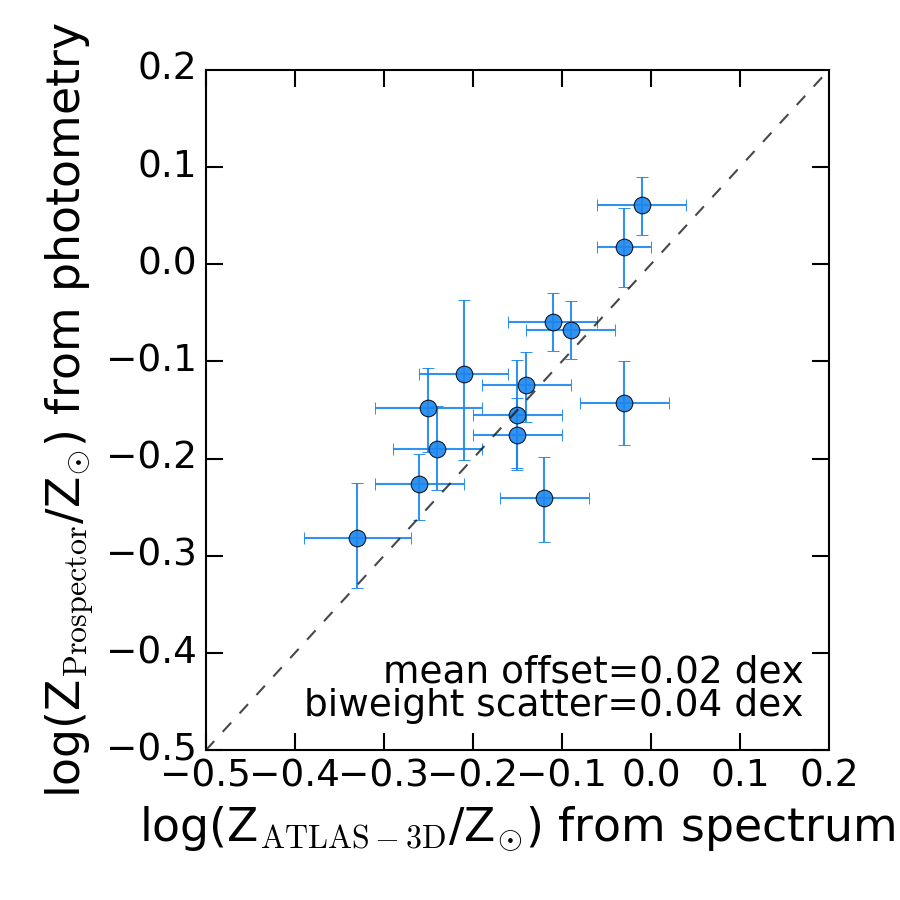}
\caption{Comparing the photometric stellar metallicities to the ATLAS-3D stellar metallicities. The ATLAS- 3D metallicities are derived from spectra using Lick indices. The Lick indices are then translated into metallicities by comparison with simple stellar population models. The \prospector{} metallicities come from fits to the broadband photometry. The agreement between the two measurements is excellent. We note that only quiescent galaxies are in the overlap with the ATLAS-3D survey.}
\label{fig:metallicity}
\end{center}
\end{figure}

We compare photometric stellar metallicities to spectroscopic metallicities from the ATLAS-3D survey ({Cappellari} {et~al.} 2011), presented in {McDermid} {et~al.} (2015). ATLAS-3D simultaneously estimates metallicities, ages, alpha abundances using {Schiavon} (2007) simple stellar populations matched to observed measurements of the Lick indices ({Worthey} {et~al.} 1994) for \hbeta{}, Fe5015, and Mg b. The overlap between the {Brown} {et~al.} (2014) sample contains only quiescent galaxies. The ATLAS-3D metallicities are measured within the effective radius, R$_\mathrm{e}$, whereas the {Brown} {et~al.} (2014) spectroscopic aperture varies substantially from galaxy-to-galaxy (see Figure \ref{fig:intro}), though it is generally larger than R$_\mathrm{e}$. This may introduce systematics on the order of $\sim$0.1 dex into this comparison, as early-type galaxies on average have negative stellar metallicity gradients with radius (e.g., {Zheng} {et~al.} 2017 measure a gradient of -0.09 $\pm$ 0.01 dex/R$_\mathrm{e}$ in nearby ellipticals). The ATLAS-3D metallicities are compared with \mname{} metallicities in Figure \ref{fig:metallicity}.

There is excellent one-to-one correlation between the photometric and spectroscopic metallicities, with little scatter or offset. The scatter is consistent with the error bars on the ATLAS-3D metallicities and the \mname{} metallicities. This agreement is encouraging, given the very different methodology used to estimate stellar metallicities. It would be useful in the future to also compare to spectroscopic metallicities from star-forming galaxies, as the photometric recovery of stellar metallicities may be a strong function of specific star formation rate.
\section{Interpretation of the Results}
\label{section:discussion}
\subsection{A Critical Look at the Balmer Emission Line Luminosities}
\label{section:balmer_line_discussion}
In this section we critically examine the factors determining the Balmer line luminosities within the \mname{} model, and show that the observed scatter is consistent with predictions from the model to within 20-30\% for \halpha{} and \hbeta{}.

Specifically, Section \ref{section:halpha_pars}, we examine in turn each of the factors in the \mname{} model affecting the Balmer line luminosity: SFR, stellar metallicity, and dust attenuation. The agreement between the observed and predicted lines validates the accuracy of these model parameters. It is also demonstrated that neglecting the effect of stellar metallicity on the \halpha{} luminosity results in a substantial increase of the scatter between the observations and model predictions. In Section \ref{section:halpha_broad}, we investigate how \halpha{} can be determined from the broadband photometry, and argue that \halpha{} is best predicted from the broadband photometry by full UV-IR SED fits. Finally, in Section \ref{section:halpha_resid}, we explore the origins of the scatter between the model and observed \halpha{} and \hbeta{} emission, showing that it is (a) not dominated by errors in the reddening curve, and (b) the scatter is consistent with the width of the posteriors in emission line luminosity from the \mname{} model, which is ultimately determined by the size of the errors in the broadband photometry.
\subsubsection{Which Model Parameters Determine Balmer Line Luminosity?}
\label{section:halpha_pars}

\begin{figure*}[ht!]
\begin{center}
\includegraphics[width=0.46\linewidth]{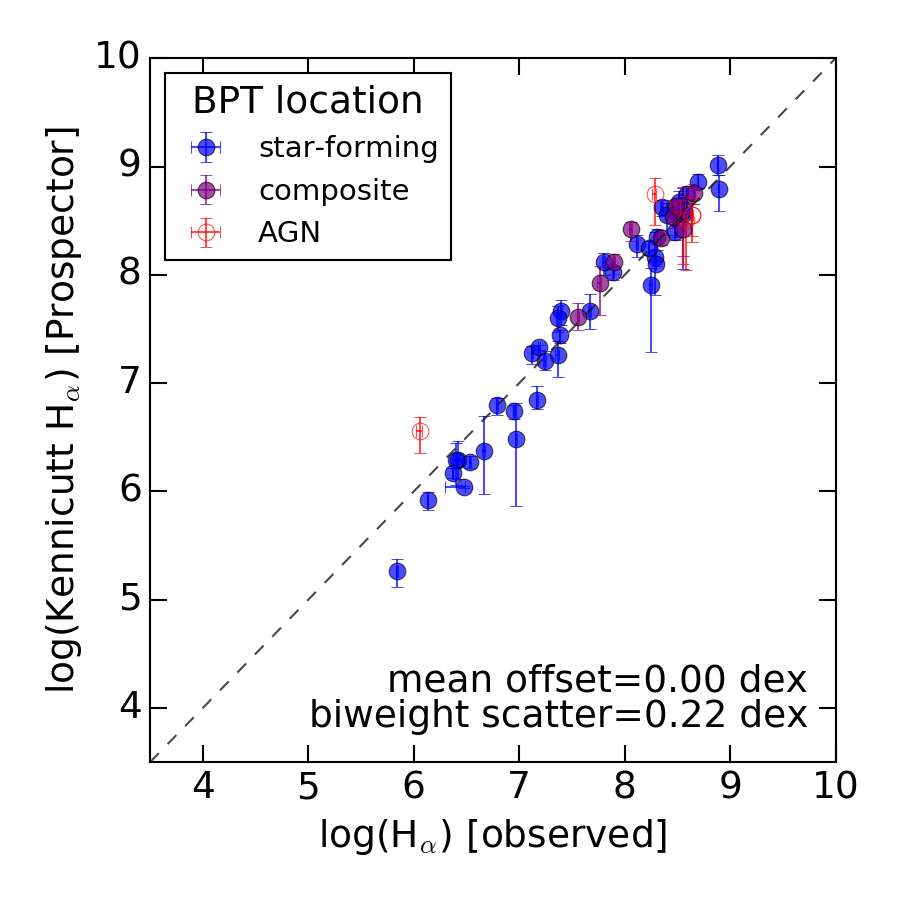}
\includegraphics[width=0.46\linewidth]{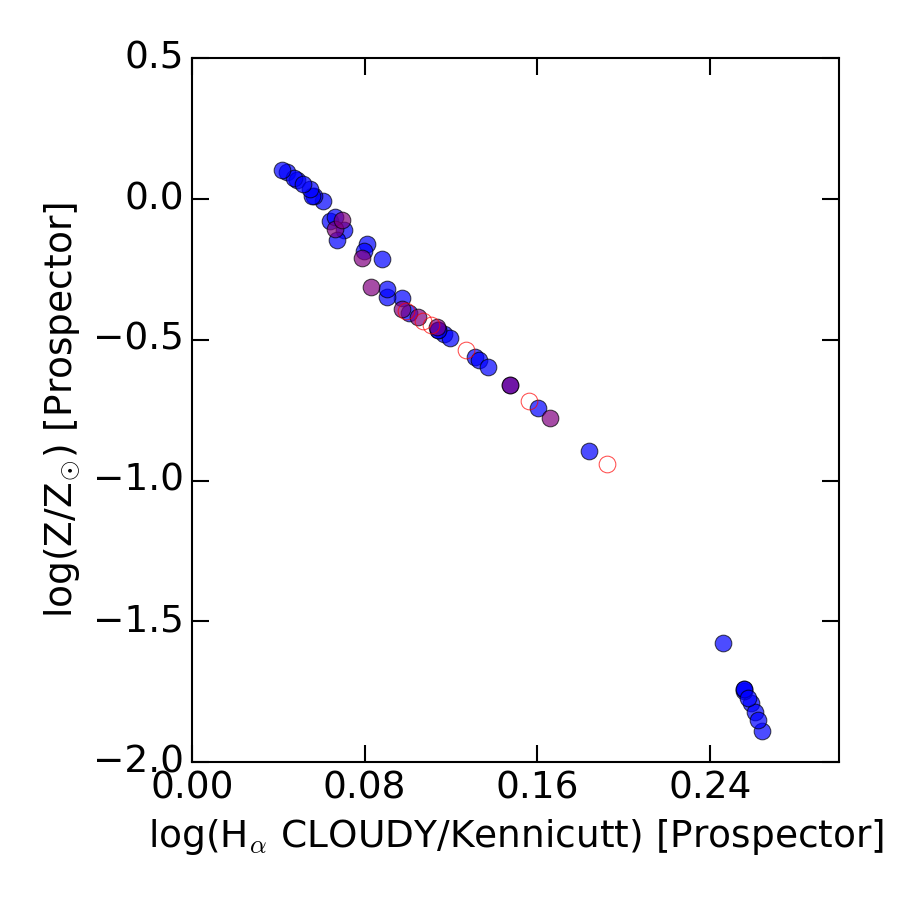}
\caption{Observed \halpha{} luminosity in units of L$_{\odot}$ compared to the \halpha{} luminosity calculated from Equation \ref{ks_eqn}, and attenuated by the model dust properties (left panel). The scatter is 0.23 dex, compared to the scatter of 0.18 dex from the \cloudy{} \halpha{} predictions which incorporate model stellar metallicities. The right panel shows the ratio between \cloudy{} versus Kennicutt \halpha{} luminosities. The residuals are nearly a monotonic function of stellar metallicity. This demonstrates that incorporating the stellar metallicity improves the prediction of \halpha{} luminosities. Galaxies are color-coded by their BPT classification. Only galaxies with S/N (\halpha{}, \hbeta{}) $> 5$ are shown.}
\label{fig:ks_halpha}
\end{center}
\end{figure*}

Balmer emission lines are produced when hydrogen recombines after being ionized by high-energy photons. The conversion from number of ionizing photons to the intrinsic \halpha{} luminosity is effectively a constant. The primary stellar source of ionizing photons is young, massive stars that live for a very short period of time ($<$10 Myr; e.g., {Kennicutt} 1998; {Conroy} 2013 and references therein). Testing model predictions of the \halpha{} luminosity is thus, in large part, a test of the recent star formation rate.

In addition to star formation rate, the dust attenuation and the stellar metallicity affect the emission line luminosity. In the \mname{} galaxy model, the dust attenuation of the line emission is determined by a combination of three parameters: the optical depth of the diffuse dust, the optical depth of the birth-cloud dust, and the shape of the diffuse dust attenuation curve. The birth-cloud dust and diffuse dust are largely degenerate in fits to the broadband photometry (see Figure \ref{fig:model_diagram2}): the ratio of the two is constrained in the model priors to mitigate the uncertainty.

Stellar metallicity affects the \halpha{} and \hbeta{} luminosities by changing the production rate of ionizing photons at a fixed star formation rate: at lower metallicities, there is an increased number of ionizing photons, resulting in a higher \halpha{} luminosity. To illustrate the effect of stellar metallicity on the \halpha{} luminosities, we use the {Kennicutt} (1998) conversion between recent star formation rate and predicted \halpha{} luminosity:
\begin{equation}
\label{ks_eqn}
\mathrm{SFR} (\mathrm{M}_{\odot} / \mathrm{yr}) = \frac{\mathrm{L(H}_{\alpha})}{1.26 \times 10^{41} \; \mathrm{erg s}^{-1}}.
\end{equation}
This is derived using a ({Salpeter} 1955) initial mass function (IMF) and assumes a fixed, solar metallicity. We correct our star formation rates from a Chabrier to a Salpeter scale by multiplying by a factor of 1.7, which accounts for the additional low-mass stars in a Salpeter IMF.

We plug the SFR inferred from the photometric fit into Equation \ref{ks_eqn} to give another estimate of \halpha{} luminosity (in addition to the \cloudy{} estimates), and attenuate the resulting \halpha{} luminosities with the \mname{} model dust attenuation.

In the left panel of Figure \ref{fig:ks_halpha}, we compare the analytical, fixed-metallicity \halpha{} luminosities from Equation \ref{ks_eqn} to the observations. There is an additional scatter of 0.05 dex over the same comparison with the \cloudy{} \halpha{} predictions (Figure \ref{fig:balmer_lineflux}). 

The difference between the two estimates is because the \cloudy{} predictions integrate the ionizing flux directly from the stellar model, thus including the effect of a variable stellar metallicity on the number of ionizing photons. We demonstrate this directly by comparing the ratio of the analytical \halpha{} luminosities to the \cloudy{} predictions in the right panel of Figure \ref{fig:ks_halpha}, as a function of model stellar metallicity. The ratio of the two is a nearly monotonic function of stellar metallicity.

Because the gas-phase and stellar metallicities are set equal in the \mname{} model (Section \ref{section:nebemission}), the right panel of Figure \ref{fig:ks_halpha} includes the effect of changing the gas-phase abundance within \cloudy{}. However, this effect is negligible compared to the effect of stellar metallicity.

The increased predictive power of the \cloudy{} \halpha{} predictions, compared to the \halpha{} predictions from Equation \ref{ks_eqn}, is indirect evidence that the \mname{} metallicities for star-forming galaxies derived from the broadband photometry are accurate.

We have shown here that neglecting the effect of stellar metallicities can dominate the scatter between model and observed \halpha{}. We thus caution against interpreting observational differences between \halpha{} SFRs and other SFR indicators as variation in SFRs over a short timescale, without first taking stellar metallicity into account.

\subsubsection{How is \halpha{} Luminosity Predicted From the Broadband Photometry?}
\label{section:halpha_broad}

\begin{figure*}[ht!]
\begin{center}
\includegraphics[width=0.9\linewidth]{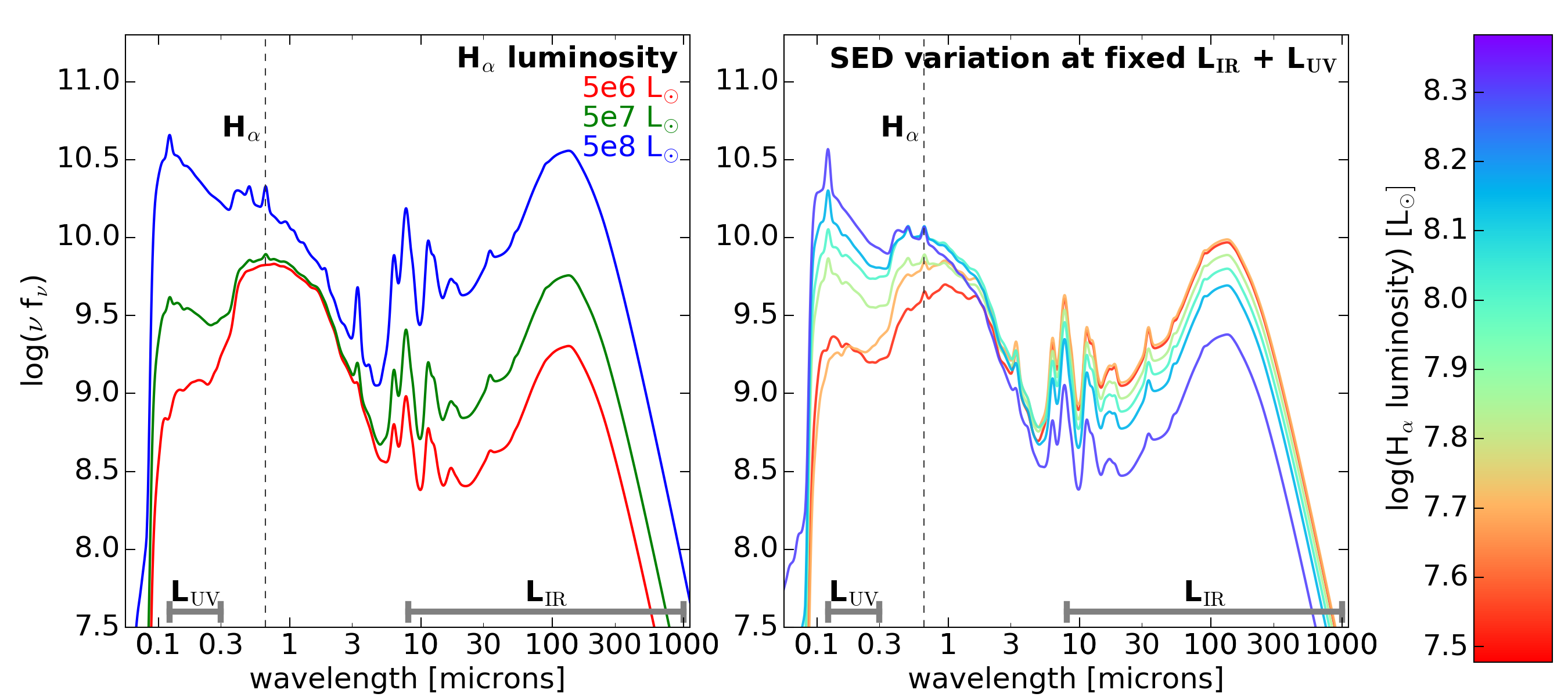}
\caption{Demonstrating how \halpha{} is inferred from the SED. The left panel shows the typical galaxy SED at three \halpha{} luminosities. The primary signature of increasing \halpha{} luminosity is increasing \lir{} + \luv{} ; however, measuring the \lir{} + \luv{} luminosity alone is insufficient. The right panel shows a series of SEDs created with fixed \lir{} + \luv{} . The SEDs are color-coded by the observed \halpha{} luminosity, after taking dust extinction into account. At fixed \lir{} + \luv{} , there is still variation in the observed \halpha{} luminosity due to dust attenuation. We demonstrate that the nebular dust attenuation can be accurately predicted from fits to the UV-MIR broadband photometry in Figure \ref{fig:reddening}.}
\label{fig:fixed_halpha}
\end{center}
\end{figure*}

In this section we demonstrate how the model \halpha{} luminosity is sensitive to the shape and normalization of the observed SED, and argue that full-SED fitting is the best way to predict \halpha{} from the photometry.

Typical mock SEDs at three different \halpha{} luminosities are generated by drawing model parameters that affect \halpha{} luminosity (dust, metallicity, and SFH) randomly within their priors, while other model parameters are held fixed. The typical SEDs at each \halpha{} luminosity are shown in the left panel of Figure \ref{fig:fixed_halpha}.

The primary signature of increasing \halpha{} luminosity in the galaxy SED is an increase in the total infrared and UV luminosities, \lir{}+\luv{}. This is because \luv{} and \lir{} are directly correlated with the recent star formation rate ({Kennicutt} 1998).

Measuring \lir{}+\luv{} alone does not uniquely determine the \halpha{} luminosity. The right panel of Figure \ref{fig:fixed_halpha} shows individual model SEDs created at a fixed \lir{}+\luv{}. The SEDs are color-coded by the resulting observed \halpha{} luminosity, which spans a range of $\sim$0.7 dex. This variation in \halpha{} luminosity comes from variation in the dust attenuation at \halpha{} wavelengths, which is visible in the right panel of Figure \ref{fig:fixed_halpha} as an gradual increase in \lir{} / \luv{} with decreasing \halpha{} luminosity.

The upshot of Figure \ref{fig:fixed_halpha} is that accurate predictions of the \halpha{} luminosity require (1) estimates of the energy budget available for \halpha{} photon production (\luv{} + \lir{}), and (2) the dust attenuation towards nebular regions at \halpha{} wavelengths. 

With only UV-MIR broadband photometry, it is possible to obtain a crude estimate of the total energy budget available for \halpha{} photon production from measurements of \luv{} and the UV slope $\beta$ by using the \luv{}/\lir{} (known as 'IRX')-$\beta$ relationship\footnote{We note that in the special case of galaxies with very high sSFRs, it is possible to measure \halpha{} directly from the broadband photometry after subtracting the stellar contribution (e.g., {Smit} {et~al.} 2016), though the scatter in this method is substantial} ({Meurer}, {Heckman}, \& {Calzetti} 1999; {Hao} {et~al.} 2011). This method will also give the nebular dust attenuation, if one assumes a shape for the dust attenuation curve and a conversion from stellar to nebular attenuation. However, the IRX-$\beta$ relationship is sensitive to variations in the intrinsic UV slope from variation in the recent SFH ({Kong} {et~al.} 2004; {Boquien} {et~al.} 2012); furthermore, we show in Section \ref{section:dustcurve} that the dust attenuation curve shows strong galaxy-to-galaxy variation. 

{Hao} {et~al.} (2011) perform the above conversion, from observed UV slope and luminosities to the dust-free UV luminosity, finding an rms scatter of 0.23 dex between the observed and predicted \halpha{}. However, they use the spectroscopic Balmer decrement to correct the \halpha{} luminosity, rather than attempting to estimate A$_{\mathrm{H}\alpha}$ directly. They find that the scatter between A$_{\mathrm{H}\alpha}$ and A$_{\mathrm{FUV}}$ without FIR fluxes to be on the order of $\sim$0.6 magnitudes or $\sim$0.25 dex (see their Figure 15): this means that using UV-MIR photometric data only would increase the scatter further.

We conclude that full UV-IR SED modeling approach that we present in this work is the most robust way to estimate \halpha{} luminosities from the broadband photometry, as it requires no assumptions about the intrinsic stellar populations or the shape of the attenuation curve. We note that having direct \lir{} measurements from \herschel{} reduces the uncertainty in both approaches, though it may be more helpful for the IRX-$\beta$ approach, as \lir{} recovery from UV-MIR imaging alone after applying the appropriate priors on the shape of the IR SED is excellent (see Appendix \ref{section:herschel}).

\subsubsection{Model Error Budget in the Emission Line Predictions}
\label{section:halpha_resid}

\begin{figure}[t!]
\begin{center}
\includegraphics[width=0.9\linewidth]{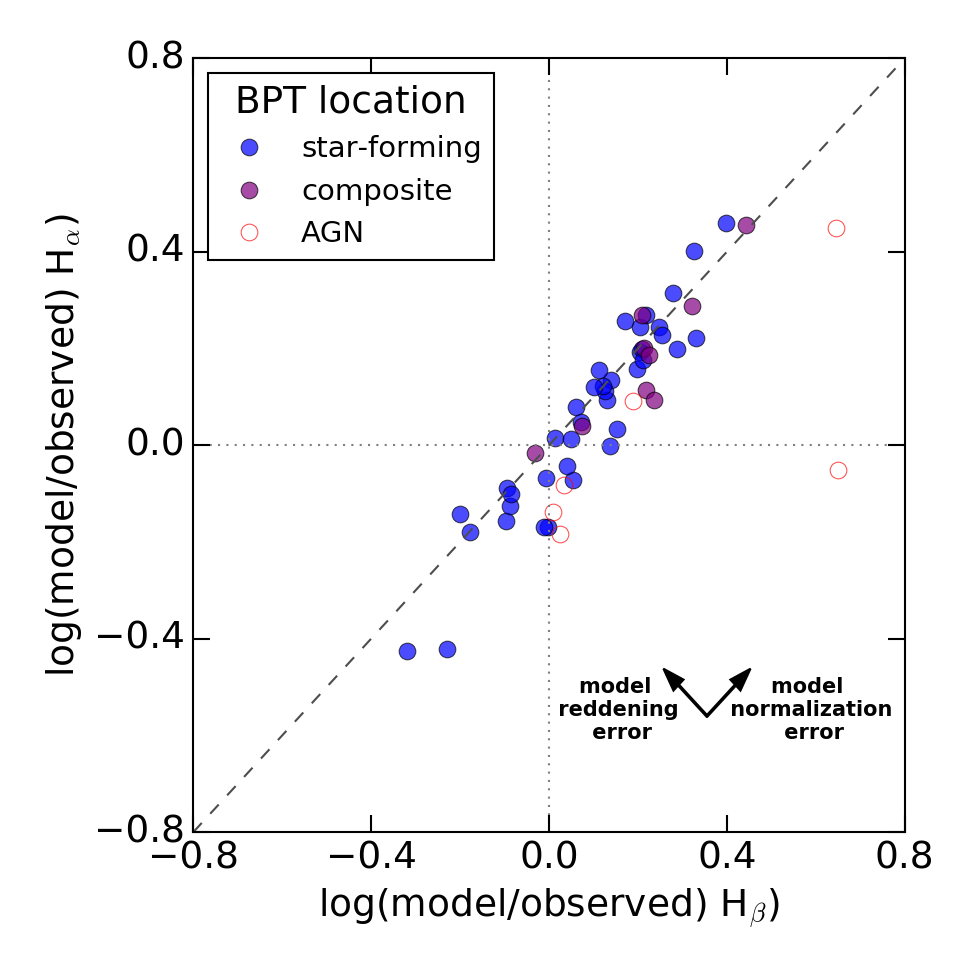}
\caption{Correlation in the residuals between \halpha{} and \hbeta{} luminosities. Since the ratio of \halpha{} and \hbeta{} photons is fixed by atomic physics, the only parameter which has a different effect on \halpha{} and \hbeta{} residuals is the reddening curve. The strong correlation in the \halpha{} and \hbeta{} residuals shows the scatter comes from errors in the production rate of Balmer photons (from errors in the star formation rate, metallicity, or other effects), which dominate over errors in the reddening curve. Galaxies are color-coded by their BPT classification. Only galaxies with S/N (\halpha{}, \hbeta{}) $> 5$ are shown.}
\label{fig:balmer_line_resid}
\end{center}
\end{figure}

\begin{figure*}[ht!]
\begin{center}
\includegraphics[width=0.85\linewidth]{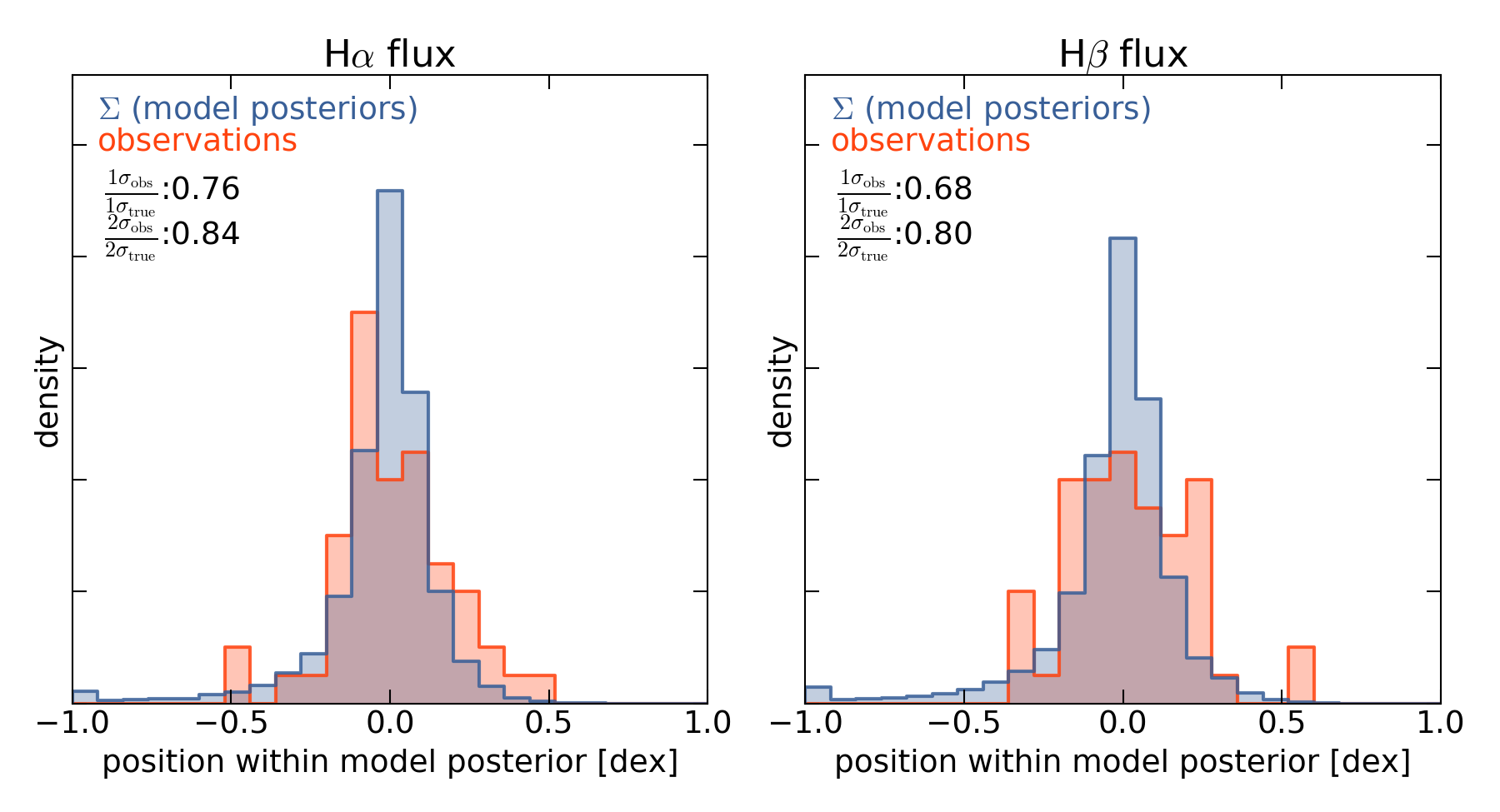}
\caption{The extent to which the model posterior probability functions accurately capture the observed distribution of \halpha{} and \hbeta{}. The red histograms show the location of the observed \halpha{} and \hbeta{} luminosities relative to the center of the model posterior, with the offset global between the observations and model removed. The blue histograms show the stacked model posterior probability functions. If the adopted \mname{} model perfectly described emission line production and attenuation, these two histograms would overlap within Poisson errors. The \mname{} error bars describe the distribution of true emission line fluxes with good accuracy: the Balmer line distributions are correct to within 20-30\%. Only galaxies with S/N (\halpha{}, \hbeta{}) $> 5$ are shown.}
\label{fig:balmer_line_posterior}
\end{center}
\end{figure*}

In this section we investigate the scatter between the predicted and observed Balmer emission line luminosities in Figure \ref{fig:balmer_lineflux}, and to what extent this scatter is consistent with the uncertainty on the model predictions.

We first compare the residuals in \halpha{} to the residuals in \hbeta{} in Figure \ref{fig:balmer_line_resid}. Since the intrinsic flux ratio between the Balmer lines is nearly invariant, the correlation between emission line residuals are informative. Correlated residuals are due to errors in normalization of the Balmer lines: intrinsic emission line strength (errors in SFR or stellar metallicity), normalization of the dust attenuation, or errors in the underlying continuum absorption. Uncorrelated errors are due to errors in the reddening curve, which is a function of $\dustone{}$, $\dusttwo{}$, and \didx{} in the \mname{} model.

Figure \ref{fig:balmer_line_resid} makes it clear that the dominant source of scatter is from the normalization of the Balmer lines, and errors in the reddening curve are a secondary contribution. This can also be calculated directly from the observed scatter in the Balmer decrements in Figure \ref{fig:reddening}: reddening errors add a scatter of $\sim$0.08 dex to the \halpha{} measurements, compared to an overall scatter of 0.18 dex.

We also investigate whether the observed scatter in emission line luminosities is consistent with predictions from the \mname{} model. Each fit to the photometry results in some posterior probability distribution for the emission line fluxes. The observational errors on emission line fluxes are much smaller than the typical width of this posterior probability function\footnote{This is not true for the other observed quantities: \hdelta{}, \dn{}, Balmer decrement, and stellar metallicity.}. Thus, we can use the observations to test the model posteriors.

In Figure \ref{fig:balmer_line_posterior}, we show the sum of the model posteriors for the \halpha{} luminosity. The posteriors are stacked such that the median of each posterior is located at the origin, and the global offset is removed (the offset is shown in the corner of Figure \ref{fig:balmer_lineflux}). We also show the location of the observations for each of these fits, again stacked relative to the median of each posterior. If the model posteriors accurately describe all relevant emission line production and attenuation processes in these galaxies, then the two histograms would match each other to within Poisson errors.

The agreement between the two histograms is good, suggesting that the posterior probability distributions are accurate. This means that the scatter in Figure \ref{fig:balmer_line_resid} is largely consistent with the errors. We quantify this statement by calculating the fraction of observations that fall within the model 1$\sigma$ range (i.e., the 16th and 84th percentiles of the posterior) and 2$\sigma$ range (the 2.5th and 97.5th percentiles of the posterior). These are shown in the upper-left corner of Figure \ref{fig:balmer_line_posterior}. This test shows that the model error bars are accurate to within 20-30\% for \halpha{} and \hbeta{}.
\subsection{Dust Properties}
\label{section:dust}
Here we explore correlations in the model dust properties. We show a strong trend in the relationship between dust optical depth and the shape of the attenuation curve, and compare to expectations from simulations in Section \ref{section:dustcurve}. In Section \ref{section:pah}, we demonstrate the recovery of the PAH mass fraction from the photometric fits, and explore trends between the stellar mass and the PAH mass fraction.

\subsubsection{Relationship Between the Dust Attenuation Curve and the Dust Optical Depth}
\label{section:dustcurve}

\begin{figure}[t!]
\begin{center}
\includegraphics[width=0.9\linewidth]{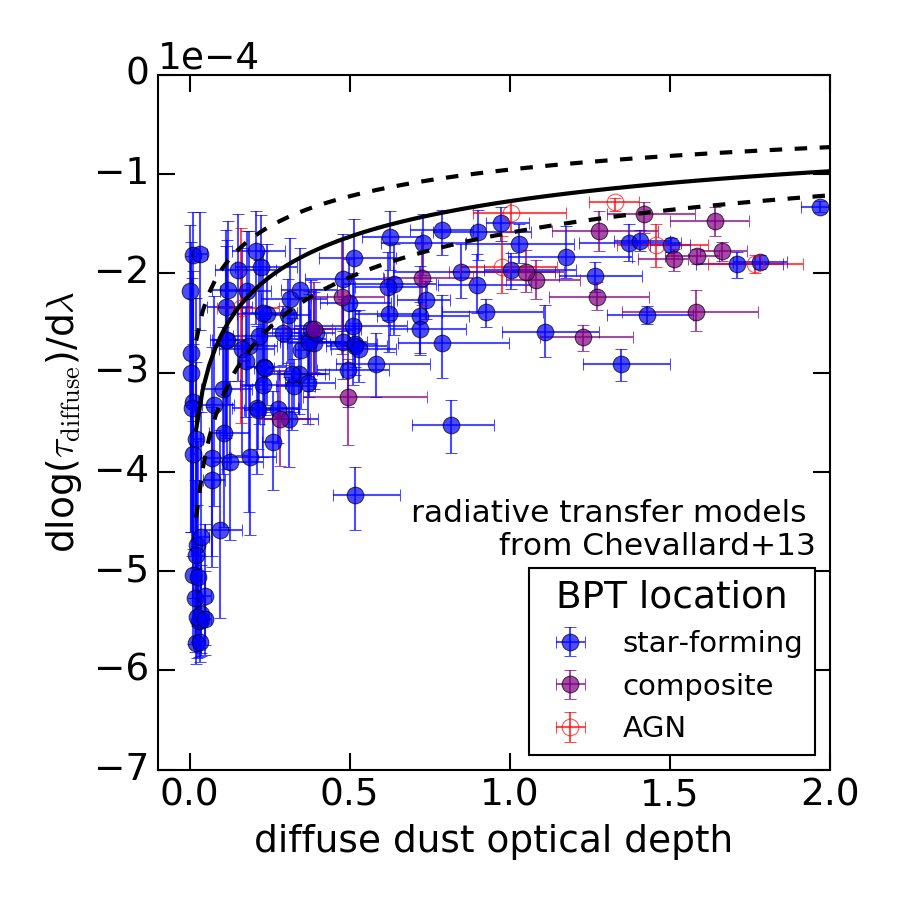}
\caption{Correlation between the optical depth of the diffuse dust and the slope of the attenuation curve for the diffuse dust. This relationship, derived solely from fits to the broadband photometry, is in good agreement with theoretical expectations from {Chevallard} {et~al.} (2013). The black solid line indicates the mean relation from {Chevallard} {et~al.} (2013), while the dashed lines indicate the dispersion about this relation. Predictions from the \mname{} fits to the {Brown} {et~al.} (2014) catalog are shown as points color-coded by their BPT classification.}
\label{fig:dust_correlations}
\end{center}
\end{figure}

A relationship between the dust attenuation optical depth and the shape of the attenuation curve in galaxies has been predicted both analytically ({Bruzual}, {Magris}, \& {Calvet} 1988) and numerically, with radiative transfer codes ({Witt}, {Thronson}, \& {Capuano} 1992; {Witt} \& {Gordon} 2000; {Gordon} {et~al.} 2001; {Chevallard} {et~al.} 2013). The wavelength-dependent scattering properties of dust and the mixed star-dust geometry both contribute to this effect. In brief, at low optical depths, red light is scattered more isotropically and escapes the galaxy, while blue light is more forward-scattered and is subjected to more absorption. This results in a net steepening of the attenuation curve. At high optical depths, in a mixed star-dust geometry, the observed radiation primarily comes from stars at an optical depth less than unity. Redder photons travel more physical distance than bluer photons before absorption or scattering; the net effect is that larger optical depths result in a greyer attenuation curve.

In Figure \ref{fig:dust_correlations}, we plot the optical depth of the diffuse dust versus the slope of the attenuation curve at 5500 \angstrom{} (dlog$\tau$ /d$\lambda$), compared to a compilation of dusty radiative transfer models from {Chevallard} {et~al.} (2013). The qualitative agreement is remarkable: as the observed dust optical depth increases, the observed attenuation curves become greyer. There is clear evidence for a systematic relationship between the dust optical depth and the shape of the attenuation curve, although the models predict a somewhat greyer attenuation curve than is observed at large optical depths. We caution that there is a correlation between the optical depth of the dust and the attenuation curve for individual galaxies which could partially mimic this relationship; however, it is not large enough to explain the observed sample trend.

The dust attenuation curve is often a fixed parameter when fitting the photometric SEDs of galaxies, even for complex SED fitters with many free parameters. Yet there is substantial evidence that the attenuation curve varies: both theoretically, as discussed above, and observationally, with serious differences between the Milky
Way ({Cardelli}, {Clayton}, \&  {Mathis} 1989), SMC ({Prevot} {et~al.} 1984; {Bouchet} {et~al.} 1985), and LMC ({Fitzpatrick} 1986), as well as in high redshift galaxies ({Buat} {et~al.} 2012; {Reddy} {et~al.} 2015; {Salmon} {et~al.} 2016).

The assumption of a fixed attenuation curve can have profound effects on the results of SED fits. For example, assuming a fixed {Calzetti} {et~al.} (2000) attenuation curve will underpredict the 1500 \angstrom{} luminosity dust correction by $\sim$2 at low optical depths, and overpredict the dust corrections by a factor of 2-5 at high optical depths ({Salmon} {et~al.} 2016). Scatter in the relationship between \lir{}/\luv{} and UV-slope could be reduced substantially by taking into account the variation in the slope of the attenuation curve ({Buat} {et~al.} 2012). Assuming the shape of the attenuation curve also affects the bulk properties of the galaxy population: {Marchesini} {et~al.} (2009) shows that the low-mass slope of the stellar mass function steepens and the normalization decreases by 20\%-50\%.

Given that the shape of the dust attenuation curve is well-recovered in photometric mock tests (Appendix \ref{appendix:mock_tests}), the excellent accuracy in which the model fits predict the reddening from the Balmer decrements (Figure \ref{fig:reddening}), and the biases introduced by a fixed attenuation curve in SED fits, it is wise to use a variable dust law when fitting galaxy SEDs.
\subsubsection{The PAH Mass Fraction}
\label{section:pah}
\begin{figure*}
\begin{center}
\includegraphics[width=0.95\linewidth]{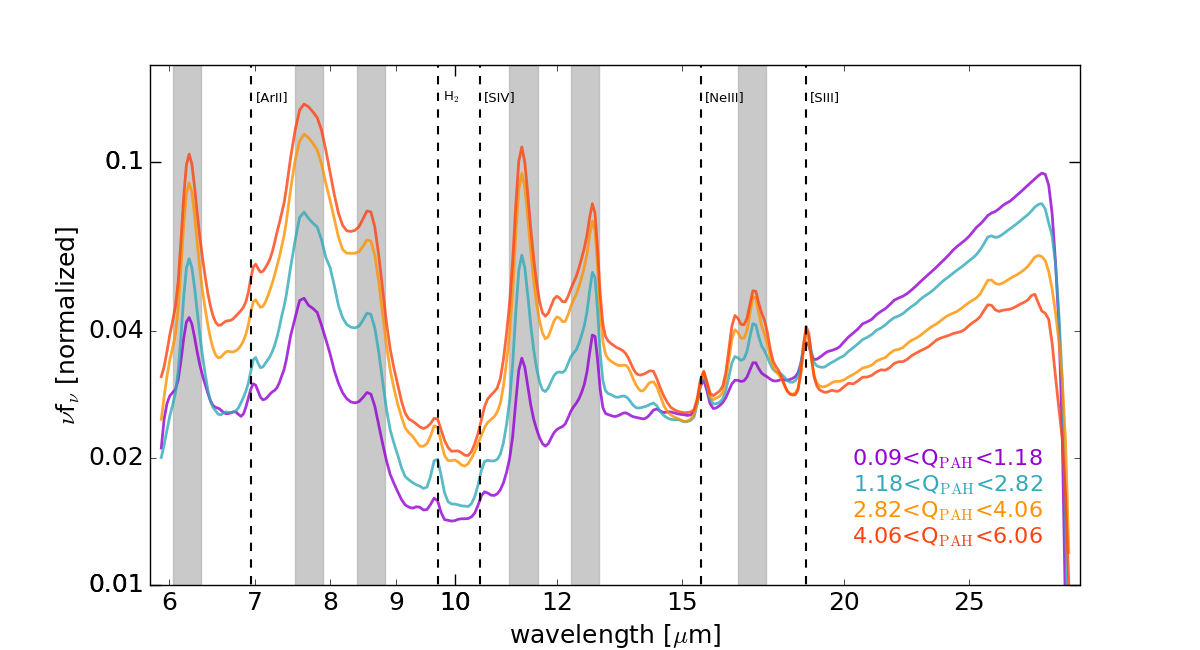}
\caption{Observed \spitzer{} IRS spectra, stacked in bins of median \qpah{}. \qpah{} is derived from the posteriors of the fits to the broadband photometry, which is a completely independent measurement from the spectra. Prominent PAH emission features are marked with dashed lines. Strong nebular emission lines are masked in grey bands. The strong correlation between the photometric \qpah{} and the observed luminosity in PAH features is an excellent validation of the photometric fits. Only galaxies with log(M) $>10$ are included, to minimize confounding trends in stellar mass.}
\label{fig:irs_stack}
\end{center}
\end{figure*}

\begin{figure}
\begin{center}
\includegraphics[width=0.95\linewidth]{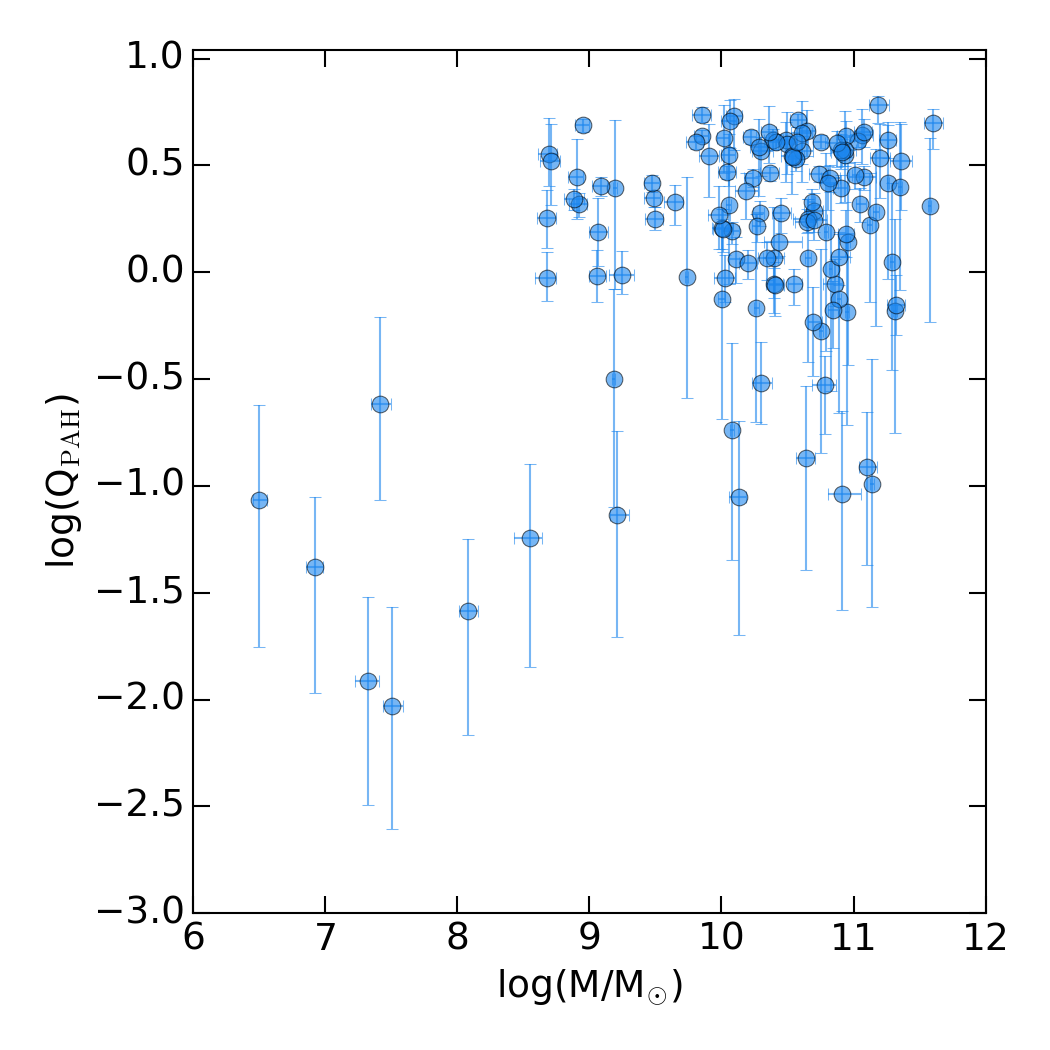}
\caption{Stellar mass as a function of the PAH mass fraction, derived from the fits to the photometry. Assuming a tight mass-metallicity relationship, stellar mass can be used as a proxy for gas-phase metallicity. The low fraction of PAHs in low-metallicity galaxies, and the large range in PAH fraction at high metallicities, is consistent with previous observations ({Roche} {et~al.} 1991; {Hunt}, {Bianchi}, \& {Maiolino} 2005; {Draine} \& {Li} 2007), suggesting that \qpah{} can be robustly determined from full SED modeling of galaxy broadband photometry.}
\label{fig:qpah_correlations}
\end{center}
\end{figure}

In this section we investigate the PAH mass fraction, parameterized in the {Draine} \& {Li} (2007) dust emission model as \qpah{}. Many studies use broadband photometry in the MIR to estimate total infrared luminosities, either by using fixed IR templates (e.g., {Daddi} {et~al.} 2007; {Wuyts} {et~al.} 2008; {Buat} {et~al.} 2009; {Whitaker} {et~al.} 2014) or by fitting IR models to a limited wavelength range (e.g., {Chang} {et~al.} 2015). The PAH mass fraction is of particular interest because PAH emission can dominate the integrated flux at MIR wavelengths. Thus, the natural variation of the PAH mass fraction in galaxies has serious consequences for estimating SFRs from photometry: unless \qpah{} can be constrained with available photometry, systematic errors in SFR estimates will be introduced by assuming a fixed IR template.

In Figure \ref{fig:irs_stack}, we investigate the recovery of \qpah{} from photometric fits to the full SED. We show the \spitzer{} IRS spectra stacked in bins of median \qpah{}, which come from the posteriors of the fits to the broadband photometry. Only galaxies with log(M) $>10$ are included, to minimize confounding trends in stellar mass. The bin sizes are chosen so that the number of galaxies is the same for each bin ($N=20$). Individual spectra are normalized such that the total flux observed in the \spitzer{} IRS spectrum is equal to unity, prior to stacking: this means that each galaxy contributes an equal amount of information to the stacks. The data are smoothed for presentation purposes. The following PAH features dominate the total PAH luminosity, and are marked with dashed lines: the 6.2$\mu$m ($\sim$10\% of \lpah{}), 7.7$\mu$m ($\sim$30\% of \lpah{}), 8.3$\mu$m+8.6$\mu$m ($\sim$15\% of \lpah{}), 11.3$\mu$m ($\sim$15\% of \lpah{}), 12.6$\mu$m ($\sim$10\% of \lpah{}, but blended with {\sc [Ne~ii]}), and 17$\mu$m ($\sim$10\% of \lpah{}) ({Smith} {et~al.} 2007). Strong nebular emission lines are masked in grey bands.

It is clear from Figure \ref{fig:irs_stack} that the EW of the PAH features increases as the \qpah{} from the photometric fits increases. This is an excellent validation of the photometric \qpah{} values derived from fitting the full SED. This result implies that PAH emission can be cleanly distinguished from other sources of MIR emission (e.g., circumstellar dust around AGB stars, or AGN; {Lange} {et~al.} 2016) with full-SED modeling.

Measuring total IR luminosity in PAH features for individual galaxies, rather than stacks, would provide additional compelling spectral evidence for the photometric \qpah{} values. However, this measurement is challenging: it is difficult to define the continuum in the MIR region, resulting in biased EWs by factors of $\sim$2-3 ({Smith} {et~al.} 2007). Furthermore, individual PAH EWs are sensitive to the intensity of the radiation field (particularly the presence of an AGN) and the galaxy metallicity ({Smith} {et~al.} 2007; {O'Dowd} {et~al.} 2009), which makes the interpretation of PAH EWs complex. A full validation of the IR SED models would involve a simultaneous fit to the observed \spitzer{} IRS, Akari spectra, and the photometry, with the {Draine} \& {Li} (2007) models and a model for the infrared emission of AGN. We defer this more careful treatment to a later investigation.

Figure \ref{fig:qpah_correlations} shows the relationship between stellar mass and \qpah{}. There is a clear link with stellar mass, such that \qpah{} is very low below a stellar mass of log(M/M$_{\odot}$) $\sim$ 8.8. Assuming a tight relationship between gas-phase metallicity and stellar mass, this is consistent with previous observations which show the PAH mass fraction is very low for galaxies below a gas-phase metallicity of log(O/H) + 12 $\sim$ 8.1, with considerable natural variation above this threshold ({Roche} {et~al.} 1991; {Hunt} {et~al.} 2005; {Engelbracht} {et~al.} 2005; {Draine} {et~al.} 2007). Full-SED modeling of the broadband photometry reproduces these trends in the PAH mass fraction. We note that the MIR is particularly well-sampled for these galaxies, with \spitzer{}/IRAC at 3-8$\mu$m, WISE at 3-22$\mu$m, and \spitzer{}/MIPS at 24$\mu$m, and that \qpah{} is very well-recovered in mock tests; see Appendix \ref{appendix:mock_tests}.

It has been demonstrated in this section that \qpah{}, and trends in \qpah{} with stellar mass, can be reproduced with a combination of MIR photometry and full-SED modeling. This, along with encouraging mock test results in Appendix \ref{appendix:mock_tests}, suggest that \qpah{} is accurately recovered by fits to the photometry. It is likely that full-SED modeling with MIR photometry can tighten the 8 $\mu$m / \lir{} relationship ({Elbaz} {et~al.} 2010, 2011), and thus refine SFR estimates, by including constraints from optical--IR energy conservation. It has recently been shown at $z = 2$ that \lir{} can be accurately estimated with the \fsps{} stellar populations combined with the {Draine} \& {Li} (2007) dust emission model ({Shivaei} {et~al.} 2016b). This is particularly pertinent in light of recent results from {Shivaei} {et~al.} (2016a), which demonstrate that systematic trends in PAH emission with mass bias global SFR estimates by $\sim$30\% at $z\sim2$. In future work, we will explore the use of \prospector{} to estimate accurate \lir{}, 8$\mu$m / \lir{}, and SFRs of $z>1$ galaxies.

\section{Past and Future Changes in the \mname{} Model}
\label{section:future}
Here, we include a short list of model SFH choices that were found to be impractical in the adopted fitting framework. This list may be useful to those building models within the \prospector{} interface, or other such interfaces, in future work. We also briefly discuss some of the planned additions to the \mname{} model. The \prospector{} parameter files for the \mname{} model used to create the results in this paper are the 'brownseds\_np' files at \url{https://github.com/jrleja/threedhst_bsfh/}, Github commit hash \texttt{84d32c0ece57b11d0c8424b44be160eb5a735537}.

\subsection{Unsuccessful SFH Parameterizations}
Fitting complex galaxy models to broadband photometry is a poorly-constrained problem, and as a result, the model posteriors often have substantial degeneracies. This lack of constraints, combined with the realities of an imperfect sampler and a finite amount of computing time, mean that the output posteriors can be sensitive to the parameterization of the model SFH. Listed here are a number of SFHs that we attempted to incorporate into the model, which, while accurately describing physical states which are known to occur in galaxies, prove difficult or impractical to implement.
\begin{enumerate}
\item Two separate galaxy SEDs, with distinct declining-$\tau$ SFH models and dust attenuations, which are then summed to create the composite SED. This model was intended to mimic the spatially-distinct bulge-and-disk system regularly seen in resolved images of galaxies. Each model component would have a unique SFH and dust attenuation model, to simulate e.g. a star-forming disk and a quiescent bulge. However, mock tests showed that this model had too many multimodalities for the \emcee{} sampler to sample efficiently in a reasonable amount of time. With only 5\% errors on the photometry, the sampler would find solutions widely separated in parameter space with significantly different masses and SFRs than the input values, and also with better $\chi^2$ values.
\item A delayed $\tau$ SFH with a truncation, followed by a linear ramp, as described in {Simha} {et~al.} (2014). This parameterization works very well in describing the observed photometry. However, the SFH posteriors in this parameterization are highly multimodal, and often have multi-dimensional flared, curving degeneracies. In order to achieve accurate posteriors on masses and star formation rates using this parameterization, the sampler must properly explore multiple widely-separated solutions with large degeneracies. This meant that while the best-fit results were often reasonable, the error bars were substantially underestimated. These truncated error bars also affected the accurate recovery of metallicity and dust parameters, which are highly sensitive to the SFH posteriors due to the age-dust-metallicity degeneracy.
\item A single burst component in the SFH, with two parameters describing the time and strength of the burst. Without data that can constrain changes in SFH on shorter timescales, such as spectra, bursts in the SFH would often create unnecessary multimodalities and would decouple SFR(10 Myr) from SFR(100 Myr) in ways that are not uniquely constrained by the broadband photometry. It is more efficient to use the adopted nonparametric SFH and tailor the time bins to the resolving power of the data, as described in Section \ref{section:sfh} and Johnson et al. in prep.
\end{enumerate}
\subsection{Future Improvements}
Here we briefly describe ways in which the \mname{} model will be expanded upon in future work. 

\begin{enumerate}
\item Galaxies with strong AGN contributions are known to have issues in the current \mname{} model; in particular, red MIR colors indicative of obscured AGN are currently instead fit with a very strong 100-300 Myr components, as the emission from circumstellar dust around AGB stars can mimic the MIR colors of AGN ({Alatalo} {et~al.} 2016). The {Nenkova} {et~al.} (2008a, 2008b) AGN models implemented in the \fsps{} source code will be tested and implemented in future versions of the \mname{} model.
\item The SFH at $\gtrsim$ 1 Gyr is poorly constrained from the broadband photometry (see the mock tests in Appendix \ref{appendix:mock_tests}). When the data do not put strong constraints on a parameter, the posteriors are prior-dominated. The current priors on the SFH parameters have a Gaussian shape on the logarithm of the specific star formation rate in each bin, log(SFR$_n$ / M$_\mathrm{total}$), as shown in Section \ref{section:sfh}. The center of the prior is at log(1/t$_{\mathrm{univ}}$) $\sim$ -10.1 yr$^{-1}$ and the FWHM is roughly 1.5 dex. However, it is known from observations of distant galaxies that the star formation rates were much higher in the past (e.g., {Madau} \& {Dickinson} 2014), so this prior which places a preference on constant SFHs does not agree with observations. The effects of the SFH prior on the model posteriors will be explored in future work.
\end{enumerate}
\section{Conclusion}
\label{section:conclusion}
In this paper, we present the \mname{} model for fitting the SEDs of galaxies, based on the \fsps{} stellar populations code and built in the \prospector{} inference framework. The \mname{} model includes a 6-component nonparametric star formation history, nebular emission, a variable attenuation curve, and dust attenuation and re-radiation. It uses the on-the-fly model generation and MCMC implementation within the \prospector{} framework to explore the model posteriors in a thorough, unbiased manner. We demonstrate the power of this framework by fitting the {Brown} {et~al.} (2014) broadband photometry, and comparing the model predictions to aperture-matched optical spectroscopy. \mname{} predicts observed \halpha{} luminosities from fits to the photometry with a scatter of 0.18 dex and an offset of $\sim$0.1 dex, a strong verification of the model star formation rates, dust attenuation, and stellar metallicities, across a wide range of galaxy types and stellar masses. We also find good agreement in SFH indicators (\dn{} and \hdelta{} absorption), direct tests of the reddening curve (Balmer decrements), stellar metallicities, and PAH mass fractions.

The accurate prediction of \halpha{} luminosities is a key component of these results, and it is explored in some detail. The \halpha{} luminosities are sensitive to the recent SFR, the stellar metallicities, and the dust attenuation in the \mname{} model. It is demonstrated that including stellar metallicities is key to achieving the tight scatter in \halpha{} comparisons. It is shown that the remaining scatter of 0.18 dex is consistent with the width of the model posteriors to within 20-30\% for \halpha{} and \hbeta{}, and that it is not dominated by errors in the reddening curve. The low scatter in this comparison implies that SFRs do not vary strongly over 100 Myr timescales.

The accuracy of recovered trends in the dust attenuation and re-emission model is also explored. In Section \ref{section:dustcurve}, it is shown that the \mname{} model, using flat priors, naturally recovers the relationship between the dust optical depth and the shape of the attenuation curve predicted in dust radiative transfer models. In Section \ref{section:pah}, it is demonstrated using stacked \spitzer{} IRS spectra that the PAH mass fractions are well constrained by the full-SED fits. 

We test the extent to which the model posteriors describe the spread in true galaxy properties by performing mock tests (Appendix \ref{appendix:mock_tests}). It is demonstrated that the \mname{} model produces accurate posteriors with well-estimated error bars, even for model parameters which are poorly constrained by the data. The \mname{} model performs better at predicting \halpha{} luminosities than SED-fitting codes of similar complexity (Appendix \ref{appendix:magphys}), largely due to different choices in the dust and SFH implementation. It is also shown that the far-infrared \herschel{} fluxes can be predicted from fits to the UV to MIR photometry alone after applying reasonable physical priors to the shape of the IR SED (Appendix \ref{section:herschel}). 

Broadband galaxy SEDs are composed of a complex mix of stellar populations, gas, and dust, but models which contain this complexity are often only somewhat constrained by the data. Thus, the parameters derived from these SEDs can be highly sensitive to the model parameterization. Model posterior checks for SED fitters are the key ingredient necessary to move the field forward towards precise, unbiased estimates of stellar masses and SFRs across a variety of redshifts. Here we have performed these tests for the stellar metallicities, star formation rates, star formation histories, dust reddening, and PAH emission, and found excellent agreement with the data. However, important components of the \mname{} model remain untested, in particular the stellar masses and the behavior of the model at higher redshifts. These aspects will be explored in future work.

The demonstrated flexibility and accuracy of the \prospector{} framework strongly motivates a fresh look at basic properties of the galaxy population, such as the stellar mass function and star-forming sequence as a function of redshift. Furthermore, \prospector{} allows broadband photometry to describe galaxy properties to a level of detail that was previously only reserved for spectra. The stellar mass-metallicity relationship, systematic variations in SFH with stellar mass and SFR, variations in the PAH fraction with stellar mass and metallicity, and variations in the total dust attenuation and attenuation curve with galaxy properties can all now be accurately explored with broadband photometry alone. This tremendously increases the sample sizes available to these types of studies. Going even further, the simultaneous fitting of broadband photometry and spectra can put tight constraints on quantities that would otherwise be highly degenerate, such as the timescale of SFH variations and the ratio between stellar and nebular attenuation. We plan to re-examine SFR-mass relations and the evolution of the stellar mass function to see if this framework is capable of producing self-consistent SFRs and masses over most of cosmic time.

\acknowledgements
We thank Marijn Franx, Eric Bell, Frank van den Bosch, and Nikhil Padmanabhan for valuable discussion and comments, and Michael Brown for the excellent spectrophotometric catalog upon which this work is based. C.C. acknowledges support from NASA grant NNX13AI46G, NSF grant AST-1313280, and the Packard Foundation. P.V.D. acknowledges support from NASA grants NNX13A146G and HST-GO-12177. This work used the Extreme Science and Engineering Discovery Environment (XSEDE, {Towns} {et~al.} 2014), which is supported by National Science Foundation grant number ACI-1053575. This time was granted under XSEDE allocations TG-AST140054 and TG-AST150015. Some of the computations in this paper were run on the Odyssey cluster supported by the FAS Division of Science, Research Computing Group at Harvard University.

\appendix
\section{Mock Tests}
\label{appendix:mock_tests}
\renewcommand\thefigure{\thesection.\arabic{figure}}    
\setcounter{figure}{0}    

\begin{figure*}
\begin{center}
\includegraphics[width=0.9\linewidth]{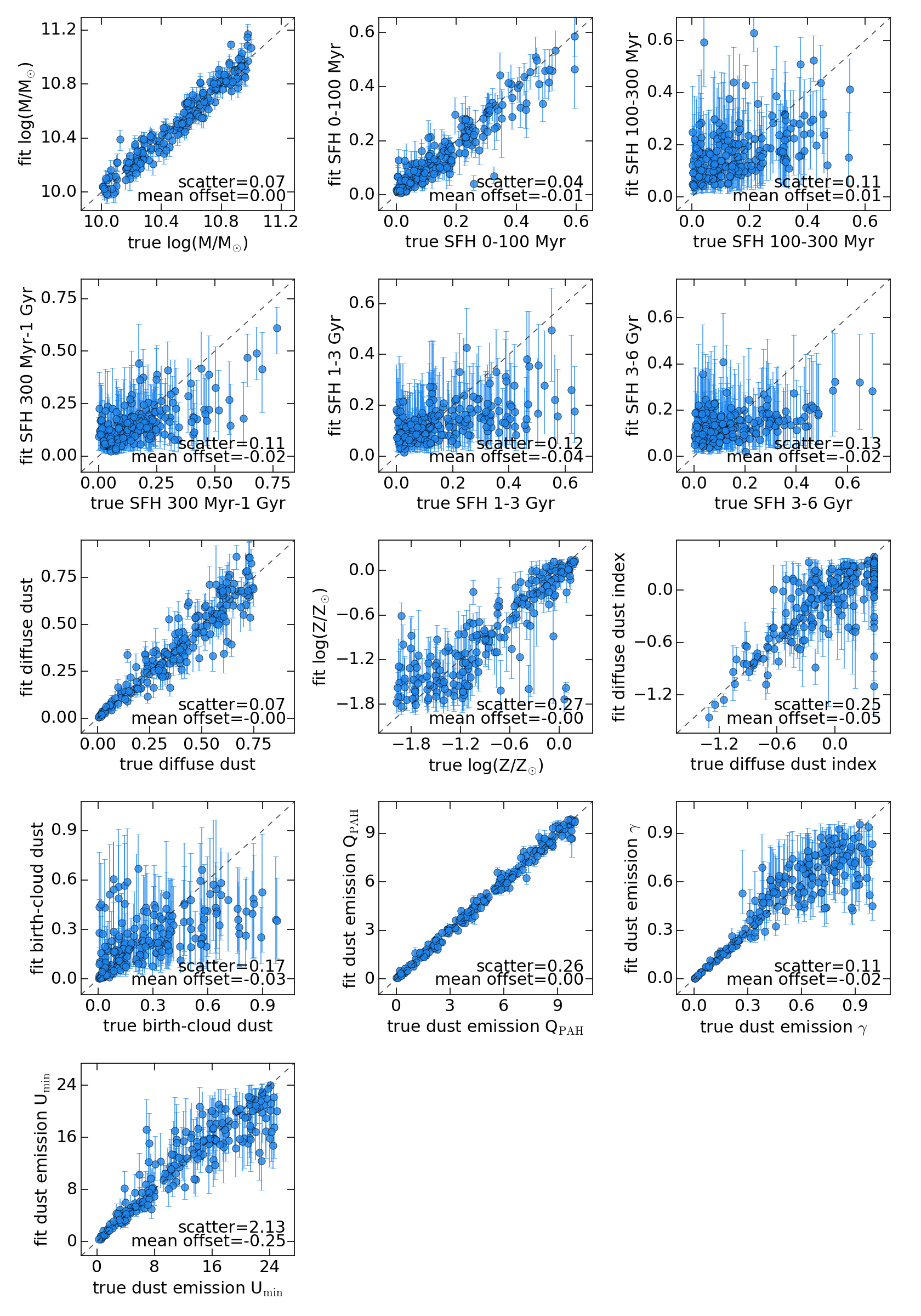}
\caption{Mock parameter recovery from fits to the broadband photometry. The x-axes shows the input parameters used to generate the model SED, while the y-axes shows the expectations from the fit. The SFH at $\gtrsim$ 1 Gyr is not well-recovered in mock tests, though this is likely quite sensitive to the adopted SFH prior, which favors a continuous SFH (see Section \ref{section:sfh}).}
\label{fig:mock_fitpars}
\end{center}
\end{figure*}

\begin{figure*}
\begin{center}
\includegraphics[trim={6cm 3cm 6cm 4cm},width=0.44\linewidth]{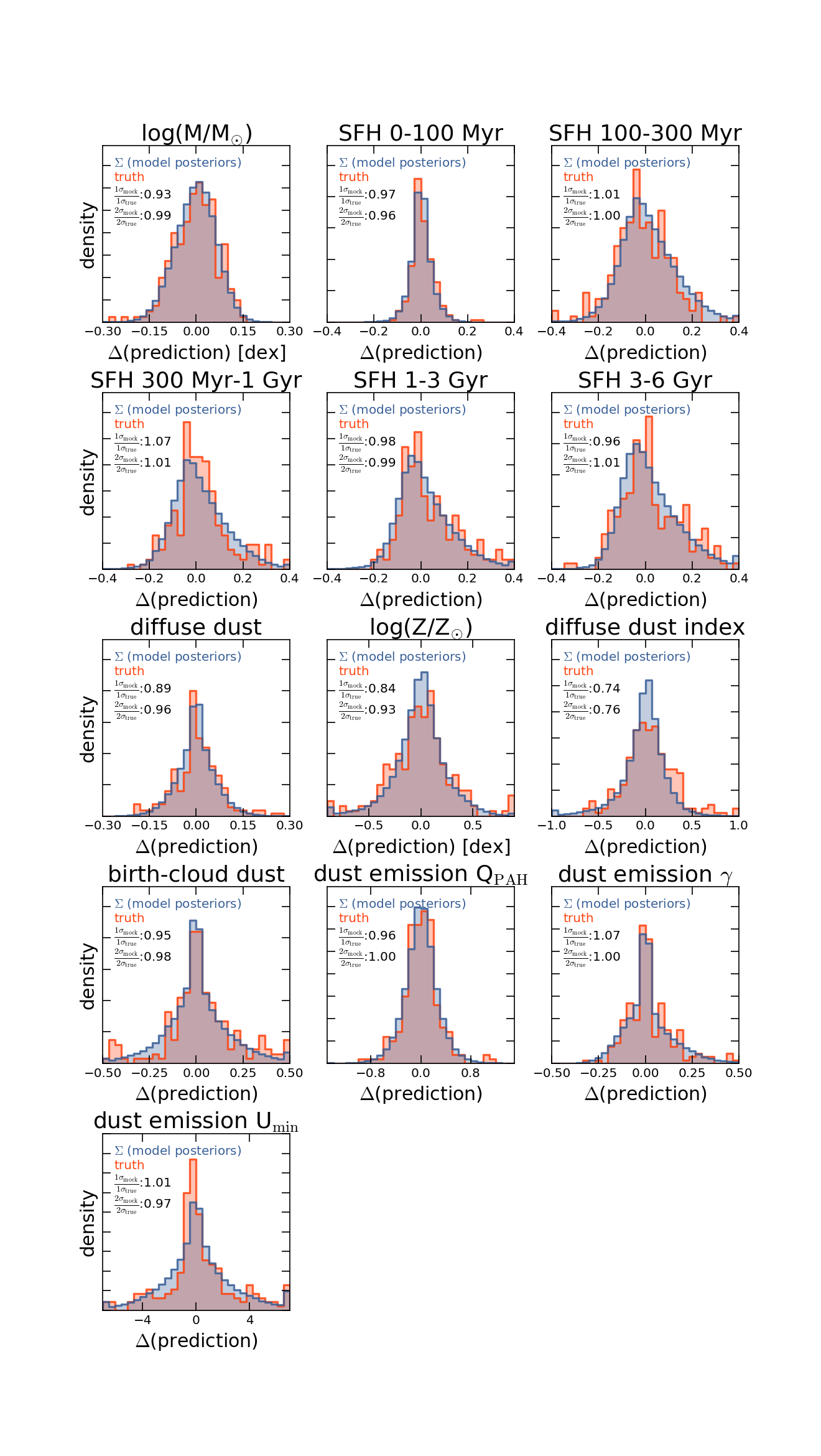}
\caption{The extent to which the model posterior probability functions accurately capture the true distribution of mock parameters. The red histograms show the location of the true mock galaxy parameters relative to the center of the model posterior. The blue histograms show the stacked model posterior probability functions. The error bars capture the model fit parameters to within 10\% for all fit parameters except for the stellar metallicity and the dust attenuation curve, where the error bars are incorrect by 15\% and 25\%, respectively. This is likely due to large-scale parameter multimodalities arising from the well-known dust-metallicity degeneracy (see Section \ref{section:posteriors}.}
\label{fig:mock_fitpdf}
\end{center}
\end{figure*}

\begin{figure*}[ht!]
\begin{center}
\includegraphics[width=0.9\linewidth]{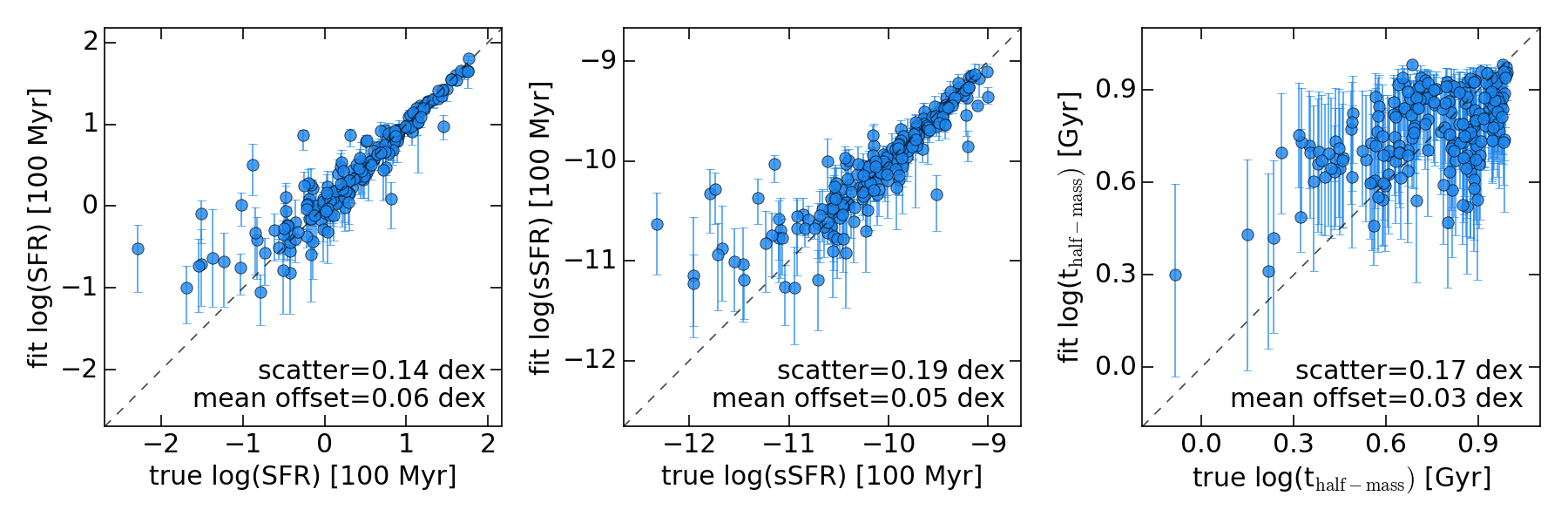}
\caption{Recovery rate for select physical properties of the mock galaxies. The x-axis shows the true physical properties of the mock galaxies, while the y-axis shows the marginalized outputs from the \prospector{} fits to the mock photometry. SFRs and sSFRs are well-recovered, while half-mass assembly times are more uncertain.}
\label{fig:mock_derpars}
\end{center}
\end{figure*}

In this section we generate and fit 200 mock galaxies with the \mname{} model, with the goal of investigating which model parameters can be accurately constrained with the available {Brown} {et~al.} (2014) photometry. 

To generate mock galaxy SEDs, free parameters from the \mname{} model are selected randomly within the model priors. The model priors are described in detail in Section \ref{section:features}. In order to produce galaxies with realistic amounts of dust, we allow the diffuse dust optical depth to only vary within $0.0 < \tau_{\mathrm{diffuse}} < 0.5$, and, for simplicity, we also constrain mass to be $10 <$ log(M/M$_{\odot}$) $< 11$. These changes are made during mock generation, but are not reflected in any adjustment of the priors during model fitting.

Photometric noise is simulated by perturbing the mock photometric fluxes. For each photometric band, the flux is adjusted by sampling from a Gaussian with a mean of zero and a width equal to 5\% of the flux value. The statistical properties of this noise are reported accurately to the sampler.

Figure \ref{fig:mock_fitpars} examines the ability of the fitter to recover the free model parameters for the mock galaxies. The stellar mass, 0-100 Myr SFH, diffuse dust content, stellar metallicity, and dust emission parameters are all well-constrained by the available photometry. The birth-cloud dust, however, is not uniquely determined from the mock photometry: this parameter has a subtle effect on the photometry even in star-forming galaxies (see Figure \ref{fig:model_diagram2}), and for quiescent galaxies with no recent star formation, will be completely unconstrained. The diffuse dust index, which controls the shape of the attenuation curve, is recovered with considerable scatter, though with no significant bias.

The SFH beyond 100 Myr is recovered modestly well: notably, the ability to accurately recover the SFH becomes increasingly poor with lookback time. This is consistent with the fact that older stellar populations have increasingly similar SED shapes, and also tend to be out-shone by younger generations of stars. These results are also consistent with mock tests of another 6-component nonparametric SFH implementation in {Zhang} {et~al.} (2012). Notably, fractional SFH parameters $\gtrsim$0.4 have poor recovery rates. This is in part due to the implicit Dirichlet prior from the fractional parameterization, which favors a constant SFH over a dominant contribution from a single SFH bin: see Section \ref{section:sfh} for an extensive discussion of this. This prior does not affect mass or SFR estimates significantly, as demonstrated by the mock tests shown here, but the SFH priors would be very relevant in any future studies attempting to recover the SFH at $\gtrsim$ 1 Gyr from broadband photometry.

Some model parameters will not be uniquely determined from the photometry alone, or may only be recoverable for particular star formation histories. As long as these parameters are assigned broad posteriors which accurately describe the constraining power of the data, then the fitter is functioning properly. Figure \ref{fig:mock_fitpdf} examines the distribution of true model parameters relative to the model posteriors. Similar to Figure \ref{fig:balmer_line_posterior}, we show the sum of the model posteriors for each model fit parameter. The posteriors are first shifted such that the median of each posterior is located at the origin. We also show a histogram of the truth values for each of these parameters. If the model posteriors are accurate, then the two histograms will match each other.

To quantify this test, we calculate the fraction of truths that fall within the posterior 1$\sigma$ range (i.e., the 16th and 84th percentiles of the posterior) and 2$\sigma$ range (the 2.5th and 97.5th percentiles of the posterior), shown in the upper-left corner of the panels in Figure \ref{fig:mock_fitpdf}. The posteriors are accurate to within $\sim$10\% for all model parameters except the stellar metallicity and the shape of the attenuation curve. The posteriors for stellar metallicity and the diffuse dust index often exhibit multimodality, which is challenging for the \emcee{} sampling algorithm to sample properly. The recovery of these parameters would likely be improved by using nested sampling algorithms such as MultiNest ({Feroz} \& {Hobson} 2008).

\begin{figure*}[ht!]
\begin{center}
\includegraphics[width=0.9\linewidth]{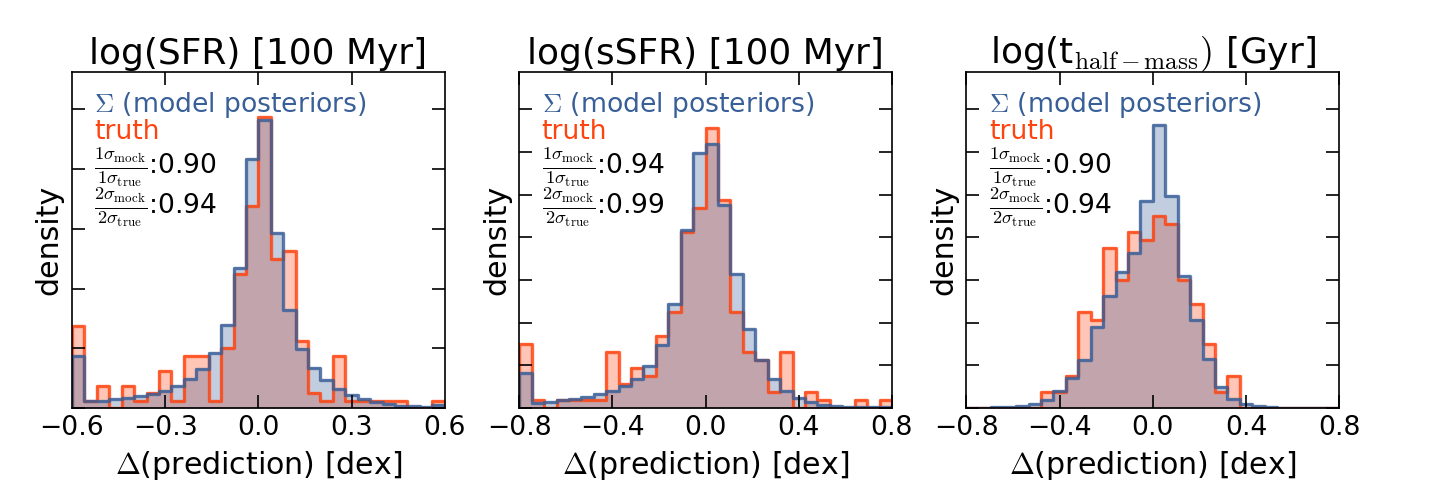}
\caption{The extent to which the model posterior probability functions accurately capture the true distribution of mock parameters. The red histograms show the location of the true mock galaxy parameters relative to the center of the model posterior. The blue histograms show the stacked model posterior probability functions. Since the fits to the mock photometry accurately describe the mock parameters, these two histograms overlap within Poisson errors. The error bars describe the distribution of mock values to within 10\% for SFR, sSFR, and half-mass time.}
\label{fig:mock_derpdf}
\end{center}
\end{figure*}

In addition to recovering the free model parameters, it is also interesting to evaluate the recovery of parameters derived from the SFH, such as SFR, sSFR, and the half- mass assembly time. The recovery rate for these derived parameters is shown in Figure \ref{fig:mock_derpars}, and a posterior test analogous to Figure \ref{fig:mock_fitpdf} is shown in Figure \ref{fig:mock_derpdf}.

SFR and sSFR are well-recovered by the mock tests, with the recovery rate increasing with increasing SFR / sSFR. This makes sense: the fraction of young stars is recovered with increasing fidelity as the young stars dominate more and more of the light. The half-mass time is poorly constrained from the available photometry, consistent with the modest recovery of the SFH parameters in Figure \ref{fig:mock_fitpars}. The posteriors for all three of these parameters are well-behaved, and the error bars are accurate to within 10\%.

\begin{figure}[ht!]
\begin{center}
\includegraphics[width=\linewidth]{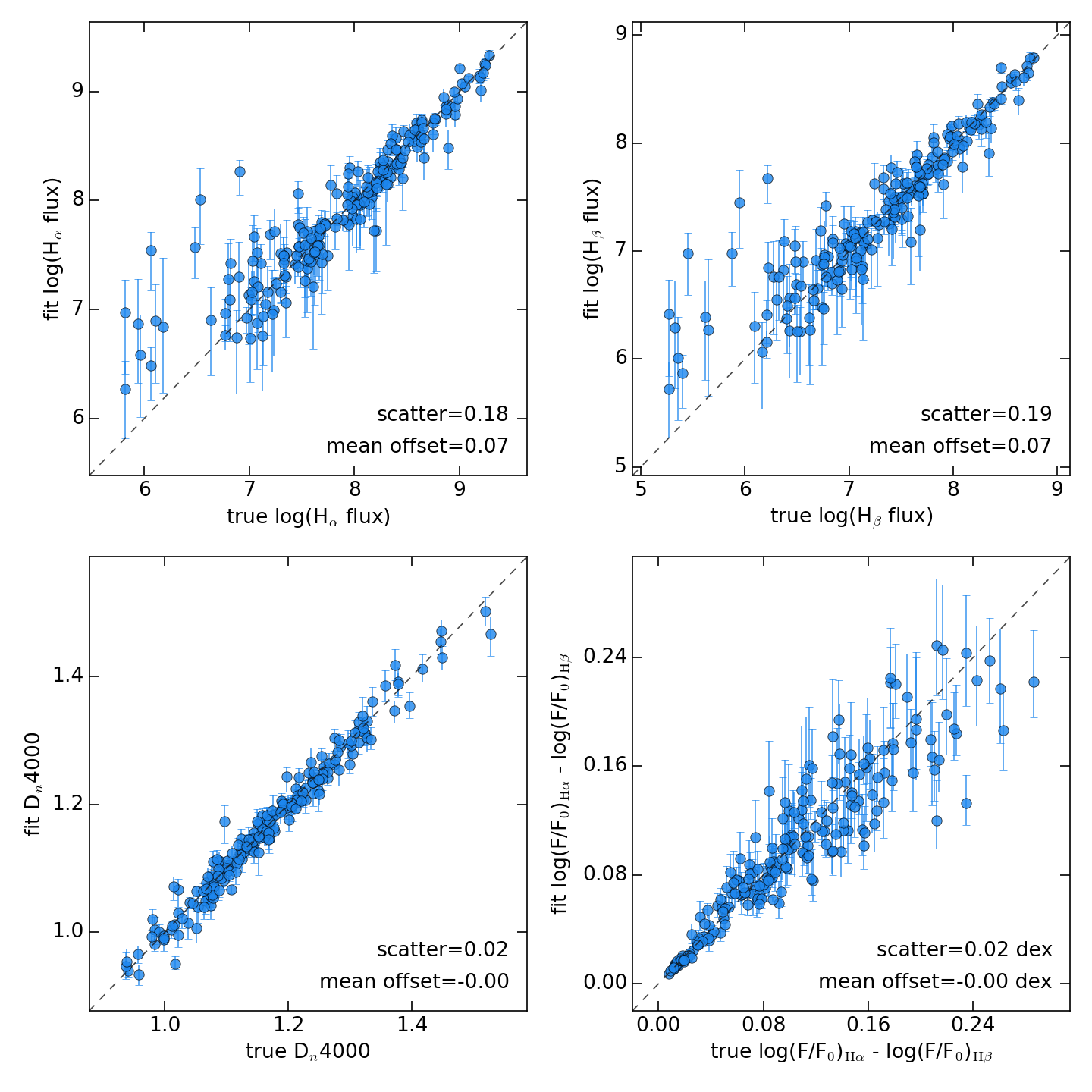}
\caption{Recovery rate for properties of the mock galaxies that are typically measured from spectra, including Balmer line luminosity, \dn{}, and the reddening curve. Here they are predicted from fits to the broadband photometry. The recovery rate for \halpha{} and \hbeta{} is slightly worse in the mocks than in the real data; this is because of the S/N limitations put on the observed \halpha{} luminosities in the real data, which select for star-forming galaxies where SFR is more well-determined.}
\label{fig:mock_specpars}
\end{center}
\end{figure}

Finally, in Figure \ref{fig:mock_specpars}, we demonstrate the capacity of the fitter to recover key spectral parameters (\halpha{} and \hbeta{} nebular emission fluxes, \dn{}, and the Balmer decrement) from fits to the photometry. The overall prediction of these spectroscopic features is of excellent quality. With 5\% photometric errors, the scatter in the \halpha{} and \hbeta{} nebular emission lines is similar to the observed scatter in the \halpha{} and \hbeta{} nebular emission (0.18 dex, Figure \ref{fig:balmer_lineflux}); both scatters are consistent with the scatter predicted from the width of the model posteriors. The observational S/N$>$5 cut is relevant to this comparison, as the ability of the model to predict \halpha{} nebular emission increases with increasing sSFR. The average sSFR of the mock galaxies is lower than the average sSFR of the {Brown} {et~al.} (2014) sample (see Section \ref{section:halpha}), so the measured scatter in mock nebular emission flux predictions is an upper limit to the scatter in a truly analogous sample. The recovery of \dn{} is slightly improved relative to the observations, with a decrease in scatter by 0.05. This is expected, as \dn{} is sensitive to the SFH. The SFH parameters are drawn directly from the SFH priors in the mock tests, whereas this is not guaranteed for observed galaxies. The Balmer decrement, parameterized by the magnitude of the differential extinction between \halpha{} and \hbeta{} wavelengths, is better recovered in the mock tests, with a scatter of 0.02 dex compared to 0.08 dex in the observations. This is likely because the dust geometry in real galaxies is considerably more complex than our model approximation.

\section{Comparison to \magphys{}}
\label{appendix:magphys}
\setcounter{figure}{0}    

\begin{figure*}[ht!]
\begin{center}
\includegraphics[width=0.95\linewidth]{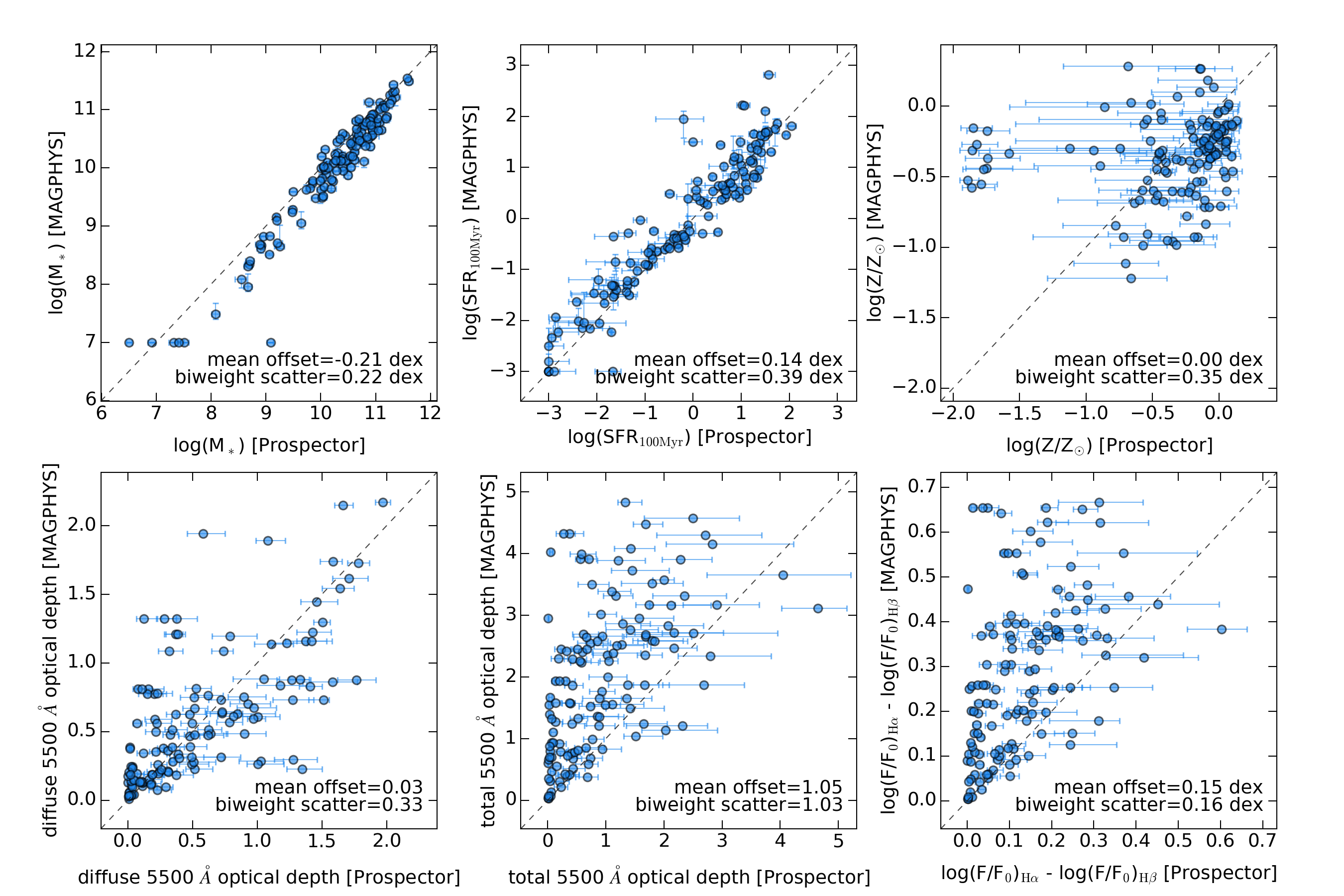}
\caption{Comparison of derived physical parameters between \magphys{} and \mname{}. The top panels, from left to right, compares the stellar masses, the star formation rates averaged over the past 100 Myr, and the stellar metallicities. The bottom panels, from left to right, show the optical depth of the diffuse dust component at 5500 $\angstrom{}$, the optical depth of the total (birth-cloud + diffuse) dust model at 5500 $\angstrom{}$, and the reddening between \halpha{} and \hbeta{} wavelengths, as calculated from the Balmer decrement.}
\label{fig:magprosp_parameters}
\end{center}
\end{figure*}

\begin{figure*}[ht!]
\begin{center}
\includegraphics[width=0.45\linewidth]{empirical_halpha_prosp.png}
\includegraphics[width=0.45\linewidth]{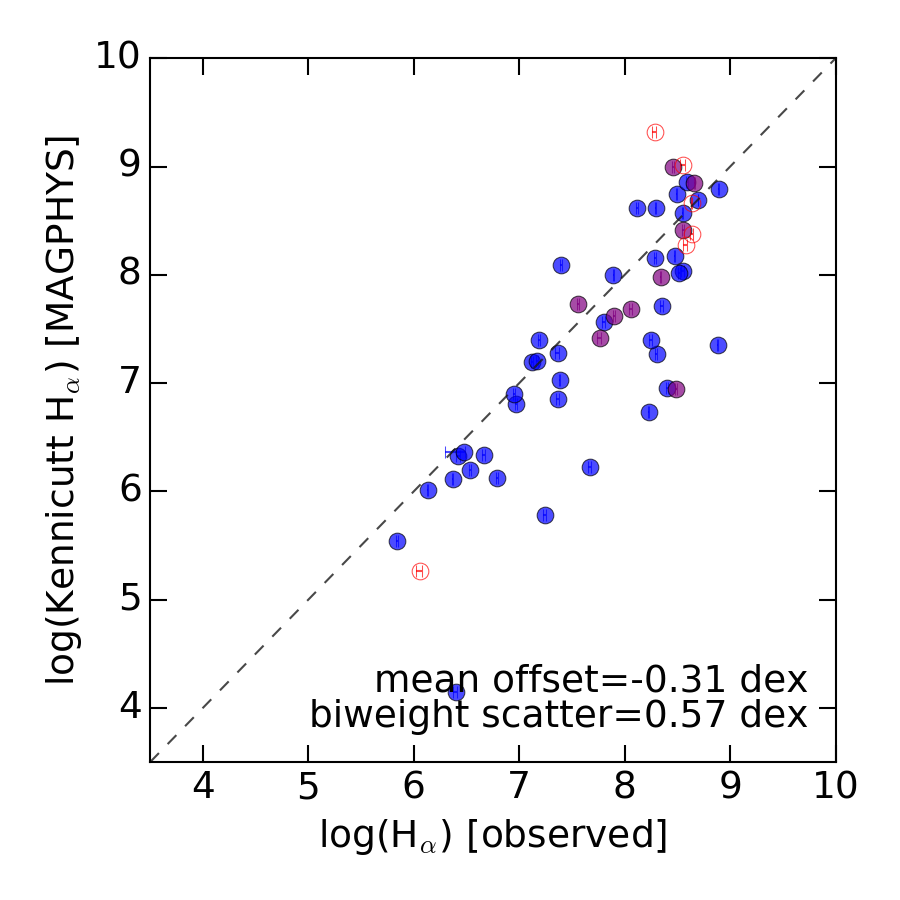}
\caption{\halpha{} luminosities in L$_{\odot}$ measured from the aperture-matched spectra compared to model \halpha{} luminosities predicted from fits to the broadband photometry. The model \halpha{} luminosity is calculated from the model SFR using Equation \ref{ks_eqn}, and attenuated by the model dust properties. The left panel shows \mname{}, the right panel shows \magphys{}. Galaxies are color-coded by their BPT classification. Only galaxies with S/N (\halpha{}, \hbeta{}) $> 5$ are shown. The left panel is a reproduction of the left panel in Figure \ref{fig:ks_halpha} for the convenience of the reader.}
\label{fig:magprosp_halpha}
\end{center}
\end{figure*}

\begin{figure*}[ht!]
\begin{center}
\includegraphics[width=0.9\linewidth]{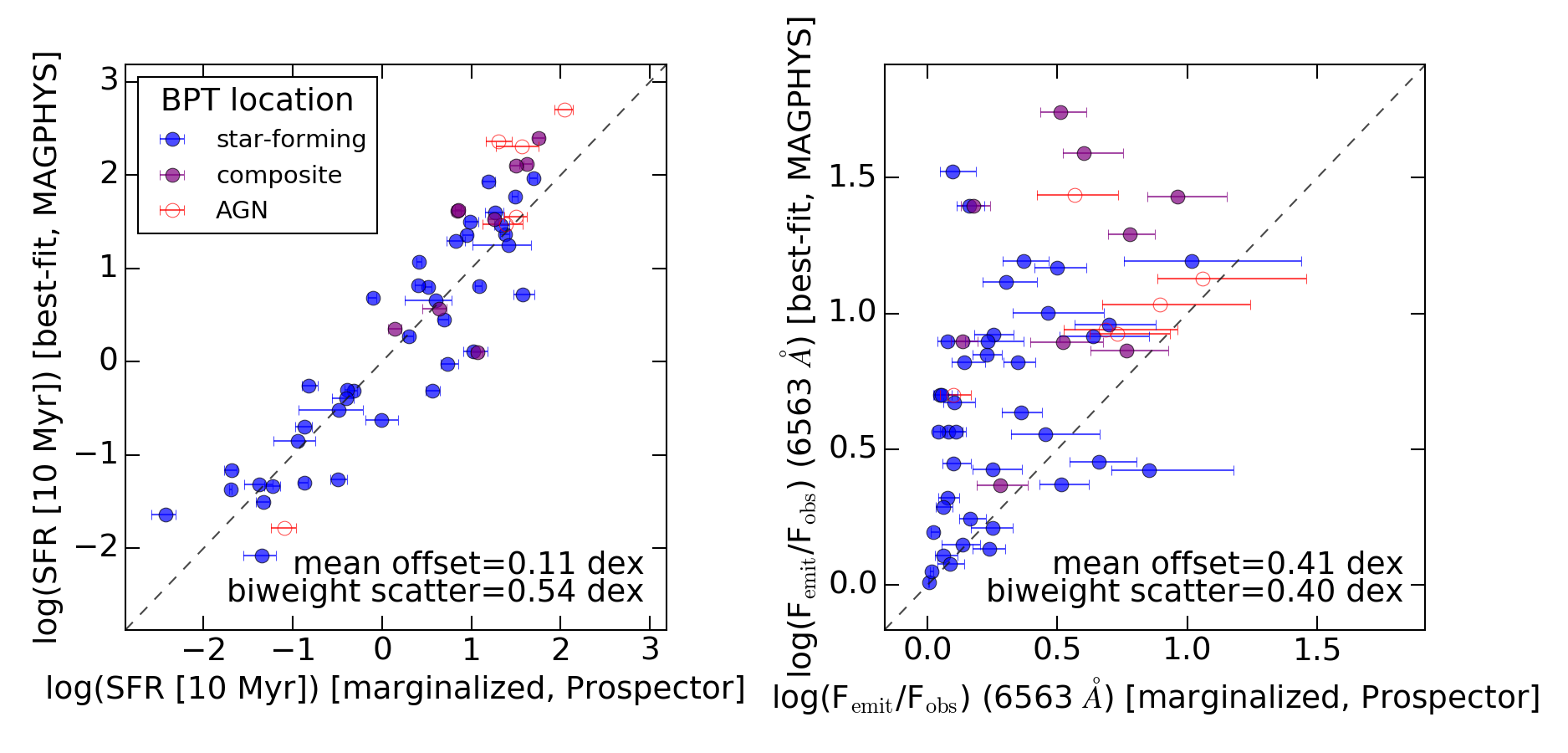}
\caption{Inputs to the empirical \halpha{} calculation from both \magphys{} and \mname{}. There is considerable scatter in the star formation rates over 10 Myr timescales, 0.2 dex more than the scatter in SFRs over 100 Myr timescales (Figure \ref{fig:magprosp_parameters}). This may be due to the random bursts superimposed on the \magphys{} SFHs. There is both scatter and a large offset in the dust attenuation at \halpha{} wavelengths, such that \magphys{} predicts much more dust attenuation. Galaxies are color-coded by their BPT classification. Only galaxies with S/N (\halpha{}, \hbeta{}) $> 5$ are shown.}
\label{fig:magprosp_halpha_input}
\end{center}
\end{figure*}

In this section, the results of running the full-SED fitting code \magphys{} ({da Cunha} {et~al.} 2008) on the {Brown} {et~al.} (2014) galaxy catalog are described. \magphys{} is chosen as it is similar to \mname{} in a number of ways: it has a similar number of free parameters, it simultaneously models the dust and stellar emission from galaxies, and it has a Bayesian implementation. In this section we compare basic physical parameters derived from the \magphys{} fits to those from \mname{}, contrast the ability of \mname{} and \magphys{} to predict \halpha{} fluxes using the {Kennicutt} (1998) prescription, and show how both model SFRs and dust attenuations contribute to the difference in \halpha{} predictions between \magphys{} and \mname{}. We note that the \magphys{} results we show in this section are very similar to the \magphys{} fits published in {Brown} {et~al.} (2014).

In brief, \magphys{} fits {Bruzual} \& {Charlot} (2003) stellar populations models to the observed SEDs of galaxies. The \magphys{} stellar templates are generated with exponentially declining star formation histories with random bursts superimposed, and a variable stellar metallicity. \magphys{} uses a two-component {Charlot} \& {Fall} (2000) dust attenuation model, with both the birth-cloud and diffuse dust components as free parameters. It employs energy balance to re-emit the energy absorbed by the dust as thermal radiation in the infrared. The shape of the infrared SED is controlled by four free parameters in the \magphys{} model, described in detail by {da Cunha} {et~al.} (2008). A Chabrier (2003) IMF is the default in \magphys{}, which is the same as used in \mname{}.

Particularly relevant physical differences between the two codes include the SFH implementation (nonparametric SFHs in \mname{}, compared to parametric SFHs with random bursts superimposed in \magphys{}), the dust attenuation curve (a free parameter in \mname{}, and a fixed parameter in \magphys{}), the parameterization of the SED from dust emission, the inclusion of nebular emission lines in the model broadband photometry (\cloudy{} models in \mname{}, but not modeled in \magphys{}). The model priors also have substantial differences: for example, \magphys{} has a template library of generated SFHs, which imposes some fixed prior on the SFH fits, while \prospector{} generates stellar populations on-the-fly and can freely explore parameter space, at the cost of increased runtime. 

Figure \ref{fig:magprosp_parameters} compares important physical parameters derived from the SED fits between \magphys{} and \mname{}. We plot the marginalized \magphys{} parameters with errors where they are available (mass, SFR); otherwise, we plot the best-fit values. On average, \magphys{} predicts that galaxies are considerably less massive (0.25 dex), with slightly higher star formation rates (0.1 dex). Despite these galaxies having excellent photometric wavelength coverage and high S/N photometry, there is a biweight scatter of a factor of 2.5 in derived star formation rates between \magphys{} and \mname{}. There is little agreement in the photometric metallicities between the two codes. On average, the optical depth of the diffuse dust components agree reasonably well, though with a substantial amount of scatter. Finally, \magphys{} assigns considerably more dust attenuation towards HII regions, attenuating the flux from young stars by an extra a factor of e$^{1.11}$ = 3.1, and producing on average 0.15 dex of additional reddening between \halpha{} and \hbeta{} wavelengths.

To compare the recovery of \halpha{}, we calculate the Kennicutt \halpha{} luminosity using Equation \ref{ks_eqn}. We adopt the average star formation rate over the past 10 Myr from each model as the SFR. We apply the model dust attenuation to the resulting \halpha{} luminosity, and compare the attenuated model \halpha{} luminosity to the observations. The results are shown in Figure \ref{fig:magprosp_halpha}. Overall, \magphys{} systematically underpredicts the \halpha{} fluxes by 0.31 dex, with a scatter of 0.54 dex.

There are systematic trends in the \magphys{} comparison, including a "parallel" set of galaxies offset from the 1:1 relationship by $\sim$ 1 dex. We examine the source of systematics in Figure \ref{fig:magprosp_halpha_input} by looking at the inputs to the \halpha{} luminosity calculation from each code. The majority of the scatter between the two models comes from the difference in model SFRs over the last 10 Myr. The offset between the predictions comes from the much larger dust attenuation towards HII regions predicted by the \magphys{} fit. The \magphys{} reddening model also systematically overpredicts the observed Balmer decrements (not shown), confirming that \magphys{} predicts too much dust attenuation in these galaxies.

The difference in model dust attenuations is likely due to the lack of a limit on the ratio of birth-cloud dust to diffuse dust in \magphys{}. The birth-cloud dust has a very small effect on the SED and is thus poorly constrained by photometry alone (see Figure \ref{fig:model_diagram2}). The dominant effect of the birth-cloud dust on the \halpha{} attenuation in the \magphys{} model can be inferred by comparing the difference between the total optical depth and the diffuse optical depth in Figure \ref{fig:magprosp_parameters}.

The origin of the difference in model star formation rates is likely the different SFH prescriptions. \magphys{} imposes a set of random bursts on top of a smoothly varying SFH, so that SFR(10 Myr) can be significantly different than SFR(100 Myr). In contrast, \mname{} enforces a constant SFH over the previous 100 Myr. While both SFH recipes reproduce the observed SED to similar levels of accuracy (as measured by the $\chi^2$ in the fit to the photometry), they predict substantial differences in SFR indicators which probe different timescales. This is relevant when trying to predict the \halpha{} fluxes ($\sim$ 5 Myr timescales) with photometric UV fluxes ($\sim$ tens of Myr) and IR fluxes ($\sim$ hundreds of Myr). This conclusion is in agreement with {Smith} \& {Hayward} (2015), which found that the best-fit \magphys{} SFH is a poor approximation of simulation SFHs. The agreement in {Smith} \& {Hayward} (2015) improves dramatically when the public version of \magphys{} is modified to marginalize over all of the library SFHs, which mitigates the effect of the random bursts.

We conclude by suggesting that the best way to resolve the substantial difference in predicted physical properties between SED fitters is with further posterior checks against spectroscopic features. This will lead to an across-the-board improvement in the recipes and prescriptions in SED models, and consequently, in the accuracy of their predicted physical properties.

\section{Effect of the \herschel{} Photometry}
\label{section:herschel}
\setcounter{figure}{0}    

\begin{figure*}[ht!]
\begin{center}
\includegraphics[width=\linewidth]{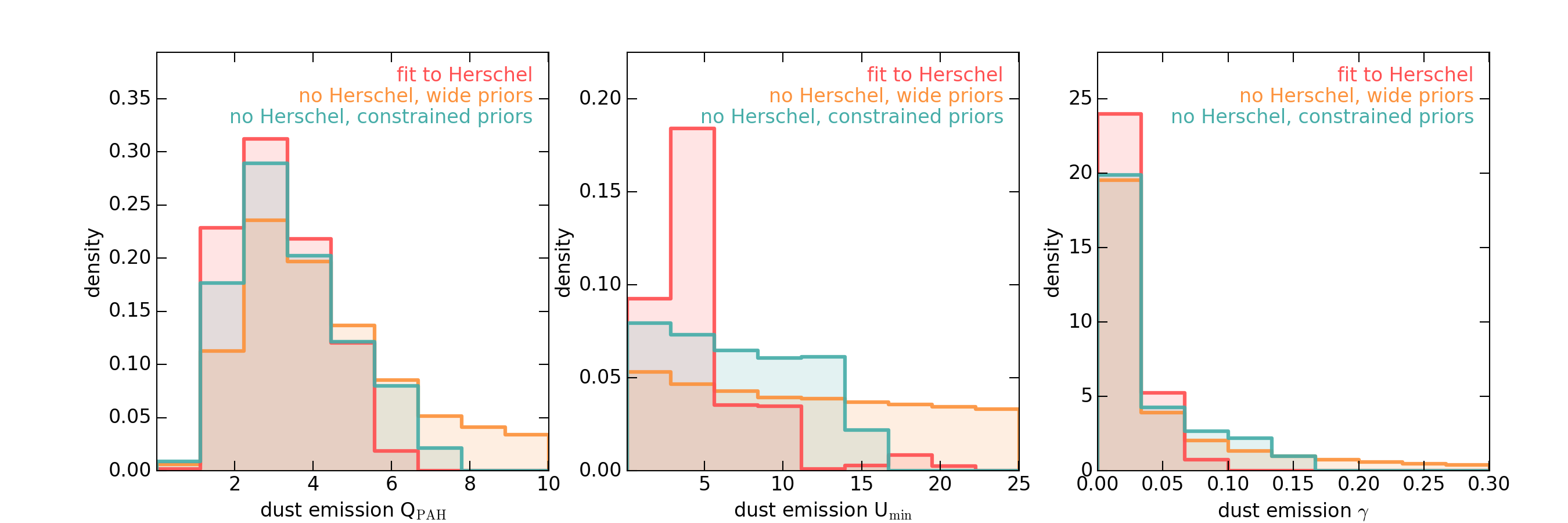}
\caption{Stacked posteriors for the dust emission parameters for all of the galaxies with FIR photometry. The left panel shows the posteriors for \qpah{}, the middle panel for \umin{}, and the right panel for \gammae{}. The posteriors from fits including the \herschel{} photometry are shown in red. These are fit with flat, wide priors on the dust emission parameters, described in the text in Appendix \ref{section:herschel}. The posteriors from fits without the \herschel{} photometry with the same wide priors are shown in orange, while fits without the \herschel{} photometry but with the narrower priors described in Section \ref{section:dustemission} are shown in blue. The narrower priors are chosen based on the range of IR parameters from the fits to the \herschel{} fluxes, and are consistent with the best-fit parameters from {Draine} {et~al.} (2007). When the fits without \herschel{} photometry are supplied with tighter priors, the posteriors more closely resemble the fits that include the \herschel{} photometry.}
\label{fig:duste_posteriors}
\end{center}
\end{figure*}

\begin{figure*}[h!]
\begin{center}
\includegraphics[width=0.8\linewidth]{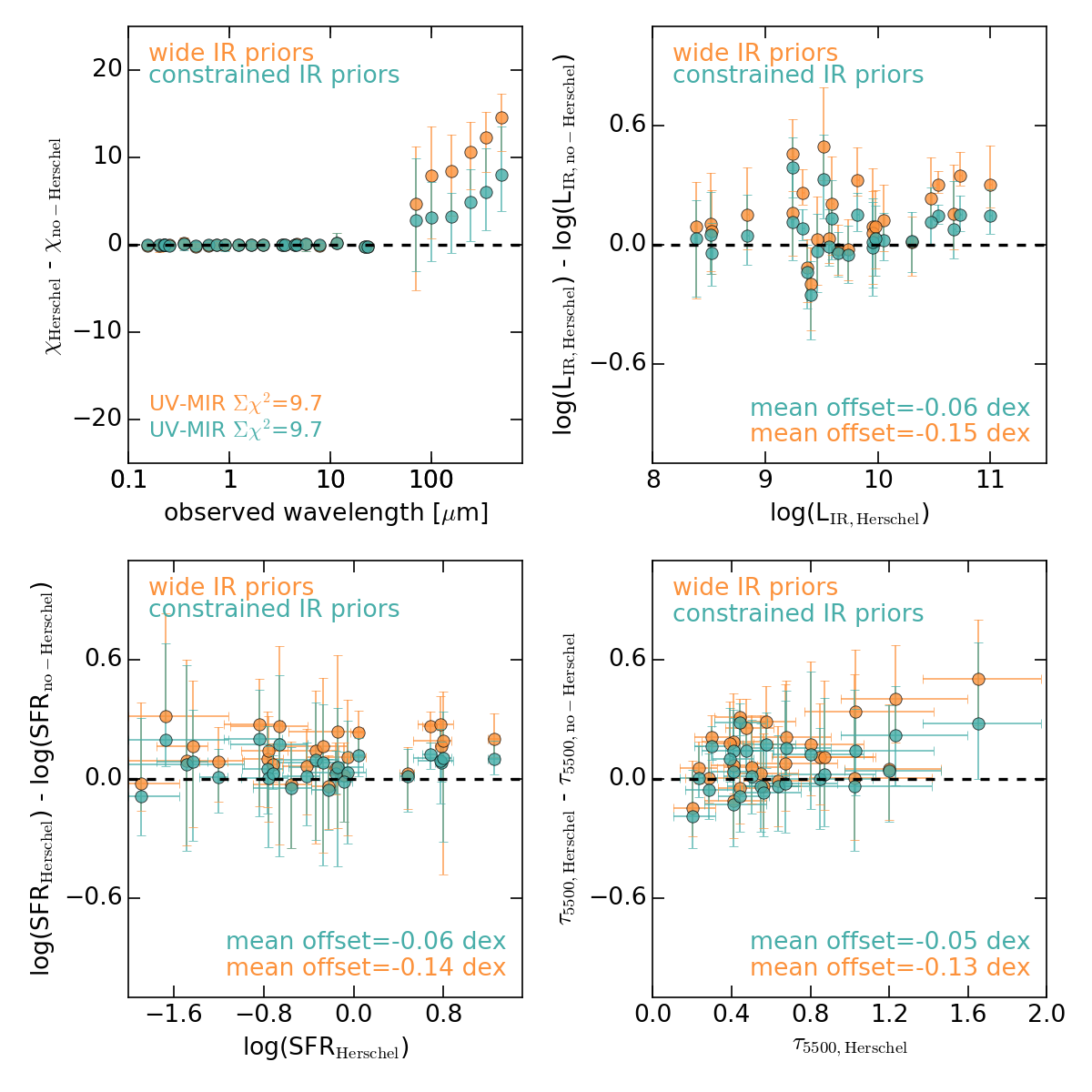}
\caption{The effect of applying appropriate priors to the IR emission parameters. The upper-left panel shows the difference in median $\chi$ in each photometric band between the models with tight IR priors, and those without. The remaining panels, starting clockwise from the upper-right, show the same effect on the total model \lir{} (8-1000 $\mu$m), the total dust optical depth at 5500 $\angstrom{}$, and the SFR averaged over 100 Myr. Using informed priors on the IR emission parameters removes the bias in the SFR, the optical depth of the dust, and the SFR. A corollary is that \lir{} cannot be uniquely determined from the UV-MIR SED to within $\sim$40\%, but tighter priors on the shape of the FIR emission can improve the recovery of \lir{} considerably.}
\label{fig:herschel_comparison}
\end{center}
\end{figure*}

In the \mname{} model, dust emission is tied to the UV-NIR SED by assuming energy balance: all energy attenuated by dust is re-radiated at infrared wavelengths. In principle, this means \lir{} can be constrained from fitting the UV-NIR SED alone. This is a crucial point, as accurate determination of \lir{} is necessary for measuring accurate total star formation rates and dust attenuations.

We investigate the extent to which \lir{} can be recovered from the UV-MIR SED. First, more permissive priors on the dust emission parameters are adopted. This prevents \lir{} from being constrained solely from a combination of the MIR fluxes and a fixed shape to the IR SED; instead, it must be determined from energy balance in the UV-NIR SED and the MIR fluxes. The full parameter space of the {Draine} \& {Li} (2007) dust emission model in \fsps{} is used: flat priors over $0.1 <$ \umin{} $<25$, $0.0 <$ \gammae{} $<1.0$, and $0.1<$ \qpah{} $<10.0$. Then, the 26 galaxies that are covered by \herschel{} PACS and SPIRE imaging are fit with the \mname{} model. We fit them in three different ways:
\begin{enumerate}
\item Including the FIR photometry, with wide IR priors.
\item Masking the FIR photometry, with wide IR priors.
\item Masking the FIR photometry, with constrained IR priors (described in Section \ref{section:dustemission}).
\end{enumerate}
In Figure \ref{fig:duste_posteriors}, the posteriors for the dust emission parameters are from each of the above fits are stacked. The fits to the \herschel{} fluxes give a tight constraint on the dust emission parameters. Fits without the \herschel{} fluxes and with broad priors have broader posteriors on the dust emission parameters, though \gammae{} and \qpah{} are constrained to some extent by the MIR photometry. Fits without the \herschel{} fluxes but with tight priors have posteriors that more closely resemble the posteriors of the \herschel{} fit.

The implications of these different dust emission posteriors are explored in Figure \ref{fig:herschel_comparison}. This figure shows the median $\chi$ in each band when the FIR photometry is not fit, for the two different sets of priors. The error floor on each photometric band is 5\%, so a $\chi$ of 10 represents a difference of 50\%. The tight priors reproduce the FIR fluxes with less offset than the broad priors, but the UV-MIR SED is almost equally well reproduced with either.

The other three panels compare the \lir{}, SFR, and the optical depth of the dust attenuation at 5500 \angstrom{} between the fits with FIR photometry, and those without. When fitting without the FIR photometry and with broad priors, the \lir{} and SFR are biased low by $\sim$ 0.15 dex, and the dust attenuation by $\sim$ 0.15 in optical depth. When fitting with tighter priors, this offset largely disappears. This has fascinating implications. The first is that the UV-MIR SED alone cannot uniquely determine \lir{}. This is clear from the upper-left panel of Figure \ref{fig:herschel_comparison}, where the UV-MIR is almost equally well fit regardless of the FIR priors, but the implied FIR fluxes have a large variance. ItÕs also visible in the upper-right panel, where the fitter assigns large error bars on \lir{} when the FIR fluxes are not fit, and very small \lir{} error bars when they are fit. This is likely related to the dust energy problem ({Bianchi}, {Davies}, \& {Alton} 2000; {Popescu} {et~al.} 2000; {Misiriotis} {et~al.} 2001; {Baes} {et~al.} 2010; {Holwerda} {et~al.} 2012; {Mosenkov} {et~al.} 2016), where far-infrared fluxes can be underpredicted by factors of 3-4 when fitting only the UV-NIR emission.

The second implication is that, given that \lir{} cannot be uniquely determine without FIR fluxes, it is critical to have tight priors on the shape of the FIR emission; otherwise, the large uncertainty on the range of possible FIR shapes is translated into a large uncertainty and, in this case, a bias in \lir{}, SFR, and dust attenuation. This may be particularly relevant when performing full-SED modeling in the case of having limited MIR and no FIR photometry, as is often true at higher redshifts.

Given these results, we have adopted the tighter priors on the dust emission parameters in our fiducial model. These tight priors are also better descriptors of the fits to the SINGS sample in {Draine} {et~al.} (2007). One potential caveat is that the \herschel{} photometry in the Brown et al. (2014) sample comes from overlap with the KINGFISH survey ({Kennicutt} {et~al.} 2011), and that the galaxies in this overlap may not be representative of the whole {Brown} {et~al.} (2014) sample.


\end{document}